\documentclass[12pt]{article}%
\usepackage{amsfonts}
\usepackage{amssymb}
\usepackage{graphicx}
\usepackage{amsmath}
\usepackage{harvard}
\usepackage{environ}
\usepackage{harvard}%
\providecommand{\U}[1]{\protect\rule{.1in}{.1in}}
\newtheorem{theorem}{Theorem}

\newtheorem{corollary}{Corollary}

\newtheorem{lemma}{Lemma}

\newtheorem{proposition}{Proposition}

\newenvironment{proof}[1][Proof]{\textbf{#1.} }{\ \rule{0.5em}{0.5em}}

\renewcommand{\cite}{\citeasnoun}
\citationmode{abbr}
\newif\ifhideproofs
\ifhideproofs
\NewEnviron{hide}{}

\fi
\begin{document}

\author{Dirk Bergemann\thanks{Yale University, dirk.bergemann@yale.edu.}
\and Tibor Heumann\thanks{Pontificia Universidad Cat\'{o}lica de Chile,
tibor.heumann@uc.cl.}
\and Stephen Morris\thanks{Massachusetts Institute of Technology, semorris@mit.edu}}
\title{Screening with Persuasion\thanks{We acknowledge financial support from NSF
grants SES-2001208 and SES-2049744, and ANID Fondecyt 1241302. We thank
seminar audiences at Boston College, Brown University, Emory University, MIT,
NBER\ Market Design, Pontificia Universidad Cat\'{o}lica de Chile, University
of Chile, the Warwick Economic Theory Workshop and the UCL memorial conference
for Konrad Mierendorff for valuable suggestions, and, in particular, Mark
Armstrong for his discussion at the latter conference. \ We benefited from
discussions with Ian Ball, Roberto Corrao, Stefano DellaVigna, Piotr Dworczak,
Fedor Sandomirskiy, Matt Gentzkow, Anton Kolotolin, Ellen Muir, Leon Musolff,
Christopher Sandmann, Philipp Strack, and Alex Wolitzky. We thank Pedro Feijo
de Moraes, Jack Hirsch, Hongcheng Li, Shawn Thacker, David Wambach, Michael
Wang and Nick Wu for excellent research assistance.} }
\date{\today }
\maketitle

\begin{abstract}
We analyze a nonlinear pricing model where the seller controls both product
pricing (screening) and buyer information about their own values (persuasion).

We prove that the optimal mechanism always consists of finitely many signals
and items, even with a continuum of buyer values. The seller optimally pools
buyer values and reduces product variety to minimize informational rents.

We show that value pooling is optimal even for finite value distributions if
their entropy exceeds a critical threshold. We also provide sufficient
conditions under which the optimal menu restricts offering to a single item.

\medskip

\noindent\textsc{Jel Classification: }D44, D47, D83, D84.


\end{abstract}

\newpage

\section{Introduction\label{sec:intro}}

\subsection{Motivation\label{subsec:mot}}

In the digital age, sellers often possess superior information about the match
between products and buyers. This informational advantage, coupled with the
ability to offer a variety of products, presents a novel challenge in the
design of optimal pricing strategies. How should a seller leverage their
informational advantage while determining the optimal variety of products to
offer? \ With fewer products offered the seller can reveal less information to
support the choice behavior of the buyer; conversely, more variety requires
additional information be provided to all the buyers to distinguish between
the product choices.

We analyze the interaction between the menu of items offered and the
information provided to the buyers in the classic nonlinear pricing
environment of \cite{muro78}. In our framework, the seller chooses the menu of
products and prices \emph{and} controls the information that buyers receive
about their own willingness-to-pay. This setting reflects key features of the
digital economy: $(i)$ sellers' informational advantage, $(ii)$\ the use of
recommendation systems, and $\left(  iii\right)  $\ the prevalence of menu
pricing over personalized pricing.\ 

We go beyond the setting of \cite{muro78} by assuming that buyers initially
only know the prior distribution of their values. The seller chooses how much
information the buyers receive about their willingness-to-pay through the
choice of an information structure. We refer to the prior distribution of
values therefore as the \emph{latent distribution} as the buyers only observe
the expected values through the signals. The seller's problem consists of
designing a public menu of products, prices and the information that buyers
receive about their value. Thus the seller screens by offering
quality-differentiated products while engaging in Bayesian persuasion to
influence buyers' decisions--\emph{screening with persuasion}. We maintain the
assumption that the seller cannot engage in personalized price discrimination
based on the information revealed about their values. \ We expand on the
interpretation of this assumption after we introduce the model.

%We analyze the interaction between the menue of items offered and the information provided to the buyers in the classic nonlinear pricing environment of \cite{muro78}. The\ buyers differ in their willingness-to-pay for quality and the seller offers many distinct products to the buyers.\ The seller cannot observe buyers' willingness-to-pay for quality. \ But the seller can screen buyers by offering all buyers a menu of products that are differentiated by their quality and price. We go beyond the setting of \cite{muro78} by allowing the seller to control how much information the buyers can obtain about their willingness-to-pay for the quality. \ Thus the seller screens while engaging in Bayesian persuasion.

Our analysis yields several striking results. First, we prove that the optimal
mechanism always consists of a finite menu, even when the underlying
distribution of buyer values is continuous. This finding contrasts sharply
with the continuous menus often found in classical screening models. The
finiteness of the optimal menu arises from the seller's ability to pool buyer
values, which reduces information rents more effectively than the traditional
approach of distorting allocations. By pooling, we refer to any information
structure that provides less than complete information to the buyer about
their value. This encompasses any deterministic or stochastic information
structure that sends the same signal for at least two distinct values.

Second, we establish sufficient conditions for the optimality of pooling.
Specifically, we show that pooling is beneficial whenever the entropy of the
value distribution exceeds log$_{2}\left(  9\right)  \ $bits. This result
provides a clear, quantifiable threshold for when information design becomes
crucial in pricing strategies.

Third, we offer sufficient conditions when a single-item menu is optimal and
hence a binary signal that supports the binary choice of the buyer--to buy or
not buy the single item on the menu. The sufficient conditions require modest
tails of the value distribution and convexity of the marginal cost of quality.

These results together advance our theoretical understanding of nonlinear
pricing and also offer practical insights for firms operating in
information-rich environments. They suggest that sellers can often improve
profits by strategically determining the information available to buyers and
offering short, that is finite menus.

The seller reduces the granularity of the buyer's information by pooling the
buyers' values into a small set of signals in the optimal mechanism, thereby
compressing the information buyers receive. \ The central trade-off is between
a suppression of the information rent that is enabled by a menu with a small
variety of products and efficiency gains from a menu with a large variety.
This trade-off is established by a variational argument that ultimately leads
us to establish the optimality of a finite menu even when the underlying
values form a continuum. \ The basic argument proceeds as follows. Suppose
that the optimal mechanism were to display an interval of values in the
support of the information structure. \ We ask what happens to profit if we
pool the allocations associated with a small interval of values in that larger
interval. By construction, distortions in the allocation from the candidate
profit-maximizing allocation only cause second-order distortions to the
virtual surplus. But if we additionally pool a small interval of values into a
single \emph{expected value, }then this causes a first-order decrease in the
information rents. Thus, locally a suppression of variety that is supported by
a reduction in information leads to an increase in profit. \ Hence, screening
across a small open set of values is never optimal. This leads to the first
main result, Theorem \ref{prop:int}, which states that the optimal mechanism
only generates a finite number of signals and a finite number of items. In
particular, the number of signals received exceeds the number of items by at
most one, to account for a signal to those buyers who optimally do not
purchase any item.

We first illustrate the trade-off between the variety of the menu and the
suppression of private information in an environment with finitely many values
and finitely many given qualities. Proposition \ref{edacc} provides a
sufficient condition when pooling the lowest two values generates a profit
improvement relative to complete disclosure. The local argument contains the
critical ingredients for the general argument and thus provides a preview of
the logic of Theorem \ref{prop:int}. The sufficient condition also enables us
to produce an upper bound on the entropy under which complete disclosure can
be optimal (Corollary \ref{dccs}), which previews the more general result in
Theorem \ref{thentrop}.

We make no assumption about the support of the distribution of buyers'
willingness-to-pay. \ In particular, it could be finite, countably infinite or
- as in the classic treatment of \cite{muro78} - a continuum. We first
establish that the optimal mechanism is discrete (Proposition \ref{prop:opi}),
thus has at most a countable number of items and signals. We then establish
that the quality increments in an optimal discrete menu must be increasing, so
that item qualities offered in the menu increase in a convex manner
(Proposition \ref{prop:id}). \ This help us to establish that the optimal menu
always consists of a finite set of items. \ In the optimal information
structure, the cardinality of items sold is always equal to the cardinality of
buyers' expected values (or one less if there is exclusion). This result
establishes that substantial pooling arises: whatever the true variety of
willingness-to-pay in the population, it is optimal to offer a finite menu and
offer only coarse information to buyers about their willingness-to-pay.

The results so far have not addressed the nature of the pooling of buyers'
values: we have argued that there will be a finite set of possible expected
values under the optimal information structure, but have not established which
values get pooled together. \ A familiar argument establishes that information
is optimally provided by a monotone partition (Corollary \ref{nonpol1}). We
take a given, and possibly pooled distribution of qualities and observe that
the information design problem is linear in the values. Thus, a maximum is
attained at an extreme point of the set of feasible distributions as
established by \cite{klms21}. \ This characterization follows after our main
results and highlights that the monotone partitional structure does not play a
critical role for our main results, Theorem \ref{prop:int} and \ref{thentrop}.

We further provide sufficient conditions for pooling to be optimal even if
there are only a finite set of values to begin with. \ We consider a setting
where buyers' willingness-to-pay are distributed on a finite grid (with equal
distances between values). \ One measure of the variation in the distribution
of values in this setting is the entropy of the distribution. \ We show that
if the entropy of the distribution is above $\log_{2}9$ bits, then the pooling
of values must be optimal (Theorem \ref{thentrop}). \ Thus if the distribution
of values has more variation than a uniform distribution on nine values, then
there must be some pooling in the optimal mechanism. This result is obtained
by constructing a best case distribution where no pooling is optimal and
establishing that its entropy has to be below $\log_{2} 9$. \ 

Our results so far are obtained for a large class of cost functions for
producing quality-differentiated products. \ In particular, we only assume
that the cost of producing a distribution of qualities is increasing with
respect to the increasing convex order. \ Thus the cost of a distribution of
qualities increases if the distribution of qualities increases or becomes more
variable. \ This class of cost functions includes standard separable cost
functions as in \cite{muro78}, where there is a convex cost of producing a
given quality and the cost of a distribution of qualities is just the sum of
the costs of individual qualities, thus \textquotedblleft variable
inventory\textquotedblright. \ But it also includes the \textquotedblleft
fixed inventory\textquotedblright\ case of \cite{lomu22}, where the seller has
an inventory of products that he can sell at zero cost but where any
distribution not covered by the fixed inventory is infeasible (e.g., has
infinite cost). \ Thus the results are insensitive to the nature of the cost
function and do not rely on any cost savings from changing the inventory of
products sold. \ 

However, to obtain further results on the number of items in the optimal menu,
we must appeal to properties of the cost function. \ We report a sufficient
condition for when a single item menu is optimal with variable inventory
(Theorem \ref{th:22}). \ In this case, the seller sells only one quality and
divides buyers only into those who buy and those who are excluded. The
sufficient condition is the convexity of the marginal cost function\ as well
as a modest tails condition on the distribution of values.

Our setting reflects three notable features of the digital economy: (i) the
sellers are better-informed than buyers about the match value; (ii) particular
items on the menu, that is, specific quality-price pairs, are recommended to
different buyers via recommendation services; (iii) personalized prices (or
more generally third-degree price discrimination) are usually not available
for the seller, but menu pricing (or second-degree price discrimination) can
occur. The lack of third degree price discrimination occurs due to the
presence of search engines and comparison sites, and business models that
discourage third-degree price discrimination (see, for example, \cite{dege19}
for the uniform price policy of national chain store and \cite{amaz00} for a
commitment by Amazon to not price on demographic characteristics).

These three features are reflected in our analysis as follows: (i) the buyers
only know the prior, or latent distribution; (ii) the seller chooses how much
information to provide to the buyers (i.e., the information that is conveyed
as a recommendation); (iii) the seller makes the products available to all
buyers with a public menu of products and prices. Thus, our analysis predicts
\emph{personalized} product recommendations from a \emph{non-personalized}
common menu. We provide a more complete interpretation and discussion after we
introduce the model formally.

\subsection{Related Literature\label{subsec:relat}}

We analyze a model of nonlinear pricing with a variable or a fixed
distributions of qualities, as in \cite{muro78} and \cite{lomu22},
respectively. In either setting, these papers show that bundling different
qualities, or randomizing the quality assignment via lotteries, can increase
the revenue in the presence of irregular value distributions.\ By contrast, in
our setting pooling is optimal for \emph{all--} regular or irregular-- distributions.

In our analysis, the seller controls the selling mechanism and the information
that the buyers receive. The seller thus has access to the tools of Bayesian
persuasion (\cite{kage11}) and mechanism design and we provide a solution to
an integrated mechanism and information design problem in a classic economic
environment. \cite{rase10} offer a sender-receiver model where the sender
offers \textquotedblleft prospects\textquotedblright\ to the receiver. Each
prospect offers a pair of distinct individual rewards for sender and receiver,
\ and those are private information to the seller. The receiver has a random
outside option that is private information to the receiver. By bundling the
prospects, the sender can increase the number and the probability of prospects
that are accepted by the receiver (and profitable for the seller.) Thus,
similar to the current environment, the sender controls how much information
the receiver has about the offered items. But distinct from the current
setting, the receiver only has unit demand. Building on \cite{rase10},
\cite{rayo13} considers a model of social status provision that also shares
some features with our model. The utility function of an agent before any
transfer is a product of their value (or an increasing function of their
value) and a social status which is equal to their \emph{expected value} given
some information structure. Thus, the allocation in \cite{rayo13} is an
information structure rather than a quality allocation.

There are a number of related papers that investigate how information may
influence the menu offered by multi-product sellers. \cite{mera22} and
\cite{ther22} consider a \cite{muro78} style model and allow the buyer to
acquire information about their willingness-to-pay. Each paper considers a
different cost of information and then derives the optimal menu that the
seller offers in anticipation of the endogenous response of the buyers. The
resulting menu offers a continuum of choices in which the buyer is implicitly
compensated for the cost of information acquisition. By contrast,
\cite{bepe07} considers a seller with many unit-demand buyers. They establish
the optimality of a finite and asymmetric allocation rule across the bidders.
They analyze finite mechanisms with an exogenous bound on the number of items,
and then show that as the bound grows, the number of items used in the optimal
mechanism eventually stays constant. The present argument identifies the
bilinear optimization problem of jointly determining information and
allocation. This allows us to obtain results regarding entropy (Theorem
\ref{thentrop}) and the optimality of single-item mechanisms (Theorem
\ref{th:22}), which has no counterparts in \cite{bepe07}.

The arguments that lead to the monotone pooling result are related to those
introduced by \cite{wils89}. While exogenously limiting the number $N$\ of
items in a nonlinear pricing environment, \cite{wils89} shows that in a
\emph{social surplus maximizing }mechanism this restriction only causes
surplus losses of order $1/N^{2}$. \cite{beyz21} extends this rate result to
multi-dimensional types and allocations. Here, we complement this argument
with the fact that the gains that come from reducing informational rents
always have a larger order of magnitude than the efficiency losses. We thus
conclude that there is always some amount of pooling in the
\emph{profit-maximizing }mechanism.

More distant, \cite{jomy03} and \cite{elli05} also ask what determines the
variety of a menu. They provide sufficient conditions under which the seller
may offer fewer or more products as a function of cost structure and the
distribution of tastes and advertising costs, respectively. The resulting
conditions are quite distinct from ours as the seller cannot control the
information in these models. Given the control of the information by the
seller, one could allow the seller to offer the information directly for the
sale to the agents, see \cite{essz03}, (2007)\nocite{essz07}, and then
possibly extract an even larger surplus. We deliberately refrain from giving
the seller this additional instrument. In the digital economy, the leading
application of this paper, the platforms are selling differentiated products
and typically bundle the information with the sale of products (through
recommendations) rather than offer the information as a separate service at a
separate price.

There is some earlier work that asks when second-degree price discrimination
may optimally resolve in a single-item menu. \cite{anda09} impose an \emph{a
priori} finite upper bound on the quality in the setting of \cite{muro78}.
They state conditions under which all values receive the same quality, namely
the quality at the upper bound. \cite{sand24} shows that a necessary condition
for a single-item menu to be profit maximizing is that the single-item menu
constitutes the socially optimal allocation. By contrast, in the current
environment a continuum of qualities is socially optimal. Hence there would be
no reason to restrict the menu and offer a bunching solution in the absence of
information design.

%\newpage

\section{Model\label{sec:model}}

\subsection{Payoffs}

A seller supplies products of varying quality $q\in\lbrack q_{l},q_{h}%
]\subset\mathbb{R}_{+}$ to a continuum of buyers with mass $1$. Each buyer has
unit demand and a willingness-to-pay, or value ${v}\in\lbrack v_{l}%
,v_{h}]\subset\mathbb{R}_{+}$, for quality $q$. The utility net of the payment
$p\in\mathbb{R}_{+}$ is:
\begin{equation}
{v\cdot}q-p.\label{u}%
\end{equation}
The buyers' values are distributed according to a common prior distribution
$F\in\Delta\left(  \lbrack v_{l},v_{h}]\right)  $, which we refer to as the
\emph{latent} distribution of values. We assume that space of values and
qualities are bounded, that is, $q_{h},v_{h}<\infty$.

The seller produces a set qualities $Q$ and incurs a cost that depends on the
distribution of qualities. A distribution of qualities $G\in\Delta\left(
\lbrack q_{l},q_{h}]\right)  $ has a cost for the seller denoted by:
\begin{equation}
C(G):\Delta\lbrack q_{l},q_{h}]\rightarrow\mathbb{R}_{+}\text{,}\label{cost}%
\end{equation}
where $Q=\operatorname*{supp}G$. We assume that $C$ is monotone increasing
with respect to the \emph{increasing convex order}. Intuitively, $G$ is
greater in the increasing convex order -- and thus more costly -- than another
distribution $\overline{G}$ if either $G$ represents higher qualities than
$\overline{G}$, or if $G$ represents more dispersed qualities than
$\overline{G}$, or both. The details of the cost function will not play a
central role in our analysis and a formal definition of the order is given in
Section \ref{discrete} (see condition (\ref{eq:ico})). Importantly, the
ordering of the cost function is flexible enough to accommodate the classic
model of \cite{muro78}: if there is a convex function $c:[q_{l},q_{h}%
]\rightarrow\mathbb{R}_{+}$ such that
\begin{equation}
C(G)=\int c(q)dG(q),\label{costmr}%
\end{equation}
then $C$ is monotone increasing in the increasing convex order. The model is
also flexible enough to accommodate a fixed inventory of goods (see
\cite{lomu22}), which is when the seller has a fixed mass of goods of various
qualities that can be pooled and discarded (at zero cost); a formal expression
is given in Section \ref{ilex}. We will be using the fixed inventory model as
a leading example, as it highlights the independence of many of our main
results from the exact nature of the cost function.

\subsection{Mechanism}

The selling mechanism $\left(  P,Q,S\right)  $\ consists of a menu $Q$\ of
products, their prices $P$ and an information structure $S$.

\paragraph{Menu Pricing}

The seller produces a set of qualities for sale $Q\subset\lbrack q_{l},q_{h}]$
and prices $P:Q\cup\left\{  0\right\}  \rightarrow\mathbb{R}_{+}$, with the
interpretation that $P\left(  q\right)  $ is the price at which every buyer
can purchase quality $q\in Q$. The pricing function must satisfy $P(0)=0$,
meaning that buyers can choose to not purchase any good (or equivalently,
acquire a zero quality good). We refer to $(P,Q)$ as the menu. By the taxation
principle, (Proposition 1, \cite{gula84}), any truthful direct mechanism can
be represented by an indirect mechanism in the form of a menu $\left(
P,Q\right)  $. For our arguments, the use of the indirect mechanism is convenient.

\paragraph{Information Structure}

The seller also chooses an information structure $S:[v_{l},v_{h}%
]\rightarrow\Delta\mathbb{R}_{+}$, with the interpretation that $S\left(
v\right)  $ is a distribution of real-valued signals observed by buyers with
latent value $v $. A buyer's expected value conditional on the signal
realization $s$ is denoted by:
\begin{equation}
\overline{v}\left(  s\right)  \triangleq\mathbb{E}[{v}\mid s].\label{w}%
\end{equation}
Frequently, we can omit the dependence of the expected value $\overline
{v}\left(  s\right)  $ on the signal $s$ and simply write $\overline{v}$ for a
generic expected value.

\subsection{Profit Maximizing Mechanism}

Now the seller chooses the mechanism $(P,Q,S)$ to maximize profits. To
calculate profits for any given mechanism, we must first identify buyers'
optimal choice rule $q:[v_{l},v_{h}]\rightarrow Q$ given the menu $\left(
P,Q\right)  $, where item $q(\overline{v})$ bought by a buyer with expected
value $\overline{v}$ is:%
\begin{equation}
q(\overline{v})\in\underset{q\in Q}{\arg\max}\left\{  \overline{v}%
q-P(q)\right\} ,\label{eq:opti}%
\end{equation}
where (as usual) ties are broken in favor of the seller.

The information structure $S$\ induces a distribution of expected values
$\overline{F}$, where
\[
\overline{F}\left(  \overline{v}\right)  =%
%TCIMACRO{\dint \limits_{v}}%
%BeginExpansion
{\displaystyle\int\limits_{v}}
%EndExpansion%
%TCIMACRO{\dint \limits_{s:\mathbb{E}[{v}\mid s]\leq\overline{v}}}%
%BeginExpansion
{\displaystyle\int\limits_{s:\mathbb{E}[{v}\mid s]\leq\overline{v}}}
%EndExpansion
dS\left(  s|v\right)  dF\left(  v\right)  .
\]
Now the distribution of qualities $\overline{G}$ needed to supply the buyers'
demand is given by:
\begin{equation}
\overline{G}(\overline{q})=\underset{\{\overline{v}:q(\overline{v}%
)\leq\overline{q}\}}{\int}d\overline{F}(\overline{v}),\label{distrq}%
\end{equation}
\ and the seller's profit is
\begin{equation}
\mathbb{E}[P(q\left(  \overline{v}\right)  )]-C(\overline{G}),\label{mr}%
\end{equation}
%
%\substack{ S:[v_{l},v_{h}]\rightarrow \Delta (\mathbb{R}_{+}) \\ P:Q\rightarrow \mathbb{R}_{+}}
where the expectation is computed using distribution $\overline{F}$. The
seller's problem is then to choose the mechanism $(P,Q,S)$ to maximize
profits. We conclude this section with a discussion of the interpretation of
our model.

\subsection{Interpretation}

\paragraph{Justifying the Common Menu Restriction}

Our model assumes that the seller has the ability to endow the buyers with any
information structure about their values. \ However, we did not specify
whether the seller did or did not observe the signal realizations of an
individual buyer. \ But this does not matter given our maintained assumption
that the seller offers a common menu of qualities and prices available to all
buyers, and that there is no personalized pricing. \ 

However, the interpretation of our model is a little different depending on
whether seller observes the buyers' signal realizations. \ In particular,
suppose the seller could not observe the realizations of the signals. \ In
this \textquotedblleft information-constrained\textquotedblright\ case, the
seller cannot offer different prices to buyers with different latent values or
signals. \ A standard argument in nonlinear pricing, the taxation principle
(\cite{gula84}), establishes that even if the seller could elicit information
from buyers, he would not be able to improve the profit compared to a public
pricing schedule. \ Thus in this \textquotedblleft
information-constrained\textquotedblright\ scenario, there is no loss in
assuming that the seller is restricted to offer a common menu. \ 

But if the seller did observe the realization of the buyers' signals, there
may be other reasons - outside our model - why the seller does not offer
personalized prices. \ We call this the \textquotedblleft
instrument-constrained\textquotedblright\ case because we have imposed a
constraint on the selling mechanism. \ As we discussed in the introduction, in
the digital economy, the \textquotedblleft
information-constrained\textquotedblright\ argument against personalized
pricing may be less relevant but there are a variety of forces pushing against
personalized pricing: (i) platform sellers may fear that personalized pricing
will scare consumers away from the platform; (ii)\ there may be legal
restrictions on personalized pricing; (iii)\ buyers may have the ability to
find others' personalized pricing by searching under alternative identities. \ 

\paragraph{Recommendation as Information}

By a standard argument, it would be without loss of generality to assume that
instead of sending abstract signals, the signals are restricted to items in
the menu and can be interpreted as recommendations of what items to buy.
\ Thus the information structure could be restricted to take the form
$S:[v_{l},v_{h}]\rightarrow\Delta\left(  Q\right)  $ and, subject to obedience
constraints, buyers could be assumed to always select the item that was
"recommended" to them. \ 

This representation of the information structure is not particularly helpful
to us analytically. \ However, we highlight this representation because we
think that recommender systems are a relevant mechanism used in the digital
economy to send information. \ In particular, when a search menu generates a
list of buying options, we know that the first or top item is more likely to
be chosen. \ One interpretation is that the top items represent
recommendations of the seller. \ 

\paragraph{Direct Allocations}

By a standard argument, it would be without loss of generality to assume a
direct allocation mechanism, i.e., letting the buyers report their expected
values to the mechanism and the mechanism assigns a quality-price pair to each
buyer in such a way that it is incentive compatible for buyers to truthfully
report their expected values. \ We use the indirect representation in part
because it fits better with our interpretations of the common menu assumption.
\ However, we emphasize that optimal choices in our mechanism $q\left(
\overline{v}\right)  $ and $P\left(  q\left(  \overline{v}\right)  \right)  $
correspond to incentive compatible and individually rational direct allocation
rules (and some readers may find it helpful reading our arguments that way).

\section{A First Example with Finitely Many Values\label{sec:ex}}

We first analyze the environment with a finite number $N\ $of values and
qualities. In this section, we do not yet seek to derive the exact nature of
the optimal mechanism. Rather, we compare the revenue of the seller under two
specific information structures. The first is \emph{complete disclosure} where
each buyer learns their value as private information\emph{. }The resulting
optimal revenue arises from the standard optimal screening mechanism. By
contrast, the second information structure pools the two lowest values of the
buyer and then derives the optimal mechanism when these adjacent values are
pooled, and we refer to this as \emph{lower pooling}\textbf{. \ }We provide
sufficient conditions for when this specific pooling of the lowest adjacent
values improves the revenue relative to complete disclosure. This will provide
key intuition for results that follow. \
%We will show that when the atoms being pooled are small, the price of qualities that are not being pooled will increase and this effect will be the dominant one.

Throughout this section, we assume that the seller has a fixed inventory of
goods, that is, a fixed distribution of qualities $G$ (as in \cite{lomu22}).
The assumption of an exogenous inventory simplifies the algebra as now revenue
equals profit, but the conclusions go through for variable inventory (as in
\cite{muro78}).

%The seller can pool values and can pool qualities to generate distributions $\overline F$ and $\overline G$ respectively. Note that the seller has no cost of production, but can only pool the exogenous inventory of goods, which satisfies our assumption on $C$. We provide the details of this mapping in Section \ref{two} but for the purpose of this section this mapping will be irrelevant.

The support of values and qualities are denoted by $\{v_{1},...,v_{N}\}$ and
$\{q_{1},...,q_{N}\}$, with $0<v_{1}<\cdots<v_{N}$ and $0<q_{1}<\cdots<q_{N}$;
both distributions have atoms of size $\{f_{1},...,f_{N}\}$. So there is an
assortative allocation where buyers of value $v_{k}$ are matched with
allocation $q_{k}$. \ The discrete version of the virtual values are denoted
by:
\begin{equation}
\phi_{i}\triangleq v_{i}-(v_{i+1}-v_{i})\frac{\sum_{j=i+1}^{N}f_{j}}{f_{i}%
}\text{.}\label{vv}%
\end{equation}
The following lemma provides a way in which the virtual values shape the
optimal mechanism in screening problems without persuasion.

\begin{lemma}
[Screening without Persuasion]\label{stscr}\quad\newline The profit-maximizing
allocation is efficient if and only if the virtual values $\phi_{i}$ are
increasing in $i$ and greater than $0$ for every $i$.
\end{lemma}

%\footnote{\cite{lomu22} analyze the optimal mechanism when the virtual values are not monotone, but for the purpose of our paper it will only be relevant when the efficient mechanism is also the revenue-maximizing one.}

For the proof see Theorem 1 in \cite{lomu22}, who characterize the optimal
screening mechanism (without persuasion) when the seller has a fixed inventory
of goods of various qualities. Throughout this section, we assume that the
virtual values are increasing in $i$ and greater than $0$ for every $i$. Thus
we focus on an environment that is most favorable towards complete disclosure
as bundling due to ironing does not occur under complete disclosure. If the
goods are sold efficiently, the optimal pricing policy is given by:
\begin{equation}
P(q_{i})=v_{1}q_{1}+\sum_{j=2}^{i}(q_{j}-q_{j-1})v_{j},\label{ics}%
\end{equation}
by a standard argument using the incentive compatibility constraints. The
seller is thus able to extract the value $v_{i}$ only for the marginal
quality, $q_{i}-q_{i-1}$, offered to each buyer $v_{i}$. The price for the
product $q_{i}$ is then the sum of the quality increments $(q_{j}-q_{j-1}%
)$\ multiplied by the value $v_{j}$ for all values below $v_{i}$. The profits
generated are given by:
\begin{equation}
\Pi=\sum_{i=1}^{N}P(q_{i})f_{i}=v_{1}q_{1}+\sum_{i=2}^{N}(q_{i}-q_{i-1}%
)v_{i}(1-\sum_{j=1}^{i-1}f_{j}),\label{cdca}%
\end{equation}
where the second expression is obtained by writing $P(q_{i})$ explicitly and
changing the order of summation.

We can interpret the profit function from the perspective of a multi-product
monopolist who sells $N$ distinct products of qualities $(q_{i}-q_{i-1})$ and
has a mass $(1-\sum_{j=1}^{i-1}f_{j})$ of each of these products. In this
interpretation, the buyers have multi-unit demand so they can buy all products
(or any ordered subset), which amounts to the same utility as buying a single
good of quality $q_{N}$ in the original interpretation of the model.

%The pricing function that sells the goods assortatively is the same term as in \eqref{cdca} but the first quality and value levels are $\overline v_1$ and $\overline q_1$:
%\[\overline P(q_i)=\overline v_1\overline q_1+(q_3-\overline q_{1})v_3+\sum_{j=4}^i(q_j- q_{j-1})v_i.\]
%Pooling implies that the qualities $\{q_1,q_2\}$ are no longer sold, and instead there is a intermediate quality $\overline q_1$ that is sold at an intermediate price. Importantly, pooling also impacts the price of qualities that are not being pooled. Furthermore, pooling does not affect the demand for these qualities as by construction this pricing function provides incentives for an assortative allocation between (pooled) values and (pooled) qualities.

Let us now consider a mechanism where the \emph{lowest two values} are pooled
so the expected value and expected quality generated are given by:
\begin{equation}
\overline{q}_{1}=\frac{f_{1}q_{1}+f_{2}q_{2}}{f_{1}+f_{2}},\ \ \ \overline
{v}_{1}=\frac{f_{1}v_{1}+f_{2}v_{2}}{f_{1}+f_{2}}.\label{cdz}%
\end{equation}
After this pooling, there are only $N-1$ distinct qualities for sale among
$N-1$ values, which we assume are sold assortatively (as before). The profits
generated by pooling, say $\overline{\Pi}$, is computed as in \eqref{cdca} but
accounting for the fact that the first two quality levels and the first two
value levels are pooled. The difference between the profits after pooling and
the profits under complete disclosure is:
\begin{equation}
\overline{\Pi}-\Pi=\left(  \overline{v}_{1}\overline{q}_{1}-\left(  v_{1}%
q_{1}+(q_{2}-q_{1})v_{2}(1-f_{1})\right)  \right)  +\left(  q_{2}-\overline
{q}_{1}\right)  v_{3}(1-f_{1}-f_{2}).\label{ccdh}%
\end{equation}
Here we wrote the difference as the sum of two terms. The first term
corresponds to the difference in the profits generated by values
$\{v_{1},v_{2}\}$ and qualities $\{q_{1},q_{2}\}$: the seller now sells a
pooled quality level $\overline{q}_{1}$ at price $\overline{v}_{1}$ to every
buyer instead of selling quality $q_{1}$ at price $v_{1}$ to every buyer and a
quality increment $(q_{2}-q_{1})$ at price $v_{2}$ to a mass of buyers
$(1-f_{1})$. The second term accounts for the fact that pooling increases the
quality difference between $q_{3}$ and the immediate quality predecessor--
which is $q_{2}$ under complete disclosure and $\overline{q}_{1}$ when
pooling-- so we get a term proportional to $\left(  q_{2}-\overline{q}%
_{1}\right)  $ and proportional to $v_{3}$ which is the price of this quality
increment. Note that pooling does not change cost (which is zero) because we
assume that there is a fixed inventory of goods, so changes in profits are the
same as changes in revenue. If there was a non-trivial production cost, then
the benefits of pooling would be even larger because pooling would also reduce
cost.
%This second term will increase the price paid by all values not being pooled, so it is multiplied times $(1-f_1-f_2)$; if we have only two values, then the second term will be absent. If we have many values and the first two have small mass, then the first term will be small relative to the second one because the second term corresponds to a price increase to every value not being pooled.

In contrast to screening without persuasion, we now have that the distribution
of qualities plays an important role to determine when pooling is optimal.
This will lead to properties that any optimal distribution of qualities must
have, and these properties will not have any counterpart in classic screening
problems. Towards these properties, we observe that the payoff difference
given by \eqref{ccdh} is linear in $q_{1}$ and strictly positive when
$q_{1}=q_{2}$. \ Thus we have that:
\begin{equation}
(\overline{\Pi}-\Pi)\mid_{q_{1}=0}>0\Longrightarrow(\overline{\Pi}%
-\Pi)>0.\label{ewc1}%
\end{equation}
Now, by evaluating the condition at $q_{1}=0$ we get a weaker condition for
when pooling generates more profits than complete disclosure. Intuitively,
pooling always increases the price at which $q_{1}$ is sold, so by evaluating
at $q_{1}=0$ we obtain a weaker condition. We provide this condition explicitly.

\begin{proposition}
[Lower Pooling Improves Profits]\label{edacc}\quad\newline Lower pooling
revenue exceeds complete disclosure revenue if:
\begin{equation}
\frac{f_{2}}{(1-(f_{1}+f_{2}))(f_{1}+f_{2})}<\frac{v_{3}-v_{2}}{v_{2}-v_{1}%
}.\label{3val0}%
\end{equation}

\end{proposition}

We have that \eqref{3val0} corresponds to the left-hand-side of \eqref{ewc1}
after re-arranging terms (the left-hand-side of \eqref{ewc1} is linear in
$q_{2}$, so \eqref{3val0} also does not depend on $q_{2}$). The trade-off is
that pooling leads to $q_{2}$ being sold at price $\overline{v}_{1}$ (instead
of $v_{2}$) but pooling also increases the quality increment $q_{3}$ minus its
predecessor (which is $\overline{q}_{1}$ instead of $q_{2}$ under pooling),
which is priced at $v_{3}$.
%This contrasts with the earlier condition of Lemma \ref{prop2} for\ the two-value model in which the condition referred to the difference in qualities.

%The main idea is that pooling adjacent values generates inefficiencies that are \textquotedblleft local\textquotedblright\ in the sense that they are small if either the values are nearby or the probability of the values are small. As the reduction of the information rent that comes with bundling values affects all higher values, the improvement however is global. Namely, the quality increment $(q_{3}-\overline{q}_{1})$ is sold at a higher price to \emph{all} values above the pool, thus $(1-f_{1}-f_{2})$, and so more surplus is extracted from all values larger than $v_{2}$. Hence, the margins of the cost-benefit analysis always tilt the balance toward pooling when the pooling involves small atoms relative to the rest of the distribution, thus if $f_{1}+f_{2}\ll 1-f_{1}-f_{2})$.

We can leverage \eqref{3val0} to provide a simpler condition for when complete
disclosure is not optimal. We say that the values are \emph{uniformly
distanced} when
\[
v_{i+1}-v_{i}=v_{i}-v_{i-1}\text{, for all }i\text{.}%
\]
Assuming the values are {uniformly distanced} will allow us to significantly
simplify the algebra.

\begin{corollary}
[Lower Pooling Improves Profit for Uniformly Distanced Values]\label{dccs}%
\quad\newline With \emph{uniformly distanced values}, lower pooling revenue
exceeds complete disclosure revenue if:
\begin{equation}
f_{2}<\sqrt{f_{1}}-f_{1}.\label{3val1}%
\end{equation}
With uniformly distanced \emph{and} uniformly distributed values ($f_{i}%
=f_{j}$ for all $i,j$), lower pooling revenue exceeds complete disclosure
revenue if $N\geq5$.
\end{corollary}

The sufficient condition (\ref{3val1}) is obtained by applying the
\emph{uniformly distanced} property to the improvability condition
\eqref{3val0} of Proposition \ref{edacc}. This corollary then provides simple
conditions for lower pooling to be optimal.

\paragraph{Pooling Fine Distributions}

If we approximate a continuous distribution with a sequence of finite
distributions on a {uniformly distanced} grid, eventually \eqref{3val1} will
be satisfied (as $\sqrt{f_{1}}$ will be larger than $f_{2}+f_{1}$), and so
complete disclosure will not be optimal. We will generalize this idea by
showing that in any optimal mechanism the induced distribution of values is
always finite (Theorem \ref{prop:int}). Hence, the optimal mechanism is coarse
even when the underlying distribution of values is very fine.

A striking feature of the analysis is that \eqref{3val1} is satisfied even
when the distribution is relatively coarse. For example, with uniformly
distanced and uniformly distributed values, lower pooling improves upon
complete disclosure as soon as $N\geq5$. We will generalize this idea by
showing that there is an upper bound on the entropy of the distribution such
that complete disclosure is optimal. This result will show that even when the
latent distribution of values is finite, pooling will be optimal unless
\textquotedblleft most\textquotedblright\ of the distribution is located in
\textquotedblleft few large atoms\textquotedblright as measured by the entropy
(Theorem \ref{thentrop}).
%To this simply note that under the conditions we have that $f_{i}<1/4$ and so the improvability condition \eqref{3val1} is satisfied.

There are no counterparts to these conditions in classic screening problems;
we now provide some intuition for the stark difference with standard screening problems.

\paragraph{Screening without Persuasion}

To contrast our analysis with screening in the absence of persuasion, we note
that \eqref{3val0} can be written as follows:
\begin{equation}
(\overline{\Pi}-\Pi)\mid_{q_{1}=0}=\frac{f_{1}f_{2}q_{2}}{f_{1}+f_{2}}%
(\phi_{1}-\phi_{2})+\frac{f_{2}q_{2}}{f_{1}+f_{2}}(\overline{v}_{1}%
-v_{1})>0.\label{dcdc5}%
\end{equation}
The first term is the change in profits due to the pooling of the allocation
only (without changing the information). Then, the first term is positive if
and only if the virtual values are strictly decreasing (as in Lemma
\ref{stscr}). The second term is the gain in profits from pooling the values
after we have already pooled the allocation. In essence, we will show that the
second term in \eqref{dcdc5} always dominates the first one when the atoms are
small enough. A heuristic argument goes as follows. Suppose the atoms are
small and the distance between values and between virtual values are small.
Both terms are proportional to $f_{2}/(f_{1}+f_{2})$, but the first term is
the product of two small factors $f_{1}(\phi_{1}-\phi_{2})$ while the second
term consists of one small factor $(\overline{v}_{1}-v_{1})$. The intuition is
that the profit losses due to pooling the allocation (without pooling
information) are limited to the locally affected values. Yet the revenue
benefits from pooling values increases the profits generated by all higher
values. Due to the differences in the extensive margin, pooling is always
beneficial when atoms are small enough. This is the reason we get distinct
conditions for the optimality of pooling in screening with persuasion relative
to screening without persuasion.

\section{The Optimality of a Finite Menu\label{discrete}}

In Section \ref{subsec:dis} we state our first main result, Theorem
\ref{prop:int}, which asserts that the optimal mechanism is finite. The main
steps of the proof are provided in Sections \ref{two}-\ref{sec:cm}. In Section
\ref{two}, we develop the properties of incentive compatible and feasible
mechanisms and state the profit maximization problem (\ref{mr}) as a joint
optimization over two distributions, the distribution of values $\overline{F}$
and the distribution of qualities $\overline{G}$. Section \ref{sec:pool}
establishes discreteness of the optimal mechanism, which is the main step in
establishing finiteness of the optimal mechanism. Here we argue that pooling
small intervals of values is always beneficial for the seller. Section
\ref{sec:cm} establishes an additional property of independent interest, the
convexity of the menu. We show that the quality differences between any two
adjacent intervals increase as the values increase. Incidentally, this allows
us to go from discreteness to finiteness by ruling out accumulation points. In
Section \ref{subsec:mono} we provide an additional property of the optimal
mechanism, a monotone partitional property, that is not used to prove Theorem
\ref{prop:int}, but will be helpful to characterize the optimal mechanism in
applications. Finally, in Section \ref{ilex} we illustrate the optimal
mechanism in two specific examples.

\subsection{Optimality of a Finite Mechanism\label{subsec:dis}}

We say a \emph{mechanism is finite} if the support of the value distribution
$\overline{F}$ is finite. Our first main result shows that every optimal
mechanism offers a finite set of distinct expected values and expected
qualities. Throughout the paper we consider mechanisms such that, if a buyer
does not purchase any positive quality, then they are provided no additional
information, thus,
\[
q(\bar{v})=q(\bar{v}^{\prime})=0\Rightarrow\bar{v}=\bar{v}^{\prime}.
\]
This only disciplines the information that the optimally excluded buyers
receive and it is without loss of generality to consider mechanisms with this property.

\begin{theorem}
[Optimality of Finite Mechanism]\label{prop:int}\quad\newline Every optimal
mechanism is finite.
%In every optimal mechanism $G^*$ is a finite pooling interval structure and a buyer with expected valuation $w_k$, generated by pooling interval $[x_{k-1}-x_k)$, is offered a bundle of all qualities in this same quantiles:
%\[q(w_k)=\frac{\int_{x_{k-1}}^{x_k}Q^{-1}(x)dx}{x_{k}-x_{k-1}}\]
%
%Consider the following function:
%\[M(t)=\int_0^t(1-s)dq(s)\]
%\begin{lemma} \hfill
%\noindent  In every optimal mechanism, $M(t)=\mathrm{conv} M(t)$ for every $t$ at which $q(t)$ is strictly increasing.
%\end{lemma}
%\begin{lemma}[Relation Between  $G$ and $q$]\label{lem:pool} \hfill
%\noindent In every optimal mechanism and for every interval interval $(v_1,v_2)$,  $q$ is constant in $(v_1,v_2)$ if and only if  $supp(G)\cap (v_1,v_2)=\emptyset$.
%\end{lemma}

\end{theorem}

The intuition for the result is that pooling small intervals generates losses
that are \textquotedblleft local\textquotedblright, meaning that they affect
only the values being involved in the pooling. On the other hand, the
reduction in informational rents affects all values higher than the interval
being pooled. Hence, by considering an interval that is small enough we get
that the benefit is an order of magnitude larger than the losses. This
argument is formally presented in Section \ref{sec:pool}, where we show that
any optimal mechanism is discrete.
%This is the main part of the proof as it argues that pooling will always arise in the optimal mechanism.

Since the space of values and qualities is bounded, to conclude the proof of
Theorem \ref{prop:int} we need to show that there are no accumulation points.
Showing that there are no accumulation points in any interior part of the
distribution can be proved in a similar way to how we prove that an optimal
mechanism is discrete. However, proving that there are no accumulation points
at the top of the distribution requires a different line of argument. We prove
this by showing that the qualities offered in any optimal mechanism satisfy
increasing differences. We provide this result after we prove that the optimal
mechanism is discrete, and we argue in due course that this result is of
independent interest. We note that if we were to relax the assumption that the
space of values and qualities is bounded, we would get that the optimal
mechanism is discrete without the property of finiteness.

Finally, we show that the endogenous distribution of values in an optimal
mechanism can be obtained from monotone partitions of the exogenous
distribution of values. This last result is not used to prove Theorem
\ref{prop:int}, but it will be useful when we characterize the optimal
mechanism (with additional assumptions on values and qualities) in Section
\ref{sec:opti}.

Before we provide the substantive and novel part of our analysis, we apply
standard tools from information and mechanism design to write the seller's
problem as a maximization problem over two distributions.

\subsection{Optimization Over Two Distributions\label{two}\ }

The seller's design problem has two components:\ $\left(  i\right)  $\ the
seller chooses the information that buyers have about their own value, and
$\left(  ii\right)  $ the menu. We now provide a more convenient
representation of these two elements by writing the seller's problem as the
maximization over two distributions. One of the distributions will be the
distribution of expected values $\overline{F}$ generated by the information
structure; the other distribution will be the mass of qualities that the
seller supplies in the optimal mechanism $\overline{G}.$

The buyers' information structure is summarized by the distribution of
expected values $\overline{F}$. \ By \cite{blac51}, Theorem 5, there exists an
information structure that induces a distribution $\overline{F}$\ of expected
values if and only if $\overline{F}$ is a mean-preserving contraction of $F$,
i.e.,
\begin{equation}
\int_{v}^{v_{h}}F({x})dx\leq\int_{v}^{v_{h}}\overline{F}({x})dx\text{,
}\forall{v}\in\lbrack v_{l},v_{h}],\label{eq:majo}%
\end{equation}
with equality for $v=v_{l}$. \ If $\overline{F}~$is a mean-preserving
contraction of $F$ (or $\overline{F}$ majorizes $F$), we write $F\prec
\overline{F}$.
%As it is standard in the mechanism design literature
%
%where $\overline{F}$ is the distribution of expected values and the second
%term inside the bracket represents the buyers' information rent. Note that
%while $\overline{F}$ may have gaps, it is without loss of generality to
%assume that the allocation $q\left( \cdot \right) $ is defined on the entire
%domain $[v_{l},v_{h}]$. Hence we can pin down the transfer payments uniquely
%with the allocation rule.

We assumed that the cost of providing a distribution $G$ of qualities is
$C(G)$\ (see (\ref{cost})). For any distributions $G,\overline{G}\in
\Delta\mathbb{R}_{+}$, we say $G$ is greater than $\overline{G}$ in the
{increasing convex order} (or $\overline{G}$ weakly majorizes $G$), and denote
this by $G\succ_{c}\overline{G}$, if
\begin{equation}
\int_{q}^{q_{h}}G({x})dx\leq\int_{q}^{q_{h}}\overline{G}({x})dx\text{,
}\forall{q}\in\lbrack q_{l},q_{h}],\label{eq:ico}%
\end{equation}
see Theorem 4.A.2 in \cite{shsh07}. Unlike the definition of mean-preserving
contractions given by \eqref{eq:majo}, we now do not require \eqref{eq:ico} to
be satisfied with equality when $q=q_{l}.$ Theorem 4.A.6 in \cite{shsh07}
states that $\overline{G}\prec_{c}G$ if and only if there exists $\widehat{G}$
such that $G$ is a mean preserving spread of $\widehat{G}$ and $\widehat{G}$
first-order stochastically dominates $\overline{G}$. Thus if $\overline
{G}\prec_{c}G$, then $\overline{G}$ is both \textquotedblleft
lower\textquotedblright\ and \textquotedblleft less variable\textquotedblright%
\ than $G$ in the sense that $\overline{G}$ corresponds to lower qualities
than $\widehat{G}$, which in turn corresponds to less variable qualities
than~$G$. Assuming that the cost function is monotone increasing in the
increasing convex order means that for any pair of distributions
$G,\overline{G}\in\Delta\mathbb{R}_{+}$, $\overline{G}\prec_{c}G$ implies that
$C(\overline{G})\leq C(G)$. So producing higher qualities and more variable
qualities is more expensive.\footnote{While the majorization order $(\prec)$
and the increasing convex order $\left(  \prec_{c}\right)  $ have nearly
identical definitions in terms of the integrals, (\ref{eq:majo}) and
(\ref{eq:ico}) respectively, we adopt the precedence symbol in the opposite
direction for each. This choice aligns with the conventions in the literature
for majorization order and for increasing convex order (as used in the
stochastic order literature), ensuring consistency within their respective
contexts.}

We recall that for any pricing function $P$, $q(\overline{v})$ denotes the
optimal choice of a buyer with expected value $\overline{v}$. Analogously, we
define the respective payment:
\[
p(\overline{v})=P(q(\overline{v})).
\]
We refer to a pair $(q(\overline{v}),p(\overline{v}))$ as a \emph{direct
menu}. We depart from the classic terminology (that is \textquotedblleft
direct mechanism\textquotedblright) because in our model a mechanism also
includes the information structure.

Optimality of the choice function \eqref{eq:opti} implies that
\begin{equation}
\overline{v}q(\overline{v})-p(\overline{v})\geq\overline{v}q(\overline
{v}^{\prime})-p(\overline{v}^{\prime}),\ \ \forall\overline{v},\overline
{v}^{\prime}\in\lbrack v_{l},v_{h}];\label{eq:ic1}%
\end{equation}
and
\begin{equation}
\overline{v}q(\overline{v})-p(\overline{v})\geq0,\ \ \forall\overline{v}%
\in\lbrack v_{l},v_{h}].\label{eq:ir}%
\end{equation}
Thus, the direct menu $\{(q(\overline v),p(\overline v))\}$ induced by $P$
satisfies the classic incentive compatibility and participation constraints.
Following standard techniques, the incentive compatibility requires that the
allocation $q(\overline{v})$ is increasing and the payment $p\left(
\overline{v}\right)  $ is determined by the allocation rule using the Envelope
condition:
\begin{equation}
p(\overline{v})=\overline{v}q(\overline{v})-\int_{v_{l}}^{\overline{v}%
}q(s)ds.\label{Eq:obj1}%
\end{equation}
Note that while $\overline{F}$ may have gaps, it is without loss of generality
to assume that the allocation $q\left(  \cdot\right)  $ is defined on the
entire domain $[v_{l},v_{h}]$. We are thus left with determining the optimal
distribution of values and the allocation function.

We thus have that for any mechanism $(P,Q,S)$ there is an induced direct menu
and a distribution of expected values $(p(\overline{v}),q(\overline
{v}),\overline{F})$. Similarly, for any $(p(\overline{v})),q(\overline
{v}),\overline{F})$ satisfying the above properties we have a mechanism
$(P,Q,S)$ that induces it.

However, rather than optimizing over the allocation function $q(\bar{v})$ we
will optimize over the distribution of qualities $\overline{G}$ offered by the
mechanism (as defined in \eqref{distrq}), and then back out what would be the
implied allocation $q$ using $\overline{F}$ and $\overline{G}$. More
precisely, we will solve the problem:
\begin{equation}
\max_{\substack{\overline{F}^{-1}\prec{F}^{-1} \\\overline{G}^{-1}\in
\Delta\mathbb{R}_{+}}}\int_{0}^{1}\overline{F}^{-1}\left(  t\right)  \left(
1-t\right)  d\overline{G}^{-1}\left(  t\right)  +\overline{G}^{-1}\left(
0\right)  \overline{F}^{-1}\left(  0\right)  -C\left(  \overline{G}\right)
.\label{eq:lin}%
\end{equation}
As in Section \ref{sec:ex}, the seller's revenue can be interpreted as the sum
(integral) over all quality increments multiplied by the revenue generated
per-unit-of-quality, \eqref{eq:lin} is the integral version of \eqref{cdca}.
By optimizing over distributions of qualities rather than the allocation
function, the production cost $C(\overline{G})$ is independent of the
information $\overline{F}.$ Of course, varying $\overline{F}$ without varying
$\overline{G}$ will change the implied allocation function $q(\overline{v}).$
We write the expressions in terms of the quantile function (the inverse of
$\overline{F}$ and $\overline{G}$) because this way the revenue is bilinear in
the maximization variables.
%
%\begin{proposition}[Optimizing Over Two Distributions]
%\label{prop2distr}\quad \newline
%A mechanism $((q(\overline{v}),p(\overline{v})),\overline{F})$ is optimal if
%and only if for some solution of \eqref{eq:lin}, say $(\overline{F},%
%\overline{G})$,
%\begin{equation*}
%q(\overline{v})=\overline{G}^{-1}(\overline{F}(\overline{v}))
%\end{equation*}%
%and $p(\overline{v})$ is given by \eqref{Eq:obj1}.
%\end{proposition}

\begin{proposition}
[Optimizing Over Two Distributions]\label{prop2distr}\quad\newline A mechanism
$(P,Q,S)$ is optimal if and only if there exists a solution of \eqref{eq:lin}
of the form $(\overline{F},\overline{G})$, such that
\[
q(\overline{v})=\overline{G}^{-1}(\overline{F}(\overline{v})),
\]
and $p(\overline{v})$ is the direct mechanism induced by $(P,Q,S)$.
\end{proposition}

We thus analyze a maximization problem over two distributions $(\overline
{F},\overline{G})$ given by \eqref{eq:lin}, and we refer to a pair
$(\overline{F},\overline{G})$ as a mechanism. When solving the problem
\eqref{eq:lin} we are only be interested in functions $(\overline{F}%
,\overline{G})$ such that at any $t$:
\begin{equation}
\label{measurcond}\text{$\overline{G}^{-1}(t)$ is increasing in }t\text{ only
if $\overline{F}^{-1}(t)$ is increasing in }t.
\end{equation}
This is without loss of generality: if $\overline{F}^{-1}(t)$ is constant in
some interval $(t_{1},t_{2})$ and $\overline{G}^{-1}(t)$ is not constant in
this interval, then we can pool qualities in this interval and thus weakly
increase revenue and weakly decrease cost. In a direct menu we construct
$q(\overline{v})$ from $\overline{G}$, so this measurability condition is
necessary and sufficient for the solution to be implementable with a direct menu.

%Thus, the constraints on the distribution $F$ of value and the distribution $G$ of qualities are nearly identical, except that the distribution\ of qualities does not have to satisfy the equality at the lower bound.

\subsection{Optimality of Discrete Mechanism\label{sec:pool}}

We say that a given distribution $H$ has discrete support if there is no open
interval $(t_{1},t_{2})\subset\lbrack0,1]$ such that $H^{-1}$ is strictly
increasing in $(t_{1},t_{2})$. We say a mechanism $(\overline{F},\overline
{G})$ is discrete if it consists of distributions with discrete support. As a
first step towards proving Theorem 1 we prove that any optimal mechanism is
discrete. We can describe a discrete mechanism by a countable collection of
quantiles $\{t_{k}\}_{k\in K}$, where at each quantile $t_{k}$ there is a
discontinuous jump in expected values and expected qualities and both
distributions are constant everywhere else. As shorthand notation, we denote
by $\overline{v}_{k}$ the expected value and by $\overline{q}_{k}$ the quality
at quantile $t_{k}$:
\begin{equation}
\overline{v}_{k}\triangleq\overline{F}^{-1}(t_{k});\ \ \ \ \ \overline{q}%
_{k}\triangleq\overline{G}^{-1}(t_{k}).\label{eq:vk}%
\end{equation}

\begin{proposition}
[Optimality of Discrete Mechanism]\label{prop:opi}\quad\newline Every optimal
mechanism $(\overline{F},\overline{G})$\ is discrete.
\end{proposition}

\begin{proof}
The measurability condition \eqref{measurcond} implies that, if $\overline F$
is discrete, then $\overline G$ (and thus the mechanism) is discrete. We
consider a candidate mechanism $\left(  {F},G\right)  $ and correspondingly
$({F}^{-1},G^{-1})$ such that $F^{-1}$ is strictly increasing on some interval
$[t_{1},t_{2})$ and show that this mechanism is not optimal. In writing the
proof, the notation suggests that ${F}$ and $G$ are the exogenous
distributions of values and qualities; this is only to make the notation more
compact and is of no substantive difference for the argument.

We first consider the case in which $({F}^{-1},G^{-1})$ are both strictly
increasing on some interval $[t_{1},t_{2})$ and show this mechanism is not
optimal. It is useful to write the interval $[t_{1},t_{2})$ in terms of its
mid-point and width:
\begin{equation}
\widehat{t}\triangleq\frac{t_{1}+{t}_{2}}{2};\quad\Delta\triangleq\frac
{{t}_{2}-t_{1}}{2}.\label{that}%
\end{equation}
So, we have that
\begin{equation}
\lbrack t_{1},t_{2})=[\widehat{t}-\Delta,\widehat{t}+\Delta)\label{that1}%
\end{equation}
and we will eventually take the limit $\Delta\rightarrow0$. Since both
$F^{-1}(t)$ and $G^{-1}(t)$ are strictly increasing in this interval, they are
differentiable at almost every $t$ in this interval. So, without loss of
generality we assume that $F^{-1}(t)$ and $G^{-1}(t)$ are differentiable at
$\widehat{t}$ and the derivative is strictly positive.

%Analogously, $P_G(t)$ is the Peano reminder of $G.$

%When the buyer has complete information, the optimal mechanism can be found
%by writing the seller's profits in terms of the virtual surplus as follows.
%The virtual surplus is formally given by:
%\begin{equation*}
%\Phi (q,{v})\triangleq ({v}q-c(q))-q\frac{1-F(v)}{f(v)}.
%\end{equation*}%
%Following standard techniques, we define the complete disclosure optimal
%qualities as follows:
%\begin{equation}
%\widehat{q}\left( v\right) \triangleq \underset{q(v)}{\arg \max }\int_{%
%\underline{v}}^{\bar{v}}\Phi (q(v),v)dF(v).  \label{fdq}
%\end{equation}%
%One can maximize the functional pointwise to obtain the complete disclosure
%optimal mechanism. (We presume here without loss of generality that the
%optimal solution does not involve any bunching.) In other words, if the
%seller does not manipulate the buyer's information, she will maximize the
%virtual surplus. The virtual surplus provides the exact tradeoff between
%social surplus and buyer's surplus that the seller faces.
%We say the information structure has finite interval structure, if there exists $\{v_0,...,v_K\}$, with $v_0=\underline v$ and $v_K=\bar v$, such that:
%\[w_k=\mathbb{E}[v\mid v\in [v_{k-1},v_k]], \text{ for all $k\in\{1,...,K\}$}.\]

We contrast the profits generated by $(F,G)$ with ones generated by a
mechanism in which values and qualities are pooled in this interval
$[t_{1},t_{2})$. Before we compute the changes in profit we introduce some
notation. We denote by $v_{1}$ and $q_{1}$ the value and the quality at the
lower limit of the interval $[t_{1},t_{2})$ and denote by $q_{2}$ the quality
at the upper limit of the interval $[t_{1},t_{2})$:
\[
v_{1}\triangleq F^{-1}(t_{1})\text{; }q_{1}\triangleq G^{-1}(t_{1})\text{;
}q_{2}\triangleq G^{-1}(t_{2}).
\]
We also denote by $\mu_{v}$ and $\mu_{q}$ the average value and average
quality provided in this interval:
\[
\mu_{v}\triangleq\frac{\int_{t_{1}}^{t_{2}}F^{-1}(t)dt}{t_{2}-t_{1}}\text{ and
}\mu_{q}\triangleq\frac{\int_{t_{1}}^{t_{2}}G^{-1}(t)dt}{t_{2}-t_{1}}.
\]
We only compute these quantities in the interval $[t_{1},t_{2}]$, and we can
therefore safely omit an index referring to the interval $\left[  t_{1}%
,t_{2}\right]  $ in the expressions.

We first consider the impact from pooling the allocation without changing the
information structure, and thus pooling the values. The distribution of
qualities can be written as follows:
\begin{equation}
G_{q}^{-1}({t})=%
\begin{cases}
\mu_{q} & \text{ if ${t}\in\lbrack{t}_{1},{t}_{2});$}\\
{G}^{-1}(t), & \text{ if ${t}\not \in \lbrack{t}_{1},{t}_{2}).$}%
\end{cases}
\label{FPOOL}%
\end{equation}
As we only change the allocation of qualities in this first step, we use the
subscript $q$ to refer to this variation in $[{t}_{1},{t}_{2})$. We denote the
profits generated by $(F,G)$ by $\Pi$ and the profits generated by $(F,G_{q})$
are denoted by $\Pi_{q}$. The change in profit can be expressed as follows:
\[
{\Pi}-\Pi_{q}=\int_{t_{1}}^{t_{2}}F^{-1}(t)(1-t)d{G}^{-1}(t)-\int_{t_{1}%
}^{t_{2}}F^{-1}(t)(1-t)dG_{q}^{-1}(t)-\left(  C(G)-C(G_{q})\right)  .
\]
We note that $F^{-1}(t)$ is differentiable at $\widehat{t}$ and so
$F^{-1}(t)(1-t)$ is also differentiable at this point. Thus, there exists
$r(t)$ such that:
\[
F^{-1}(t)(1-t)=F^{-1}(\widehat{t})(1-\widehat{t})+\frac{d\left[
F^{-1}(t)(1-t)\right]  }{dt}\big|_{t=\widehat{t}}(t-\widehat{t})+{r}%
(t)(t-\widehat{t}),
\]
with ${r}(t)\rightarrow0$ as $t\rightarrow\widehat{t}$. We thus get that:
\[
{\Pi}-\Pi_{q}=\int_{t_{1}}^{t_{2}}{r}(t)(t-\widehat{t})d{G}^{-1}%
(t)-\int_{t_{1}}^{t_{2}}{r}(t)(t-\widehat{t})dG_{q}^{-1}(t)-\left(
C(G)-C(G_{q})\right)  .
\]
Here we used that:
\[
\int_{t_{1}}^{t_{2}}dG^{-1}(t)=\int_{t_{1}}^{t_{2}}dG_{q}^{-1}(t)=(q_{2}%
-q_{1})\text{ and }\frac{\int_{t_{1}}^{t_{2}}G^{-1}(t)dt}{t_{2}-t_{1}}%
=\frac{\int_{t_{1}}^{t_{2}}G_{q}^{-1}(t)dt}{t_{2}-t_{1}}=\mu_{q},
\]
so when we compute the difference in revenue the first two terms of the Taylor
expansion of $F^{-1}(t)(1-t)$ cancel out. To obtain an upper bound on the
difference we first note that $G_{q}\prec_{c}G$ so $C(G_{q})\leq C(G)$ and we
can also bound the value of each integral by:
\begin{align*}
\int_{t_{1}}^{t_{2}}{r}(t)(t-\widehat{t})d{G}^{-1}(t) &  \leq(q_{2}%
-q_{1})\Delta\max_{t\in\lbrack\widehat{t}-\Delta,\widehat{t}+\Delta]}|r(t)|,\\
\int_{t_{1}}^{t_{2}}{r}(t)(t-\widehat{t})dG_{q}^{-1}(t) &  \geq-(q_{2}%
-q_{1})\Delta\max_{t\in\lbrack\widehat{t}-\Delta,\widehat{t}+\Delta]}|r(t)|.
\end{align*}
%We thus have that:
%\begin{equation*}
%{\Pi }-\Pi _{q}\leq 2 (G^{-1}(t_2)-G^{-1}(t_1)) \max_{t\in[\widehat t-\Delta,\widehat t+\Delta]}P_F(t)(t-\widehat t).
%\end{equation*}%
We thus have that:
\[
{\Pi}-\Pi_{q}\leq2(q_{2}-q_{1})\Delta\max_{t\in\lbrack\widehat{t}%
-\Delta,\widehat{t}+\Delta]}|r(t)|.
\]
We now note that:
\[
\lim_{\Delta\rightarrow0}\frac{(q_{2}-q_{1})}{\Delta}=2\frac{dG^{-1}%
(\widehat{t})}{dt}.
\]
Thus:
\[
\lim_{\Delta\rightarrow0}\frac{{\Pi}-\Pi_{q}}{\Delta^{2}}\leq4\frac
{dG^{-1}(\widehat{t})}{dt}\lim_{\Delta\rightarrow0}\ \left\{  \max
_{t\in\lbrack\widehat{t}-\Delta,\widehat{t}+\Delta]}|r(t)|\right\}  =0.
\]
The limit is because $r(t)\rightarrow0$ as $t\rightarrow\widehat{t}$.

We now compute the profit when the allocation is $G_{q}$ and we now
additionally pool the values in this interval, which we denote by $F_{v}$:
\[
F_{v}^{-1}({t})=%
\begin{cases}
\mu_{v} & \text{ if ${t}\in\lbrack{t}_{1},{t}_{2});$}\\
{F}^{-1}(t), & \text{ if ${t}\not \in \lbrack{t}_{1},{t}_{2}).$}%
\end{cases}
\]
The corresponding profits are denoted by $\Pi_{vq}.$
%If information about the values is pooled then the whole quality increment between $[t_{1},t_{2})$ is priced at $\mu _{v}$ instead of $v_{1}$.
The change in profits is:
\begin{align}
\Pi_{vq}-\Pi_{q}= &  \int_{t_{1}}^{t_{2}}F_{v}^{-1}(t)(1-t)dG_{q}^{-1}%
(t)-\int_{t_{1}}^{t_{2}}F^{-1}(t)(1-t)dG_{q}^{-1}(t)\\
= &  {({\mu}_{q}-q_{1})({\mu}_{v}-v_{1})(1-t_{1})}{}.\label{ccddz}%
\end{align}
The production cost is the same in both mechanisms, so the cost does not
appear in the difference. We obtain the second equality by noting that the
only quantile in which $F^{-1}$ and $F_{v}^{-1}$ differ and $dG_{q}^{-1}(t)$
is not zero, is the discrete quality increment $\Delta q=(\mu_{q}-q_{1})$ at
$t=t_{1}.$ Intuitively, if information about the values is pooled then the
whole quality increment is priced at $\mu_{v}$ instead of $v_{1}$. We now note
that:
\[
\lim_{\Delta\rightarrow0}\frac{({\mu}_{q}-q_{1})}{\Delta}=\frac{dG^{-1}%
(\widehat{t})}{dt};\quad\lim_{\Delta\rightarrow0}\frac{({\mu}_{v}-v_{1}%
)}{\Delta}=\frac{dF^{-1}(\widehat{t})}{dt}%
\]
Hence,%

\begin{equation}
\lim_{\Delta\rightarrow0}\frac{\Pi_{vq}-\Pi_{q}}{\Delta^{2}}=\frac
{dG^{-1}(\widehat{t})}{dt}\frac{dF^{-1}(\widehat{t})}{dt}(1-\widehat{t}%
)>0,\label{ccdc}%
\end{equation}
Hence, for a small enough $\Delta$:
\[
(\Pi_{vq}-\Pi_{q})+(\Pi_{q}-\Pi)>0.
\]
Hence, the mechanism $\left(  F,G\right)  $ is not optimal.

Finally, we consider the case in which $F^{-1}(t)$ is strictly increasing in
$[t_{1},t_{2}]$ and $G^{-1}$(t) is not strictly increasing in $[t_{1},t_{2}]$.
Without loss of generality we consider interval $[t_{1},t_{2}]$ to be such
that $G^{-1}(t)$ is constant in $(t_{1},t_{2})$ and either $t_{1}=0$ or
$G^{-1}(t)$ is increasing at $t_{1}$. If $G^{-1}(t)$ is continuously
increasing in some interval $[t_{1}-\epsilon,t_{1}]$ the first part of the
proof applies to this interval so, without loss of generality, if $t_{1}>0$ we
assume that $G^{-1}(t) $ is discontinuously increasing at $t_{1}$. Recall that
whenever two values are excluded they have the same expected valuation so we
must also have that $G^{-1}(t_{1})\neq0$. Since $G^{-1}$ is constant, pooling
information does not change the allocation. Pooling information, analogously
to \eqref{ccddz}, generate revenue gains equal to:
\[
\Pi_{vq}-\Pi_{q}= {({\mu}_{q}-q^{-}_{1})({\mu}_{v}-v_{1})(1-t_{1})}>0,
\]
where now $q_{1}^{-}\triangleq0$ if $t_{1}=0$ and $q_{1}^{-}\triangleq
\lim_{t\uparrow t_{1}}G^{-1}(t)$ if $t_{1}>0$. Hence, the mechanism is not
optimal. This concludes the proof.
\end{proof}

Before we proceed, we discuss the possibility of accumulation points in the
optimal mechanism. Accumulation points in the interior of the distribution can
be ruled out in an analogous way to the argument we just presented. However,
when the accumulation point is at the top of the distribution the arguments no
longer go through. Technically, we can see in \eqref{ccdc} that it was
necessary to take an interval bounded away from the top of the distribution
(that is, $\widehat{t}<1$). Conceptually, we need that the interval being
pooled is small relative to the survival function. In the next subsection we
use a different type of argument to rule out accumulation points at the top of
the distribution, which will also provide novel insights into the optimal mechanism.

\subsection{Optimality of Convex Menus\label{sec:cm}}
%
%We say that a mechanism is finite if the collection of quantiles $\{t_{k}\}_{k\in K}\ $is finite, and thus $K<\infty $.
%In a discrete we define the quality increments by:%
%\begin{equation}
%\Delta \overline{q}_{k}\triangleq \overline{q}(t_{k})-\overline{q}(t_{k-1}).
%\label{eq:qi}
%\end{equation}%
We previously defined a discrete menu by the expected values $\overline{v}%
_{k}$\ and the quality $\overline{q}_{k}$ at quantile $t_{k}$\ (see
\ref{eq:vk}):
\[
\overline{v}_{k}\triangleq\overline{F}^{-1}(t_{k});\ \ \ \ \ \overline{q}%
_{k}\triangleq\overline{G}^{-1}(t_{k}).
\]
We say that the quality increments display increasing differences if
\begin{equation}
\overline{q}_{k+1}-\overline{q}_{k}>\overline{q}_{k}-\overline{q}%
_{k-1},\label{eq:con}%
\end{equation}
for all $k<K$ and where $\overline{q}_{0}=0$. If the menu of offered qualities
display increasing quality increments then we say that the \emph{menu is
convex}.
%We say a pair of distributions $(G^{\ast },R^{\ast })$  are on a common quantile support if at any $t\in [0,1]$

\begin{proposition}
[Optimality of Convex Menus]\label{prop:id}\quad\newline Every optimal
mechanism generates a convex menu.
\end{proposition}

Thus, the quality increments between menu items increase as we move up the
quality ladder. This finding suggests implications for how sellers should
structure their product lines. The convexity of the menu arises with some
frequency in nonlinear pricing. In mobile phone pricing, the data packages
offered are frequently convex. Similarly, mobile phones themselves are offered
with a variety of memory chips with convex structure.\footnote{For example,
ATT offers three options : 3 GB, 15GB, 50GB for hotspot data packages.
(https://www.att.com/plans/wireless/) and apple offers memory at levels of,
128, 256, 5124GB (https://www.apple.com/shop).}

The source of the convexity can be explained as follows. Suppose we have only
two values and two quality levels and we compare the revenue generated by
separating the values and qualities relative to the revenue generated when the
values and the qualities are pooled (this trade-off is computed in
\eqref{ccdh}, with $f_{1}+f_{2}=1$ so the last term is 0). Before pooling,
quality $q_{1}$ is sold at price $v_{1}$ and the quality increment
$(q_{2}-q_{1})$ is sold at price $v_{2}.$ After pooling, the average quality
is sold at the average value. Hence, pooling increases the price at which the
quality $q_{1}$ is being sold and decreases the price at which the quality
increment $({q}_{2}-{q}_{1})$ is being sold. One can then show that if
$q_{1}>(q_{2}-q_{1})$ we have that pooling will be optimal. When pooling any
two values $v_{k},v_{k+1}$ the same trade-off appears: pooling decreases the
price at which the quality increment $(\overline{q}_{k+1}-\overline{q}_{k})$
is being sold and increases the price at which the quality increment
$(\overline{q}_{k}-\overline{q}_{k-1})$ is being sold (when $k=1$, $q_{k-1}%
=0$). Hence, a necessary condition for pooling not to be optimal is that
$(\overline{q}_{k+1}-\overline{q}_{k})>(\overline{q}_{k}-\overline{q}_{k-1}),$
which is the convex menu condition. Note that there is no counterpart to this
result in classic screening models without persuasion. In particular, when
there is no persuasion the optimality of pooling only depends on the
distribution of values and not on the distribution of qualities (see Lemma
\ref{stscr}).

The property of increasing differences allows us further to exclude the
possibility of an accumulation point in the menu, not only at the top of the
distribution, but at any point of the distribution except possibly at the
bottom of the distribution; we rule out accumulation points at the bottom of
the distribution using arguments similar to the ones presented in Section
\ref{sec:pool}. Since there are no accumulation points and the space of values
and qualities is compact, every optimal mechanism is finite. When the seller
has a fixed inventory, we can go further and bound the number of items offered
in any mechanism.

\begin{corollary}
[Finite Upper Bound on the Number of Items]\label{bi} \quad\newline With a
fixed inventory on support $\left[  q_{l},q_{h}\right]  $, the number of items
$K$\ offered by an optimal mechanism is bounded above by
\[
K<\frac{q_{h}}{q_{l}}.
\]

\end{corollary}

The lowest quality item will be at least $\overline{q}_{1}\geq q_{l}.$ Using
the convexity result of Proposition \ref{prop:id}, the $k$-th item must have
quality strictly exceeding $kq_{l}$. However, the last item, the $K$-th item,
cannot have a quality higher than $q_{h},$ so we have that $Kq_{l}<q_{h}$,
which proves the result.

The property that the menu satisfies increasing quality increments is of
interest as it informs us about the structure of the menu independent of the
distribution of values. The result predicts that in any multi-item menu the
distance between any item and its next lower ranked item is increasing as one
moves up the quality ladder, thus establishing that the menu offers qualities
that are increasing in a convex manner.

\subsection{Optimality of Monotone Partition\label{subsec:mono}}

So far we have shown that the optimal mechanism consists of a finite menu. We
now use the linear structure of the preferences to provide a sharper
characterization of the optimal information structure. More precisely, we show
that the discrete jumps of the distributions are obtained by pooling intervals
of values. The arguments employed in this section are orthogonal to those used
in the previous subsections.

A distribution of values $\overline{F}\ $is said to be \emph{monotone
partitional} if $\left[  0,1\right]  $ is partitioned into countable intervals
$\left[  t_{i},t_{i+1}\right)  _{i\in I}$ and each interval either has
complete disclosure, i.e., all buyers with values corresponding to quantiles
in that interval know their value; or pooling, i.e., buyers know only that
their value corresponds to a quantile in the interval $\left[  t_{i}%
,t_{i+1}\right)  $, and so their expected value is
\[
\overline{v}_{i}\triangleq\mathbb{E}[v\mid F(v)\in\left[  t_{i},t_{i+1}%
\right)  ]\text{.}%
\]
The expectation can be written explicitly in terms of the quantile function as
follows:
\[
\overline{v}_{i}=\frac{\int_{t_{i}}^{t_{i+1}}F^{-1}\left(  t\right)
dt}{t_{i+1}-t_{i}}.
\]
Thus writing $J$ for the labels of intervals with complete disclosure, we
have
\begin{equation}
\overline{F}^{-1}\left(  t\right)  \triangleq%
\begin{cases}
F^{-1}\left(  t\right)  , & \text{ if }t\in\left[  t_{i},t_{i+1}\right)
\text{ for some }i\in J;\\
\overline{v}_{i}, & \text{ if }t\in\left[  t_{i},t_{i+1}\right)  \text{ for
some }i\notin J.
\end{cases}
\label{eq:ep}%
\end{equation}
Proposition~1 in \cite{klms21} shows that the set $\{\overline{F}%
^{-1}:\overline{F}^{-1}\prec{F}^{-1}\}$ is a convex and compact set, and their
Theorem 1 shows that the extreme points of this set are given by \eqref{eq:ep}.

%In Figure \ref{fig1} and \ref{fig2} we illustrate a monotone partitional and a monotone pooling distribution $G$ of the same underlying distribution $F$ (represented by the dashed curve).

\begin{corollary}
[Monotone Partition of Values]\label{nonpol1}\quad\newline There exists an
optimal mechanism $(\overline{F},\overline{G})$ in which $\overline{F}$ is
monotone partitional.
\end{corollary}

The corollary shows that the optimal information structure can be constructed
in a straightforward manner: the value space is partitioned into intervals,
and buyers are only told to which interval their value belongs. We note that
the results in the previous subsections, in particular Theorem \ref{prop:int}
and Proposition \ref{prop2distr}-\ref{prop:id} do not make use of the monotone
partitional structure of the optimal information structure. \ The result of a
monotone partitional information structure relies on the linear utility in
$v$, while versions of our main results continue to hold in nonlinear utility
environment, see Theorem 5 in the working paper version (\cite{behm24}).
%The source of pooling in our model differs to those that arise in the literature. First, pooling can arise in allocation problems to increase profits (see, for example, \cite{muro78}, \cite{myer81} or \cite{lomu22} ). Second, pooling information can increase profits even when the seller is constrained to using an efficient mechanism (see, for example, \cite{bhms22}).

\subsection{Two Examples of Optimal Mechanisms\label{ilex}}

\begin{figure}[ptb]%
\centering
\includegraphics[
%natheight=11.497800in,
%natwidth=10.909000in,
height=3.9597in,
width=3.7597in
]%
{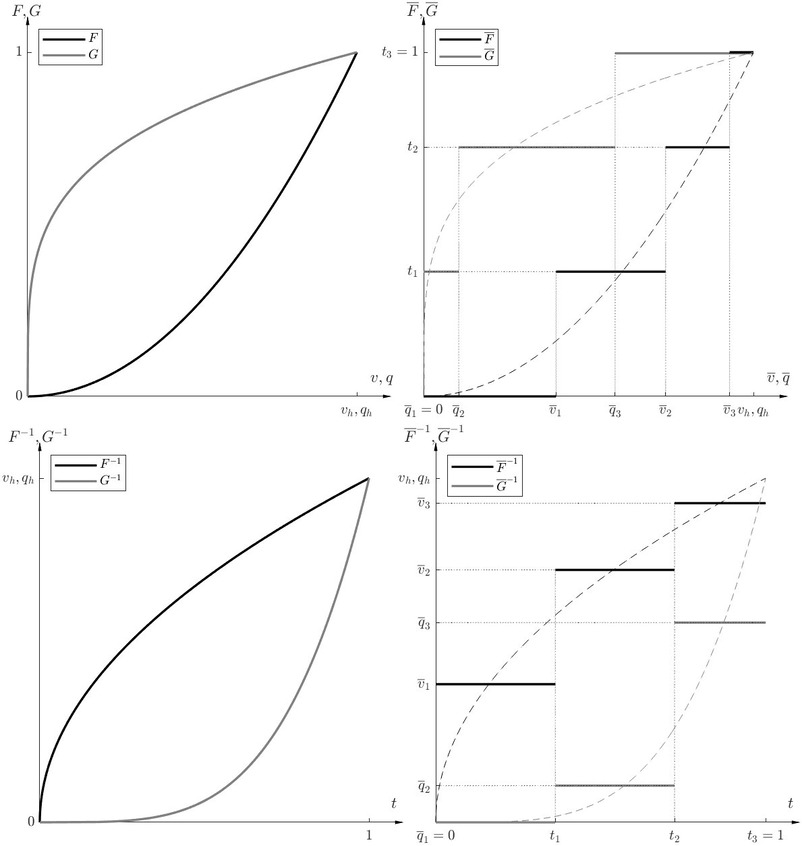}%
\caption{\label{fig3bb} The given distributions of values $F\left(  v\right)
=v^{2}$ and qualities $G\left(  q\right)  =q^{1/4}$ are depicted on the top
left. The associated optimal distributions $\overline{F}$ and $\overline{G}$,
which are monotone partitional distributions are depicted on the top right.The
corresponding quantile distributions $F^{-1}\left(  t\right)  =t^{1/2}$ and
$G^{-1}\left(  t\right)  =t^{4}$ are at the bottom left. The optimal quantile
distributions $\overline{F}^{-1}$ display jumps at the same quantiles at the
bottom right.}%
\end{figure}

We now illustrate the optimal solution in a mechanism with exogenous
inventory. A model with exogenous inventory (as in Section \ref{sec:ex})
corresponds to a cost function of the form:
\begin{equation}
C(\overline{G})=%
\begin{cases}
0, & \text{if }\overline{G}\prec_{c}G;\\
\infty, & \text{otherwise}.
\end{cases}
\label{costlm}%
\end{equation}
In this case, the cost in the optimal solution of \eqref{eq:lin} is always 0
but there is a constraint on the set of feasible qualities. That is, the
feasible distribution of qualities are $\overline{G}\prec_{c}G$. The
constraints on qualities $\overline{G}$ is almost the same as the constraint
on values $\overline{F}$, except that the seller can discard goods but cannot
discard values. In Figures \ref{fig3bb} we illustrate the optimal mechanism in
an example. The top panel illustrates the exogenous distributions of values
and qualities $\left(  F,G\right)  $\ that the seller can pool in the optimal
mechanism $(\overline{F},\overline{G})$ and the endogenous distributions of
expected values and expected qualities generated by the optimal mechanism
$(\overline{F},\overline{G})$; the bottom panel illustrates the respective
inverses.%
%TCIMACRO{\FRAME{ftbpFU}{3.7597in}{3.9597in}{0pt}{\Qcb{\label{fig3bb} The given
%distributions of values $F\left(  v\right)  =v^{2}$ and qualities $G\left(
%q\right)  =q^{1/4}$ are depicted on the top left. The associated optimal
%distributions $\overline{F}$ and $\overline{G}$, which are monotone
%partitional distributions are depicted on the top right.The corresponding
%quantile distributions $F^{-1}\left(  t\right)  =t^{1/2}$ and $G^{-1}\left(
%t\right)  =t^{4}$ are at the bottom left. The optimal quantile distributions
%$\overline{F}^{-1}$ display jumps at the same quantiles at the bottom right.}%
%}{}{Figure}{\special{ language "Scientific Word";  type "GRAPHIC";
%maintain-aspect-ratio TRUE;  display "USEDEF";  valid_file "T";
%width 3.7597in;  height 3.9597in;  depth 0pt;  original-width 10.909in;
%original-height 11.4978in;  cropleft "0";  croptop "1";  cropright "1";
%cropbottom "0";  tempfilename 'SSPTPE00.wmf';tempfile-properties "XPR";}} }%
%BeginExpansion
%EndExpansion

\begin{figure}[ptb]%
\centering
\includegraphics[
%natheight=11.497800in,
%natwidth=10.652300in,
height=3.9597in,
width=3.6716in
]%
{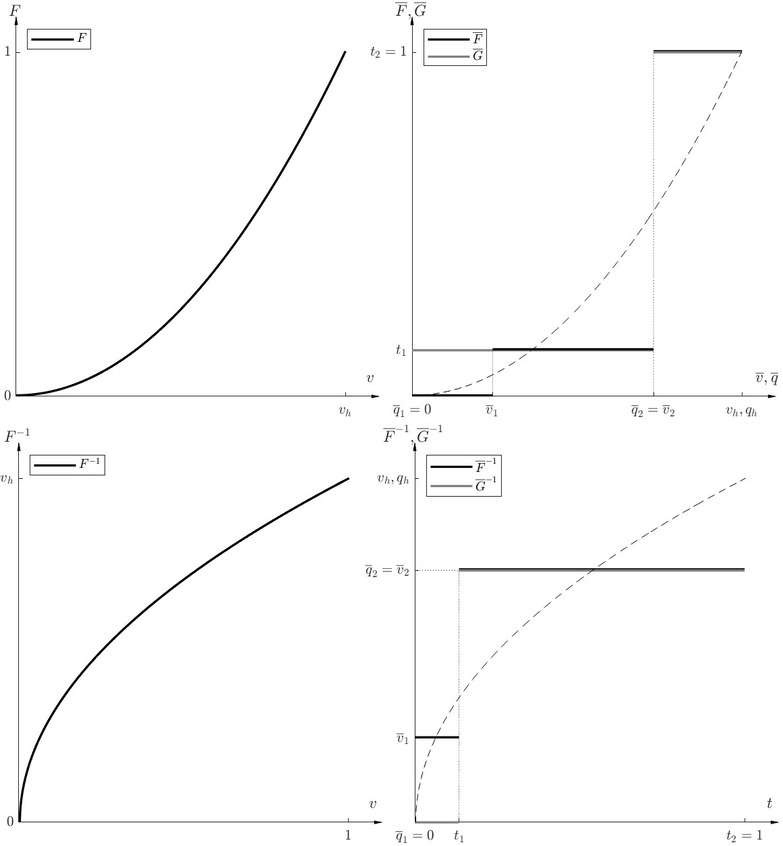}%
\caption{\label{fig4bb} The given distribution of values $F\left(  v\right)
=v^{2}$ and the optimal distributions $\overline{F}$ and $\overline{G}$ given
the cost function $c\left(  q\right)  =(1/2)q^{2}$ in the upper panel. \ The
quantile distribution $F^{-1}\left(  t\right)  =t^{1/2}$ and optimal quantile
distributions $\overline{F}^{-1}$ and $\overline{G}^{-1}$ given the cost
function $c\left(  q\right)  =(1/2)q^{2}$ in the lower panel.}%
\end{figure}

The determination of the optimal menu with a variable supply follows the same
logic. Below we display the optimal solution when the latent distribution is
again given by a quadratic distribution of values, $F\left(  v\right)  =v^{2}%
$, but the supply of qualities is determined endogenously. We compute the
solution under a quadratic cost function $c\left(  q\right)  =\left(
1/2\right)  q^{2}$. In Figures \ref{fig4bb} we illustrate the optimal
information structure and allocations. In Section \ref{sec:opti}, Theorem
\ref{th:22} establishes the optimality of a single item to be offered in the
menu under sufficient conditions on distribution and cost. The current example
with a quadratic distribution function and a quadratic cost function satisfies
these sufficient condition. As might have been expected, the solution with a
variable supply offers fewer items as it is costly to produce the supply.
Perhaps less intuitively, the solution with variable quality supply can
exclude fewer buyers than the solution with a fixed supply. But as we might
expect, conditional on being offered a positive quality level, the quality
offered is lower in the variable supply than in the fixed supply case for all
latent values.%
%TCIMACRO{\FRAME{ftbpFU}{3.6716in}{3.9597in}{0pt}{\Qcb{\label{fig4bb} The given
%distribution of values $F\left(  v\right)  =v^{2}$ and the optimal
%distributions $\overline{F}$ and $\overline{G}$ given the cost function
%$c\left(  q\right)  =(1/2)q^{2}$ in the upper panel. \ The quantile
%distribution $F^{-1}\left(  t\right)  =t^{1/2}$ and optimal quantile
%distributions $\overline{F}^{-1}$ and $\overline{G}^{-1}$ given the cost
%function $c\left(  q\right)  =(1/2)q^{2}$ in the lower panel.}}{}%
%{Figure}{\special{ language "Scientific Word";  type "GRAPHIC";
%maintain-aspect-ratio TRUE;  display "USEDEF";  valid_file "T";
%width 3.6716in;  height 3.9597in;  depth 0pt;  original-width 10.6523in;
%original-height 11.4978in;  cropleft "0";  croptop "1";  cropright "1";
%cropbottom "0";  tempfilename 'SSPTPE01.wmf';tempfile-properties "XPR";}} }%
%BeginExpansion
%EndExpansion

\section{Pooling and Entropy\label{sec:ope}}

In Section \ref{discrete} we provided a variety of results concerning the
optimal mechanism, most notably, Theorem \ref{prop:int} that establishes that
every optimal mechanism is finite. These results hold independent of the
latent distribution of values $F$. An immediate corollary is that, if the
latent distribution of values has infinite support, then the optimal mechanism
will by necessity pool information.
%Hence, a takeaway of   Theorem \ref{prop:int} is that concealing information will be part of the optimal mechanism.

Of course, if the latent distribution of values had finite support, we
wouldn't be able to reach the same conclusion. So a natural question is: can
we still assert that, in an optimal mechanism, the seller pools information
regarding the value of the buyer? We now show that the answer is positive as
long as the original distribution is not \textquotedblleft too
coarse\textquotedblright. More precisely, we show that pooling is optimal as
long as the entropy of the latent distribution of values is larger than
$\log_{2} (9)$ bits. Note that, like Section \ref{sec:ex}, but unlike Section
\ref{discrete}, we will not pursue general properties about the optimal
mechanism, but simply explore when pooling some information is optimal. And
thus, the answer will inevitably depend on the latent distribution of values.

Throughout this subsection we analyze the situation in which ${F}$ has a
distribution with support on a uniformly distanced (or equidistant)\ \ grid
$\{v_{1},...,v_{k},...\}$, with atoms denoted by $\{f_{1},...,f_{k},...\}$,
which we assume are strictly positive. The entropy of a discrete distribution
$F$ is defined by:
\begin{equation}
-\sum_{k}f_{k}\log_{2} (f_{k}).\label{eq:entr}%
\end{equation}
The entropy is a measure of the amount of information that is contained in a
random variable; we expand on the interpretation after we provide the result.
The entropy of a random variable is independent of the domain of the random
variable, so our assumption that the distribution is on a equidistant grid
does not affect how entropy is measured. We define entropy using logarithms
with base 2 so that it is measured in bits. We find an upper bound on the
entropy generated by any distribution of values (on a equidistant grid) that
satisfies a necessary condition for complete disclosure to be optimal, which
we provide next.

\begin{proposition}
[Sufficient Condition for Optimality of Pooling]\label{cor:impr}\quad\newline
Pooling improves revenue over complete disclosure if for some $k\in
\{1,2,.....\}$:
\begin{equation}
\frac{f_{k+1}}{\sum_{i\geq k}f_{i}}<\sqrt{\frac{f_{k}}{\sum_{i\geq k}f_{i}}%
}-\frac{f_{k}}{\sum_{i\geq k}f_{i}}.\label{eq:centrop22}%
\end{equation}

\end{proposition}
This condition is a generalization of Corollary \ref{dccs} and the earlier
condition \eqref{3val1}. In particular, the condition now covers the value of
pooling of \emph{any two adjacent} atoms, not only the lowest two values. We
maximize \eqref{eq:entr} subject to:
\begin{equation}
\frac{f_{k+1}}{\sum_{i\geq k}f_{i}}\geq\sqrt{\frac{f_{k}}{\sum_{i\geq k}f_{i}%
}}-\frac{f_{k}}{\sum_{i\geq k}f_{i}}.\label{eq:centrop}%
\end{equation}
which is a necessary conditions for complete disclosure to be optimal. We show
that the entropy-maximizing distribution is defined on an infinite grid and
every constraint \eqref{eq:centrop} is binding. Hence, we consider the family
of distributions in which the first element $f_{1}$ is arbitrarily chosen, and
each successive probability is found using that the constraints
\eqref{eq:centrop} are satisfied with equality.

\begin{lemma}
[Distribution with Binding Constraints]\quad\newline For any $f_{1}$, there
exists a unique distribution on an infinite support $\{\widehat{f}%
_{1},\widehat{f}_{2},...\}$ such that every constraint \eqref{eq:centrop} is
satisfied with equality and $\widehat{f}_{1}=f_{1}.$
\end{lemma}

\begin{figure}[ptb]%
\centering
\includegraphics[
%natheight=5.307400in,
%natwidth=10.256700in,
height=3.0701in,
width=5.9081in
]%
{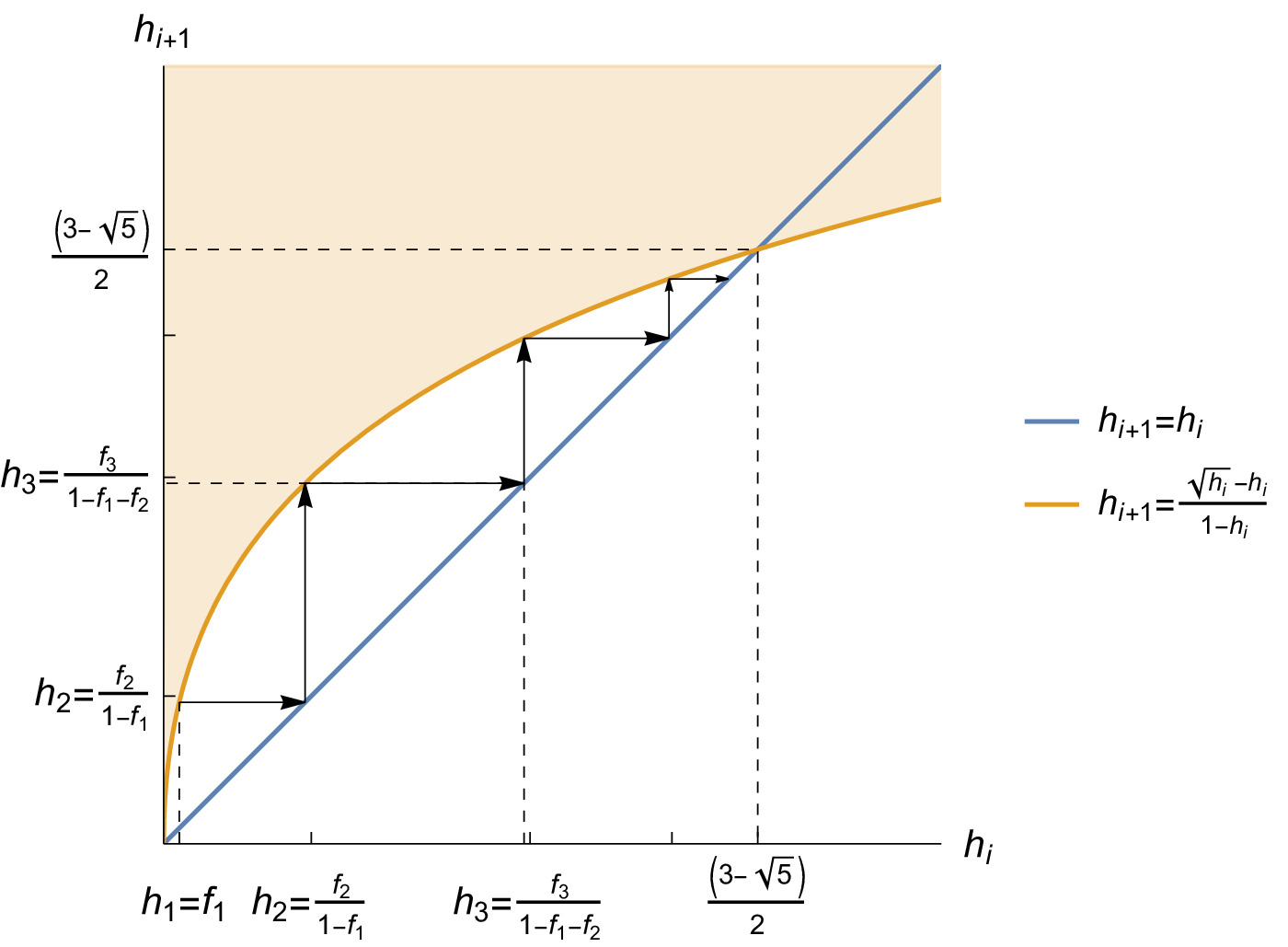}%
\caption{\label{fig0} Behavior of the hazard rate $h_{i}$ for the critical
distribution $\protect\widehat{f}$}%
\end{figure}

We write the atoms of such distribution as $\widehat{f}_{k}(f_{1})$, to
emphasize that the atoms depend on the initial condition $f_{1}$. To prove
this lemma, we write the inequality \eqref{eq:centrop} in terms of the hazard
rate $h_{i}:$
\[
h_{i}\triangleq\frac{f_{i}}{\sum_{j\geq i}f_{j}},
\]
where $h_{i}$ is the hazard rate of $i$-th atom.$\,$With this the
inequality\eqref{eq:centrop} can be written as follows:
\begin{equation}
h_{i+1}\geq\frac{\sqrt{h_{i}}-h_{i}}{1-h_{i}}.\label{dchz}%
\end{equation}
The advantage of this expression is that the set of feasible hazard rates
$h_{i+1}$ depends only on the value of $h_{i}$, and we have that $f_{1}=h_{1}%
$. Starting from the initial atom $f_{1}$, we find each successive atom using
that the hazard rate is given by \eqref{dchz} (satisfied with equality). Since
the hazard rates are always in $[0,1]$ and the hazard rates converge to a
unique fixed point at $h=\left(  3-\sqrt{5}\right)  /2$, such a distribution
can obviously be constructed for any initial probability $f_{1}$. In Figure
\ref{fig0} we illustrate \eqref{dchz}: the shaded area represents the set of
feasible hazard rates $(h_{i},h_{i+1})$; the arrows show how the hazard rates
evolve when \eqref{dchz} is binding. The hazard rates give an alternative
definition of the critical distribution that maximizes entropy.%

%TCIMACRO{\FRAME{ftbpFU}{5.9081in}{3.0701in}{0pt}{\Qcb{\label{fig0} Behavior of
%the hazard rate $h_{i}$ for the critical distribution $\protect\widehat{f}$}%
%}{}{Figure}{\special{ language "Scientific Word";  type "GRAPHIC";
%maintain-aspect-ratio TRUE;  display "USEDEF";  valid_file "T";
%width 5.9081in;  height 3.0701in;  depth 0pt;  original-width 10.2567in;
%original-height 5.3074in;  cropleft "0";  croptop "1";  cropright "1";
%cropbottom "0";  tempfilename 'SSPTPE02.bmp';tempfile-properties "XPR";}} }%
%BeginExpansion
%EndExpansion

The entropy generated by the distribution that maintains all constraints
\eqref{eq:centrop} binding-- parametrized by $f_{1}$-- is given by:
\[
E(f_{1})\triangleq-\sum_{k\geq1}\widehat{f}_{k}(f_{1})\log_{2} (\widehat{f}%
_{k}(f_{1})).
\]
In the proof of Theorem \ref{thentrop} we show that the $E\left(
f_{1}\right)  $ is maximized by taking the limit as $f_{1}\rightarrow0$. We
also show that:
\[
E^{\ast}\triangleq\lim_{f_{1}\rightarrow0}E(f_{1})
\]
is bounded between:
\[
\log_{2} (8)<E^{\ast}<\log_{2} (9)<\infty.
\]
As reference, the entropy of a uniform distribution with support on $N$
different values is given by $\log_{2} (N)$. Hence, complete disclosure is
optimal only if $F$ has an entropy below the entropy of a uniform distribution
with support on 9 different values.

\begin{theorem}
[Entropy and Optimality of Pooling Values]\label{thentrop}\quad\newline The
optimal mechanism has to pool values if the entropy of $F$\ is larger than
$E^{\ast}\in\left(  \log_{2}8,\log_{2}9\right)  $.
\end{theorem}

%The theorem shows that complete disclosure is optimal only in distributions with finite entropy and the bound is determined by $E\left( 0\right) $.
The intuition for the result is that the atoms of the individual values in the
distribution cannot have too small hazard rate. Otherwise, pooling the
allocation of adjacent atoms generates small distortions to revenue while
pooling information generates an order of magnitude larger revenue gains. With
a uniform distribution the result is particularly stark, as illustrated in
Section \ref{sec:ex}, and the same intuition goes through as long as both
atoms being pooled are of similar size. However, when the hazard rate of the
first atom is much smaller than the hazard rate of the second atom, then
pooling can decrease revenue, even if both hazard rates are small. Hence, it
may be possible to have \textquotedblleft many small\textquotedblright\ atoms
and still have complete disclosure being optimal. However, the sum of all
these small atoms will end up having a negligible amount of mass, which is
formalized by the fact that the entropy must be small.

To gain a more quantitative intuition for the result, consider a distribution
with constant hazard rate $h$, that is, a Geometric distribution. The entropy
of the Geometric distribution is:
\[
E=-\frac{h\log_{2}(h)+(1-h)\log_{2}(1-h)}{h}.
\]
As $h$ converges to 0, the entropy diverges to infinity. Let's now contrast
this with the dynamics in Figure \ref{fig0}. While the hazard rate might be
initially small, the constraints guarantee that the hazard rates quickly
increase. This leads to a bounded and relatively small entropy.
%The intuition for the result is that atoms cannot have too small hazard
%rate, as otherwise, pooling the allocation of adjacent atoms generates small
%distortions to revenue while pooling information generates an order of
%magnitude larger revenue gains. With a uniform distribution the result is
%particularly stark, as illustrated by Corollary \ref{dccs} in Section \ref%
%{sec:ex} To gain a more quantitative intuition for the result, consider the
%class of distribution with constant hazard rate $h\in \left[ 0,1\right] $
%and discrete values, that is the class of geometric distributions (the
%equivalent to the exponential distribution with a continuous variable). The
%entropy of the geometric distribution is given by:
%\begin{equation*}
%E=-\frac{h\log (h)+(1-h)\log (1-h)}{h}.
%\end{equation*}%
%As $h$ converges to 0, the entropy diverges to infinity. For the entropy of
%a geometric distribution to be smaller than $E(0)$, the constant hazard rate
%has to be larger than $0.25.$

Theorem \ref{thentrop} states that there is an upper bound on the entropy
distributions that induce complete disclosure to be optimal. The proof can be
directly extended to show that in the optimal mechanism, there cannot be
complete disclosure of any adjacent subset of values that have a
\emph{conditional entropy} larger than $E^{\ast}$. While we have assumed that
the grid is uniform, the results would remain unchanged if the grid had
increasing differences:
\[
v_{i+1}-v_{i}\geq v_{i}-v_{i-1}.
\]
Intuitively, we can see from \eqref{3val0} that in this case the conditions
for pooling continue to hold (where one obviously needs to do the appropriate
adjustments to account for all atoms and not only the first two).

\section{Optimality of a Single-Item Menu\label{sec:opti}}

So far we have established that the optimal menu only invokes a finite number
of items. We further proved that the pooling of values and thus the
compression of information is optimal when the original distribution of values
is not too coarse (as expressed by the entropy bound). A natural follow-up
question is, does the optimal mechanism offer \textquotedblleft
few\textquotedblright\ items? (Naturally, the qualifier \textquotedblleft
few\textquotedblright\ would have different meanings according to the
application.) An ideal answer would provide a sharp characterization of the
number of items offered by the optimal mechanism as a function of the cost
function and distribution of values. Such a general result is too ambitious,
so we will instead pursue a more modest goal. In this section, we provide
sufficient conditions under which the optimal menu only contains a single
item. The conditions include many natural parametrizations (for example,
uniform distribution and separable quadratic cost function) so this is a
plausible outcome in many applications. Besides being a plausible outcome, the
result will suggest that the optimal mechanism offers few items even when the
conditions are not satisfied, and the intuition for the result will be
informative of how properties of the cost function and distribution of values
can lead to more or less pooling.

The results in previous sections were derived independently of the cost
function, but a sharper characterization of the optimal mechanism necessarily
depends on the cost function. Throughout this section, we assume that the
seller has a separable cost function (see \eqref{costmr}). There are two
reasons to focus on this case. First, it is the most widely studied model of
second-degree price discrimination (see \cite{muro78}). Hence, studying how
properties of the cost function determine the number of items offered by a
mechanism is of particular importance. Second, it allow us to build a
connection between the values and the qualities (as these are endogenously
determined), which provides enough structure to characterize when the optimal
mechanism offers just one item.

To provide our single-item result we assume that the distribution of values is
absolutely continuous and its density $f$ is quasi-concave, and strictly
concave on the decreasing part of the density. Formally, we say a distribution
has \emph{modest tails} if
\begin{equation}
f^{\prime}(v)<0\Rightarrow f^{\prime\prime}(v)\leq0.\text{ }\label{eq:qc}%
\end{equation}
This condition states that $f$ must be concave when it is decreasing. For
example, any distribution with (weakly) increasing density satisfies
\eqref{eq:qc}. It is also satisfied if the density is linearly decreasing. In
contrast, the condition cannot be satisfied by any distribution with unbounded
support. We say a mechanism offers a single-item menu if the range of
$q(\overline{v})$ has at most two values and only one strictly positive value.
Hence, every buyer is either offered a standard quality $q(\overline{v})>0$ or
is excluded altogether $q(\overline{v})=0$.
%for all $t\in \lbrack 0,1]:$
%\begin{equation}
%\mathbb{E}[v^{\prime }\mid F(v^{\prime })\geq t]\leq \mu ^{\ast }+2\left( 1-%
%\sqrt{\frac{1-t}{1-F(v^{\ast })}}\right) (\mu ^{\ast }-v^{\ast }).  \tag{MT}
%\label{conda2}
%\end{equation}%
%The condition imposes an upper-bound on the conditional expected value of
%the tail of the distribution (i.e., values that are above some quantile $t$)
%based only on the threshold $v^{\ast }$ and the conditional mean $\mu ^{\ast
%}$\ of the optimal single-item mechanism. Observe that the right hand side
%is equal to $\mu ^{\ast }$ when $t=F(v^{\ast })$, and thus the inequality
%holds with equality. \ The left hand side and right hand side are both
%increasing in $t$. \ Thus the condition requires that the conditional
%expected value of the tail does not increase too fast. \ The right hand side
%equals $3\mu ^{\ast }-2v^{\ast }$ when $t=1$, and so the condition puts an
%upper bound on the support: $\bar{v}\leq 3\mu ^{\ast }-2v^{\ast }.$ From a
%technical perspective, the condition guarantees that the distribution is a
%mean-preserving contraction of an appropriately constructed distribution
%that has linear density. We give our second main result and then provide
%additional interpretation of \eqref{conda2}.

\begin{theorem}
[Optimality of Single-Item with Modest Tails]\label{th:22}\quad\newline If the
distribution $F$\ satisfies modest tails and $c^{\prime\prime\prime}(q)\geq0$,
then the optimal mechanism is a single-item menu.
\end{theorem}

Theorem \ref{th:22} shows that in large class of environments the optimal
mechanism is a single-item menu. The condition of the Theorem states that only
a single-item is offered, but it does not imply that the seller finds it
optimal to sell to all values, i.e. to pool all values. In other word, for
some distribution and cost function the optimal mechanism consists of a
single-item under which some buyers (with some values) are excluded (and hence
do not buy any positive quality). Theorem \ref{th:22} can be restated in the
value space: It is optimal to provide a binary information structure, where
the higher signal (higher expected value) suggests to buy the single item on
offer, and the lower signal suggest to not buy. To gain some intuition we
explain how the two assumptions of Theorem \ref{th:22}$\ $lead to substantial
pooling of values.

We begin by explaining why imposing a bound on the convexity of the marginal
cost is expected to guarantee that a single-item mechanism is optimal.
Consider some pooling of $F$ into finitely many atoms. The quality offered to
$\overline{v}_{i}$ is:
\[
\overline{q}_{i}={c^{\prime}}^{-1}(\phi_{i}),
\]
where $\phi_{i}$ is the discrete virtual value associated to $\overline{v}%
_{i}$ (as defined earlier in \eqref{vv}). A necessary condition for a
mechanism to be optimal is that the qualities generated by the mechanism
satisfy the increasing difference property (see Proposition \ref{prop:id});
this reflects that pooling is more likely to be optimal when the qualities
offered to different buyers are more homogeneous.
%The virtual values are increasing, so the increasing difference property is more likely to be satisfied when the marginal cost is more concave (that is, when ${c^{\prime}}^{-1}$ is more convex).
Hence, we can see how a \textquotedblleft more convex\textquotedblright%
\ marginal cost pushes the trade-off to more pooling because the qualities
generated by the mechanism (for any fixed information structure) are more
homogeneous (since ${c^{\prime}}^{-1}$ becomes more concave as $c^{\prime}$
becomes more convex).

The second condition of the Theorem is the modest tails condition, which
guarantees that the distribution of expected values generated by the mechanism
cannot be too spread out. To gain intuition we provide a necessary condition
for full pooling to be optimal. More precisely, we examine under what
circumstances it is optimal to separate an infinitesimal sliver of values at
the top of the distribution, starting from a mechanism that pools all values.

We denote by $\overline{q}_{\mu}$ the quality offered to the buyers when they
are pooled (that is, when $\overline{v}=\mu_{v}$) and $q_{h}$ the efficient
quality offered to the highest value (that is, when $\overline{v}=v_{h}$):
\[
\overline{q}_{\mu}\triangleq{c^{\prime}}^{-1}(\mu_{v})\text{ and }\overline
{q}_{h}\triangleq{c^{\prime}}^{-1}(v_{h}).
\]
When all values are pooled the seller can leave the buyers with no surplus,
but separating a small interval of values will generate some (infinitesimal)
informational rents. The informational rents gained by the buyers in the
interval that is being separated are:
\[
U\triangleq\overline{q}_{\mu}(v_{h}-\mu_{v}).
\]
That is, buyers being separated have a value of $v_{h}$ and can buy
$\overline{q}_{\mu}$ at price $p=\overline{q}_{\mu}\mu_{v}$, which gives $U $.
Separating a sliver of high values also (infinitesimally) increases total
surplus because the seller can offer them the efficient quality. The gains in
total surplus from the separation are:
\[
\Delta S\triangleq v_{h}(\overline{q}_{h}-\overline{q}_{\mu})-(c(\overline
{q}_{h})-c(\overline{q}_{\mu})).
\]
Here we assume that the separation is small enough so that it does not affect
the quality offered to the remaining pooled values.\footnote{Separating a
non-infinitesimal sliver would decrease the quality offered to the rest of the
values, hence making $\Delta S$ small. This additional effect would tilt the
balance towards more pooling, which would relax the necessary conditions for
pooling to be optimal.} The proposed separation increases profits if and only
if $U<\Delta S.$

It is easy to check that $U$ is linear in $v_{h}$ while $\Delta S$ is convex
in $v_{h}$ with:
\[
\frac{d\Delta S}{dv_{h}}\mid_{v_{h}=\mu_{v}}=0.
\]
This condition is essentially saying that the surplus loss from pooling is of
second order in the size of the distortions. Thus, when the highest value is
sufficiently separated from the mean, then separating the infinitesimal sliver
around the top increases profits, and a fortiori, full pooling will not be
optimal. The analysis suggests (although it does not prove) that when the
highest value is sufficiently close to the mean, then pooling will be optimal.

We can now relate the above intuition to the condition on the distribution of
values. If $f$ is non-decreasing, we have that $v_{h}/\mu_{v}\leq2$ (where the
inequality is tight when the distribution is uniform on $[0,v_{h}] $); if the
distribution $f$ is linearly decreasing, we then have that $v_{h}/\mu_{v}%
\leq3$. Hence, the modest tails condition imposes a bound on the difference
between the mean and the highest value of the distribution. Theorem
\ref{th:22} implies that $v_{h}/\mu_{v}\leq3$ is not a sufficiently large
difference for separating a sliver of high values to be optimal. For the
separation to be optimal, we require $f$ to be decreasing and convex, which is
when the distribution does not have modest tails. If $f$ is decreasing and
sufficiently convex $v_{h}/\mu_{v}$ may be arbitrarily large, and some
separation will always be optimal.

Theorem \ref{th:22} provides general conditions under which a single-item
mechanism is optimal. We know of distinct sufficient conditions under which a
single-item mechanism is optimal and have provided additional results on how
the cost function determines the optimal mechanism in \cite{behm24}. For
example, if the cost function is a power function $c(q)=q^{\eta}$ and the
distribution of values has narrow support, more precisely $v_{h}/v_{l}<\eta$,
then the optimal mechanism will pool all values (see Proposition 10 in
\cite{behm24}). When the cost is a power function we were also able to show
that full pooling is approximately optimal if $\eta$ is large enough and
complete disclosure is approximately optimal if $\eta$ is close enough to 1.
All these results support the intuitions we have provided in this section.
Further, in an environment with a fixed inventory of qualities (see
\eqref{costlm}), pooling all values is optimal if the distribution of
qualities has increasing density (see Theorem 3 in \cite{behm24}).
%We thus have that the conditions in Proposition \ref{th:22} are by no means also necessary for a single-item mechanism to be optimal.% However, to provide a complete characterization of the optimal mechanism, it is necessary to provide more structure on the problem

\section{Conclusion}

In the digital economy, the sellers and intermediaries working on their behalf
frequently have a substantial amount of information about the quality of the
match between their products and the preferences of the buyers. Motivated by
this, we considered a canonical nonlinear pricing problem that gave the seller
control over the disclosure of information regarding the value of the buyers
for the products offered.

We showed that in the presence of information and mechanism design, the seller
offers a menu with only a small number of items. In considering the optimal
size of the menu, the seller balances conflicting considerations of efficiency
and surplus extraction. The socially optimal menu would provide a menu with a
continuum of items to perfectly match quality and taste. By contrast, the
profit-maximizing seller seeks to limit the information rent of the buyers by
narrowing the choice to a few items on the menu. We provided sufficient
conditions for a broad class of distributions under which this logic led the
seller to offer only a single item on the menu. While we obtained our results
in the model of nonlinear pricing pioneered by \cite{muro78}, we showed that
the discrete menu result remained a robust property in a larger class of
nonlinear payoff environments.

We analyzed a canonical model of second-degree price discrimination as in
\cite{muro78} or \cite{mari84}. These models largely consider (pure)\ vertical
differentiation among the buyers and in consequence in the choice and price of
products. While vertical differentiation captures an important economic
aspect, other specifications, in particular horizontal differentiation, might
be of interest as well. Towards this end, we briefly discuss why horizontal
differentiation is likely to lead to very different implications regarding the
optimal information policy. Thus, consider a model of pure horizontal
differentiation where there are many varieties of the product, and for each
value of buyer there is some variety that attains the maximum value and all
other varieties generate a lower surplus. Thus for example a utility function
$u\left(  v,q\right)  =u-\left(  v-q\right)  ^{2}$ would represent such a
model of pure horizontal differentiation where the quadratic loss function
expresses the fact that for every value$\ v$, there is an optimal variety,
namely $q\left(  v\right)  =v$, and any deviation leads to a lower utility. In
this setting of pure horizontal differentiation, the optimal information
policy would be to completely disclose the information about the preferences,
and then provide the optimal variety $q^{\ast}\left(  v\right)  =v$ at a
constant price $p^{\ast}=u$ that would indeed extract the efficient social
surplus from all values of buyers. This admittedly stark model of pure
horizontal differentiation thus leads to a very different information policy
than the model of vertical differentiation that we analyzed. For example,
movie and TV series recommendations on Netflix and similar streaming services
would seem to mirror the implications that a model of horizontal
differentiation would predict. By contrast, service level agreements for
utilities and telecommunications \ or tiered memberships for services would
seem to be more directly related to the predictions from the vertical model we analyzed.

In related work, \cite{mcaf02} matches two given distributions of, say,
consumer demand and electricity supply, and shows how discrete matching by
pooling adjacent levels of demand and supply can approximate the socially
optimal allocation. \ In this analysis, a range of different products are
offered in the same class and with the same price. From the perspective of the
buyers, the product offered is therefore $\emph{opaque}$, as its exact
properties are not known to the buyers who is only guaranteed certain
distributional properties of the product. This practice is sometimes referred
to as \emph{opaque pricing, }see \cite{jian07} and \cite{shsh08} for
applications to services and transportation and Bergemann et al.
(2022)\nocite{bhms22} for auctions, in particular for digital advertising. Our
analysis regarding the optimality of discrete menus would equally apply if we
were to take the distribution of qualities as given and merely determine the
partition of the distribution of the qualities. The novelty in our analysis is
that the seller renders the preferences of the buyers opaque to find the
optimal trade-off between efficient matching of quality and taste against the
revenues from surplus extraction.

\newpage

\section{Appendix}

The appendix contain all auxiliary results and the proofs omitted in the main
body of the text.

\begin{proof}
[Proof of Proposition \ref{prop2distr}]Using the change of variables
\[
t=\overline{F}(\overline{v})\Leftrightarrow\overline{F}^{-1}\left(  t\right)
=\overline{v},
\]
and, given that $q(\overline{v})$ is non-decreasing, we can write the
distribution of qualities in \eqref{distrq} in terms of the quantile
$t\in\left[  0,1\right]  $:
\[
\overline{G}^{-1}(t)\triangleq q(\overline{F}^{-1}(t)).
\]
Using the expression for the payments \eqref{Eq:obj1} we have that:
\begin{align}
\mathbb{E}[p(\overline{v})]= &  \int_{v_{l}}^{v_{h}}\bigg(\overline
{v}q(\overline{v})-\int_{v_{l}}^{\overline{v}}q(s)ds\bigg)d\overline
{F}(t)\nonumber\\
= &  \int_{0}^{1}\bigg(\overline{F}^{-1}(t)\overline{G}^{-1}(t)-\int_{0}%
^{t}\overline{G}^{-1}(s)d\overline{F}^{-1}(s)\bigg)dt.\nonumber
\end{align}
Integrating by parts twice, the revenue generated by any incentive compatible
mechanism is given by:
\[
\mathbb{E}[p(\overline{v})]=\int_{0}^{1}\overline{F}^{-1}(t)(1-t)d\overline
{G}^{-1}(t)+\overline{F}^{-1}(0)\overline{G}^{-1}(0).
\]
Of course, the above is only the revenue, and to write the profit we need to
include the cost. Since $F$ must satisfy the majorization constraint, we have
that \eqref{eq:lin} is an upper bound for the profits that the seller can attain.

Consider $(\overline{F},\overline{G})$ that solve \eqref{eq:lin}. If
$\overline{G}^{-1}$ is measurable with respect to $\overline{F}(\overline{v}%
)$, then the mechanism in the proposition attains the upper bound. Suppose the
measurability condition is not satisfied, then there exists $\{t_{1},t_{2}\}$
such that $\overline{F}^{-1}(t)$ is constant in $[t_{1},t_{2}]$ and
$\overline{G}^{-1}(t)$ is not constant in $[t_{1},t_{2})$. We now define:
\[
\widehat{G}^{-1}(t)=%
\begin{cases}
\overline{G}^{-1}(t), & \text{if }t\not \in \lbrack t_{1},t_{2});\\
\frac{\int_{t_{1}}^{t_{2}}\overline{G}^{-1}(t)dt}{t_{2}-t_{1}}, & \text{if
}t\in\lbrack t_{1},t_{2}).
\end{cases}
\]
By construction we have that $\widehat{G}\prec_{c}\overline{G}.$ We thus have
that by construction $\widehat{G}$ is less costly to produce and it is easy to
check that it generates (weakly) higher revenue. Hence, we can find a solution
to \eqref{eq:lin} that satisfies the measurability condition and thus
generates an optimal mechanism.
\end{proof}

\begin{proof}
[Proof of Proposition \ref{prop:id}]In Proposition \ref{prop:opi} we showed
that the optimal mechanism is discrete. Hence, the mechanism consists of a
countable collection of values, qualities and atom sizes $(\overline{v}%
_{i},\overline{q}_{i},\overline{f}_{i})$ such that value $\overline{v}_{i}$ is
allocated quality $\overline{q}_{i}$. We now provide a lemma that provides
conditions for when a discrete mechanism is improvable.

\begin{lemma}
[Improvable Discrete Mechanism]\label{cor:impr0}\quad\newline A discrete
mechanism $\left(  \overline{F},\overline{G}\right)  $\ is not optimal if for
any two consecutive atoms $\{k,k+1\}$:%
\begin{equation}
\frac{\overline{f}_{k+1}}{(\overline{f}_{k}+\overline{f}_{k+1})\left(
1-\frac{(\overline{f}_{k}+\overline{f}_{k+1})}{\sum_{j\geq k}\overline{f}_{j}%
}\right)  }\left(  1-\frac{\overline{f}_{k}+\overline{f}_{k+1}}{\overline
{f}_{k}}\frac{\overline{q}_{k}-\overline{q}_{k-1}}{\overline{q}_{k+1}%
-\overline{q}_{k}}\right)  <\frac{\overline{v}_{k+2}-\overline{v}_{k+1}%
}{\overline{v}_{k+1}-\overline{v}_{k}}.\label{bounds23}%
\end{equation}

\end{lemma}

\begin{proof}
We introduce the language of a discrete mechanism with expected values
$\overline{v}_{k}$\ and qualities $\overline{q}_{k}$ at quantile $t_{k}$ used
earlier in (\ref{eq:vk}):
\[
\overline{v}_{k}\triangleq\overline{F}^{-1}(t_{k});\ \ \ \ \ \overline{q}%
_{k}\triangleq\overline{G}^{-1}(t_{k}).
\]
Incentive compatibility implies that $\overline{q}_{k}\leq\overline{q}_{k+1}.$
We denote by $\overline{F}_{k}$ the probability that a value is at most
$v_{k}$:
\[
\overline{F}_{k}=\sum_{\{j:\overline{v}_{j}\leq\overline{v}_{k}\}}\overline
{f}_{j}.
\]
We can write the profits as follows:
\[
\overline{\Pi}=\overline{v}_{1}\overline{q}_{1}+\sum_{k\geq2}(\overline{q}%
_{k}-\overline{q}_{k-1})\overline{v}_{k}(1-\overline{F}_{k-1}).
\]
This is the same expression as in \eqref{cdca}. If atoms $\{k,k+1\}$ are
pooled the difference in profits is given by:
\begin{align*}
\Delta\Pi= &  \left(  \frac{\overline{f}_{k}\overline{q}_{k}+\overline
{f}_{k+1}\overline{q}_{k+1}}{\overline{f}_{k}+\overline{f}_{k+1}}-\overline
{q}_{k-1}\right)  \left(  \frac{\overline{f}_{k}\overline{v}_{k}+\overline
{f}_{k+1}\overline{v}_{k+1}}{\overline{f}_{k}+\overline{f}_{k+1}}\right)
(1-\overline{F}_{k-1})\\
&  -\left(  \overline{v}_{k}(\overline{q}_{k}-\overline{q}_{k-1}%
)(1-\overline{F}_{k-1})+\overline{v}_{k+1}(\overline{q}_{k+1}-\overline{q}%
_{k})(1-\overline{F}_{k})\right) \\
&  +\left(  \overline{q}_{k+1}-\frac{\overline{f}_{k}\overline{q}%
_{k}+\overline{f}_{k+1}\overline{q}_{k+1}}{\overline{f}_{k}+\overline{f}%
_{k+1}}\right)  \overline{v}_{k+2}(1-\overline{F}_{k+1}).
\end{align*}
We have that pooling these atoms is strictly optimal only if $\Delta\Pi>0$.
Re-arranging terms, we get the above inequality (\ref{bounds23}).
\end{proof}

To conclude the proof of Proposition \ref{prop:id}, we note that if
${\ \overline{q}_{k}-\overline{q}_{k-1}}>{\ \overline{q}_{k+1}-\overline
{q}_{k}}$, then the left-hand-side of \eqref{bounds23} is negative. Thus, in
this case, the mechanism is not optimal.
\end{proof}

\medskip

\begin{proof}
[Final Step of the Proof of Theorem \ref{prop:int}]In Proposition
\ref{prop:opi} we established that the optimal mechanism is discrete and in
Proposition \ref{prop:id} we proved that any optimal mechanism generates a
convex menu of qualities. We now show that there are no accumulation points.
Proposition \ref{prop:id} implies that there cannot be any accumulation
points, except possibly at some quantile $\hat{t}$ satisfying $G^{\ast-1}%
(\hat{t})=0$. Hence, it is a decreasing accumulation point (that is, the limit
of expected values converges to $\hat{t}$ from the right) and we recall
$\phi_{i}$\ is the discrete virtual value defined earlier in \eqref{vv}.
Before we proceed, we provide a sufficient condition for a mechanism to be suboptimal.

\begin{lemma}
[Improvable Discrete Mechanism II]\label{0dc}\quad\newline A discrete
mechanism $\left(  \overline{F},\overline{G}\right)  $\ is not optimal if for
any two consecutive atoms $\{k,k+1\}$:%
\begin{equation}
\frac{\overline{f}_{k}(\phi_{k+1}-\phi_{k})(\overline{f}_{k}+\overline
{f}_{k+1})}{\overline{f}_{k+1}(\overline{v}_{k+1}-\overline{v}_{k})\sum_{j\geq
k}\overline{f}_{j}}<1.\label{bounds2}%
\end{equation}

\end{lemma}

\begin{proof}
If we omit the term in the parenthesis in \eqref{bounds23}:%
\[
\left(  1-\frac{\overline{f}_{k}+\overline{f}_{k+1}}{\overline{f}_{k}}%
\frac{\overline{q}_{k}-\overline{q}_{k-1}}{\overline{q}_{k+1}-\overline{q}%
_{k}}\right)  ,
\]
which is smaller than $1$, then we obtain the following weaker bound:
\[
\frac{\overline{f}_{k+1}}{(\overline{f}_{k}+\overline{f}_{k+1})(1-\frac
{(\overline{f}_{k}+\overline{f}_{k+1})}{\sum_{j\geq k}\overline{f}_{j}}%
)}<\frac{\overline{v}_{k+1}-\overline{v}_{k}}{\overline{v}_{k}-\overline
{v}_{k-1}}.
\]
That is, if this condition is satisfied then the mechanism is not optimal.
Re-arranging terms, we obtain \eqref{bounds2}.
\end{proof}

We now consider any \emph{infinite} optimal mechanism and show that it cannot
be optimal by proving that \eqref{bounds2} is satisfied. To make the notation
more compact, we denote the left-hand-side of the inequality as follows:
\[
B_{k}\triangleq\frac{\overline{f}_{k}(\phi_{k+1}-\phi_{k})(\overline{f}%
_{k}+\overline{f}_{k+1})}{\overline{f}_{k+1}(\overline{v}_{k+1}-\overline
{v}_{k})\sum_{j\geq k}\overline{f}_{j}}.
\]
We denote by $\{t_{k}\}_{k\leq K}$ the discontinuity quantiles, with
$t_{k}<t_{k+1}$. The index runs from $-\infty$ to $K$. We denote the expected
value in the limit by $\widehat{v}\triangleq\lim_{k\rightarrow-\infty
}\overline{v}_{k}.$That is, the accumulation point is at $\widehat{v} $. We
now define:
\[
h_{k}\triangleq\frac{\overline{f}_{k}}{\overline{v}_{k+1}-\overline{v}_{k}},
\]
and note that we can write $\phi_{k}$ as follows:
\[
\phi_{k}=\overline{v}_{k}-\frac{(1-t_{k+1})}{h_{k}}.
\]
Since $\phi_{k}$ is increasing in $k$, we must have that the limit of
$\phi_{k}$ and $h_{k}$ exists and denote the respective limits as follows:
\[
\widehat{\phi}\triangleq\lim_{k\rightarrow-\infty}\phi_{k};\quad
\widehat{h}\triangleq\lim_{k\rightarrow-\infty}h_{k}.
\]
Since $\phi_{k}$ is positive we must have that $\widehat{h}>0.$ We must also
have that $\widehat{h}<\infty.$ To prove this, we note that for every $k$:
\[
(\overline{v}_{k}-\widehat{v})(1-t_{k})=\sum_{\ell\leq k}g_{l}(\widehat{v}%
-\phi_{l})\leq(\widehat{v}-\widehat{\phi})\sum_{\ell\leq k}g_{k}.
\]
The equality is obtained by simple algebraic manipulation of the terms while
the inequality follows from the fact that the virtual values $\phi_{k}$ must
be non-decreasing. Since $(\overline{v}_{k}-\widehat{v})(1-t_{k})$ is strictly
positive for every $k$, we must have that $\widehat{v}>\widehat{\phi}$, which
implies that $\widehat{h}<\infty.$

%\begin{equation*}
%\frac{(1-t_k)g_{k+1}}{(g_k+g_{k+1})(1-t_k-(g_k+g_{k+1}))}<\frac{v_{k+1}-v_k}{%
%v_{k}-v_{k-1}}
%\end{equation*}

We now prove the result using that $\widehat{h}<\infty.$ For this, we write
the bound as follows:
\[
B_{k}\leq\frac{h_{k}(\phi_{k+1}-\phi_{k})(\overline{v}_{k+2}-\overline{v}%
_{k})\max\{h_{k},h_{k+1}\}}{(\overline{v}_{k+2}-\overline{v}_{k+1}%
)h_{k+1}(1-t_{k-1})}.
\]
Since $(\overline{v}_{k+1}-\overline{v}_{k})$ converges to 0 in the limit
$k\rightarrow-\infty,$ we can consider intervals such that $(\overline
{v}_{k+1}-\overline{v}_{k})<(\overline{v}_{k+2}-\overline{v}_{k+1})$. We thus
have that:
\[
B_{k}\leq\frac{h_{k}(\phi_{k+1}-\phi_{k})2\max\{h_{k},h_{k+1}\}}%
{h_{k+1}(1-t_{k-1})}.
\]
Since $\phi_{k}$ is monotonic in $k$ and positive, we must have that
$(\phi_{k+1}-\phi_{k})$ converges to 0 in the limit $k\rightarrow-\infty$.
Taking the limit $k\rightarrow-\infty$, and using that $h_{k}$ converges to
$\widehat{h}$, we obtain:
\[
\lim_{k\rightarrow-\infty}B_{k}\leq\lim_{k\rightarrow-\infty}\frac{h_{k}%
(\phi_{k+1}-\phi_{k})2\max\{h_{k},h_{k+1}\}}{h_{k+1}(1-t_{k-1})}=0.
\]
Thus, in the limit $k\rightarrow-\infty,$ we can find two consecutive
intervals that can be pooled and increase revenue.
\end{proof}

\begin{proof}
[Proof of Corollary \ref{nonpol1}]The choice of information structure
$\overline{F}$ must be optimal if we hold fixed a distribution $G^{\ast}$\ of
qualities. \ So we consider the problem of choosing $\overline{F}$ to
maximize
\[
\Pi=\underset{\{\overline{F}^{-1}:{F}^{-1}\prec\overline{F}^{-1}\}}{\max}%
\int_{0}^{1}\overline{F}^{-1}(t)(1-t)dG^{\ast-1}(t)+G^{\ast-1}(0)F^{\ast
-1}(0).
\]
This optimization problem is an upper semi-continuous linear functional of
$\overline{F}^{-1}$. Upper semi-continuity can be verified by noting that the
quantile function $\overline{F}^{-1}$ is (by definition) upper
semi-continuous. Hence, if $\widehat{F}^{-1}\rightarrow\overline{F}^{-1}$
(taking the limit using the $L^{1}$ norm), we have that $\lim\sup
\widehat{F}^{-1}(t)\leq\overline{F}^{-1}(t)$ for all $t\in\lbrack0,1]$. Hence,
$\lim\sup\int_{0}^{1}\widehat{F}^{-1}(t)(1-t)dG^{\ast}{}^{-1}(t)\leq\int%
_{0}^{1}\overline{F}^{-1}(t)(1-t)dG^{\ast}{}^{-1}(t)$.

Following Bauer's maximum principle (\cite{baue58}), the maximization problem
attains its maximum at an extreme point of $\{\overline{F}^{-1}:{F}^{-1}%
\prec\overline F^{-1}\}$. By Theorem \ref{prop:int}, we excluded the
possibility of intervals of complete disclosure.
\end{proof}

\begin{proof}
[Proof of Proposition \ref{cor:impr}]To obtain this result we omit the
parenthesis in \eqref{bounds23} and re-arrange terms.
\end{proof}

\begin{proof}
[Proof of Theorem \ref{thentrop}]An upper bound on the entropy generated by
any discrete distribution that induces complete disclosure to be optimal can
be found by finding the distribution that maximizes entropy subject to
\eqref{eq:centrop}:
\begin{equation}
E_{K}^{\ast}=\max_{\{f_{1},...,f_{K}\}}\ -\sum_{k\geq1}f_{k}\log_{2}
(f_{k}),\text{ subject to: \eqref{eq:centrop}}.\label{maximi}%
\end{equation}
Since \eqref{eq:centrop} is a necessary condition for complete disclosure to
be optimal, we have that this is an upper bound on the total entropy of any
distribution that induces complete disclosure to be optimal among all
distributions with a grid size of $K$. Throughout the analysis we assume that
$5\leq K<\infty$.\footnote{We disregard the cases with grid sizes less than
$5$ because the solution in these cases have a different structure; when
$K\leq4$ the entropy-maximizing distribution is the uniform distribution, so
the constraints do not bind.}

We characterize the supremum over all grid sizes
\[
E^{\ast}=\sup_{K\in\mathbb{N}}\ E_{K}^{\ast}.
\]
This is an upper bound on entropy across all discrete-distributions that
induce complete disclosure to be optimal. We will prove that the upper bound
$E^{\ast}$ is finite and also provide a way to compute the upper bound.

We begin the proof by analyzing the solution to \eqref{maximi} when $K<\infty
$. We recall that the constraints are given by \eqref{eq:centrop}, which we
can write as follows:
\begin{equation}
\frac{f_{k+1}}{\sum_{i\geq k}f_{i}}+\frac{f_{k}}{\sum_{i\geq k}f_{i}}%
-\sqrt{\frac{f_{k}}{\sum_{i\geq k}f_{i}}}\geq0.\label{eq:centrop34}%
\end{equation}
We first show that \textquotedblleft most\textquotedblright\ of the
constraints must bind.

\begin{lemma}
[Binding Constraints]\label{binding}\quad\newline If $\{f_{i}^{\ast}%
\}_{i\in\{1,...,K\}}$ solves \eqref{maximi}, then there exists $\iota
\in\{K-5,K-4,K-3,K-2,K-1\}$ such that all the constraints with index
$i\leq\iota$ in \eqref{eq:centrop} bind.
\end{lemma}

\begin{proof}
We begin with some useful definitions. We define:
\[
\mathcal{C}(x,y)\triangleq{(x+y)}-\sqrt{x},
\]
as a compact representation of the constraints given by (\ref{eq:centrop34}).
For any distribution of values $\{f_{1},...,f_{K}\}$ we define:
\[
f_{i,1}\triangleq\frac{f_{i}}{\sum_{j=i-1}^{K}f_{j}}\text{ and }%
f_{i,2}\triangleq\frac{f_{i}}{\sum_{j=i}^{K}f_{j}}.
\]
Note that $f_{i,2}$ is the hazard rate, and $f_{i,1}$ is a modified hazard
rate in which the denominator is evaluated one event before $i$. We can then
write \eqref{eq:centrop34} as follows:
\[
\mathcal{C}(f_{i,2},f_{i+1,1})\geq0\text{ for every $i\in\{1,....,K-1\}$}.
\]
We show the feasible set in Figure \ref{fig1}, where we also illustrate other
curves that will be useful in the proof. We begin by providing some results
about the feasible set.%
%TCIMACRO{\FRAME{ftbpFU}{5.9081in}{3.0701in}{0pt}{\Qcb{\label{fig1} Geometric
%representation of the constraints via $\QTR{cal}{C}\left(  x,y\right)  .$}}%
%{}{Figure}{\special{ language "Scientific Word";  type "GRAPHIC";
%maintain-aspect-ratio TRUE;  display "USEDEF";  valid_file "T";
%width 5.9081in;  height 3.0701in;  depth 0pt;  original-width 10.2567in;
%original-height 5.3074in;  cropleft "0";  croptop "1";  cropright "1";
%cropbottom "0";  tempfilename 'SSPTPE03.bmp';tempfile-properties "XPR";}} }%
%BeginExpansion
\begin{figure}[ptb]%
\centering
\includegraphics[
%natheight=5.307400in,
%natwidth=10.256700in,
height=3.0701in,
width=5.9081in
]%
{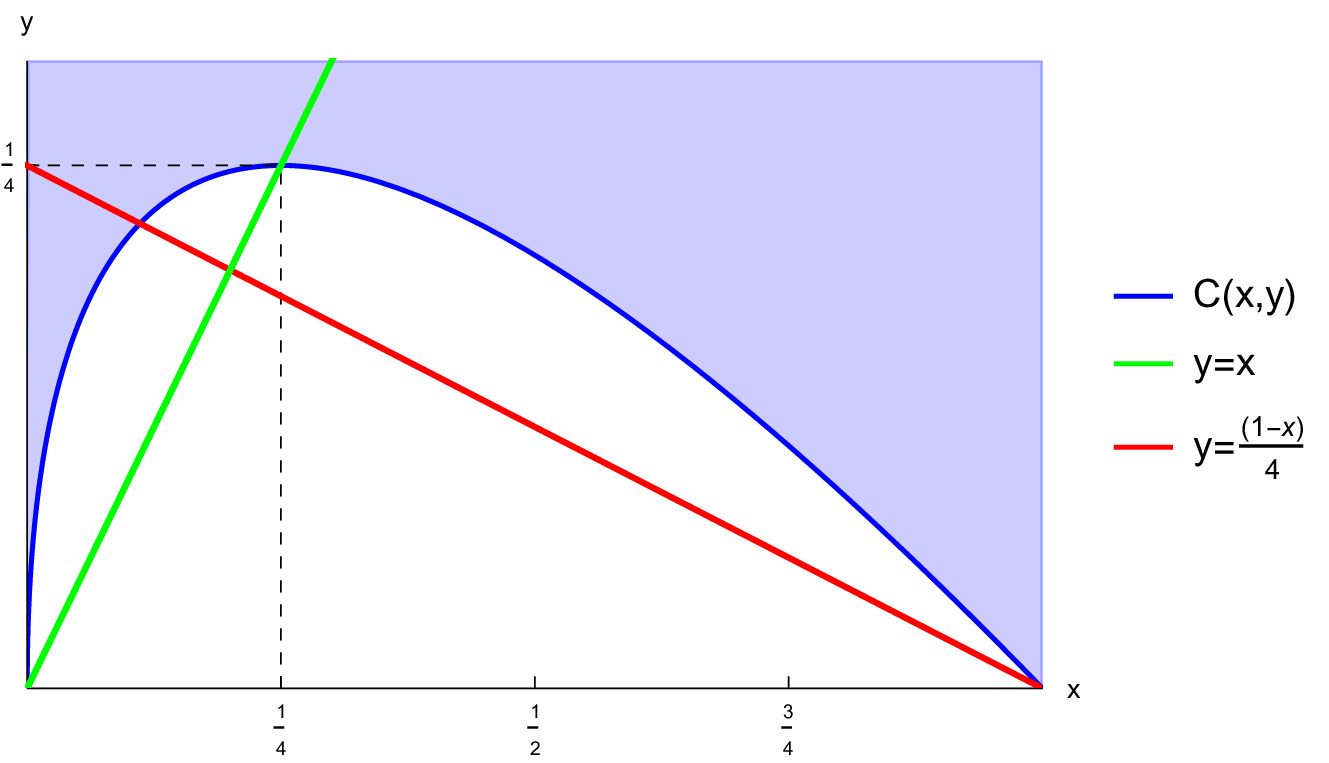}%
\caption{\label{fig1} Geometric representation of the constraints via
$\mathcal{C}\left(  x,y\right)  .$}%
\end{figure}
%EndExpansion

\bigskip

\begin{lemma}
[Geometry of the Constraints]\label{lemconst}\quad

\begin{enumerate}
\item \label{itema} If $\mathcal{C}(x,y)\geq0$, then $\mathcal{C}(x,y^{\prime
})\geq0$ for all $y^{\prime}\geq y$.

\item \label{itemb} If $\mathcal{C}(x,y)\geq0$ and $x\leq1/4$, then
$\mathcal{C}(x^{\prime},y^{\prime})>0$ for all $x^{\prime}\leq x$ and
$y^{\prime}\geq y$.

\item \label{itemc} If $\mathcal{C}(x,y)\geq0$ and $y<(1-x)/4$, then $x<y.$

\item \label{itemd} If $\mathcal{C}(x,y)\geq0$ and $x\geq1/4$ then
$y/(1-x)>1/4$.
\end{enumerate}
\end{lemma}

We now prove that some of the final constraints must bind.

\begin{lemma}
[Last Constraint is Binding]\label{eq:cx}\quad\newline There exists constraint
$i\in\{K-5,K-4,K-3,K-2\}$ such that
\[
\mathcal{C}(f_{i,2}^{\ast},f_{i+1,1}^{\ast})=0.
\]

\end{lemma}

\begin{proof}
We prove by contradiction. Suppose that all these constraints are satisfied
with slack. We then must have that:
\[
f_{K-4}^{\ast}=f_{K-3}^{\ast}=f_{K-2}^{\ast}=f_{K-1}^{\ast}=f_{K}^{\ast}.
\]
We then must have that $f_{K-4,2}^{\ast}=f_{K-3,1}^{\ast}=1/5.$ But in this
case the constraint is not satisfied:
\[
C(f_{K-4,2}^{\ast},f_{K-3,1}^{\ast})<0,
\]
so we reach a contradiction.
\end{proof}

We thus have that some constraint must bind, and we denote by $\iota
\in\{K-5,K-4,K-3,K-2\}$ the last binding constraint. That is,
\begin{equation}
C(f_{i,2}^{\ast},f_{i+1,1}^{\ast})>0,\text{ for all $i>\iota$.}\label{iota}%
\end{equation}
We now proceed to show that the constraints $i\leq\iota$ must also bind. We
assume that there exists $i\leq\iota$ such that $\mathcal{C}(f_{i,2}^{\ast
},f_{i+1,1}^{\ast})>0$ and reach a contradiction.
%Without loss of generality we can consider index  $i$ that additionally satisfies:
%\begin{equation*}
%\mathcal{C}(f_{i+1,2}^*,f_{i+2,1}^*)=0.
%\end{equation*}
%To verify it is indeed without loss of generality, it suffices to consider
%the last non-binding constraint with index lower than $\iota.$ We now

\bigskip

\emph{Case 1. } We first consider the case that
\begin{equation}
f_{i+1,1}^{\ast}\leq\frac{1-f_{i,2}^{\ast}}{4}.\label{ccd}%
\end{equation}
We note that:
\begin{equation}
f_{i+1,2}^{\ast}=\frac{f_{i+1,1}^{\ast}}{1-f_{i,2}^{\ast}},\label{cdx}%
\end{equation}
which can be obtained by simply using the definitions and manipulating the
expressions. Hence, \eqref{ccd} implies that $f_{i+1,2}^{\ast}\leq1/4$. We
then consider the following distribution:
\[
\tilde{f}_{j}=%
\begin{cases}
f_{j}^{\ast}+\epsilon, & \text{if }j=i;\\
f_{j}^{\ast}-\epsilon, & \text{if }j=i+1;\\
f_{j}^{\ast}, & \text{otherwise;}%
\end{cases}
\]
where $\epsilon$ is small enough such that $f_{i+1}^{\ast}-\epsilon>0$ and
$\mathcal{C}(\tilde{f}_{i,2},\tilde{f}_{i+1,1})>0.$

We first prove that $\{\tilde{f}\}$ is feasible. For this, we note that
\[
\mathcal{C}(\tilde{f}_{j,2},\tilde{f}_{j+1,1})=\mathcal{C}(f_{j,2}^{\ast
},f_{j+1,1}^{\ast}),\text{ for all }j\not \in \{i-1,i,i+1\},
\]
since these constraints do not change with $\epsilon$. We assumed that
$\epsilon$ is small enough such that constraint with index $j=i$ continues to
be satisfied with slack. Hence, need to prove that
\begin{equation}
\mathcal{C}(\tilde{f}_{j,2},\tilde{f}_{j+1,1})\geq0,\text{ for all }%
j\in\{i-1,i+1\}.\label{prfdS}%
\end{equation}
We first note that $\tilde{f}_{i-1,2}=f_{i-1,2}^{\ast}$ and $\tilde{f}%
_{i,1}>f_{i,1}^{\ast}$, so $\mathcal{C}(\tilde{f}_{i-1,2},\tilde{f}_{i,1})>0 $
(see item \ref{itema} of Lemma \ref{lemconst}). We now note that $\tilde
{f}_{i+1,2}<f_{i+1,2}^{\ast}$ and $\tilde{f}_{i+2,1}>f_{i+2,1}^{\ast}$, so
$\mathcal{C}(\tilde{f}_{i+1,2},\tilde{f}_{i+2,1})>0$ (see item \ref{itemb} of
Lemma \ref{lemconst}). Thus \eqref{prfdS} is also satisfied.

We now prove that $\{\tilde{f}\}$ generates higher entropy. We have that
\eqref{ccd} and $\mathcal{C}(f_{i,2}^{\ast},f_{i+1,1}^{\ast})>0$ imply that
$f_{i}^{\ast}<f_{i+1}^{\ast}$ (see item \ref{itemc} of Lemma \ref{lemconst}).
Hence, we have that
\[
|\tilde{f}_{i}-\tilde{f}_{i+1}|<|f_{i}^{\ast}-f_{i+1}^{\ast}|.
\]
Hence $\{\tilde{f}_{j}\}$ is feasible and generates higher entropy, thus
reaching a contradiction.

\bigskip

\emph{(Case 2) } We now consider the case that
\begin{equation}
f_{i+1,1}^{\ast}>\frac{1-f_{i,2}^{\ast}}{4}.\label{ccd0}%
\end{equation}
We have that \eqref{cdx} continues to be satisfied so in this case we have
that $f_{i+1,2}^{\ast}>1/4$. Furthermore, $f_{j,2}^{\ast}\geq1/4$ for all
$j>i+1$ (see item \ref{itemd} in Lemma \ref{lemconst}). Without loss of
generality, we can assume $f_{i+1}^{\ast}$ is such that:
\begin{equation}
f_{i+1}^{\ast}=\underset{j\in\{i+1,....,K\}}{\arg\max}f_{j}.\label{fcvf}%
\end{equation}
If multiple maximizers exist, then we take the lowest index that that
maximizes the atom's size. To show this is without loss of generality, let
$j^{\ast}$ be the maximizing index. We now show that we must also have that
$\mathcal{C}(f_{j^{\ast}-1,2},f_{j^{\ast},1})>0$, so we can relabel $i$ to be
$i=j^{\ast}-1$. To check this, note that $f_{j^{\ast}-1,2}>1/4$ and
$f_{j^{\ast}}>f_{j^{\ast}-1}$ so the constraint must slack.

We then consider the following distribution:
\[
\tilde{f}_{j}=%
\begin{cases}
f_{j}^{\ast}-\epsilon. & \text{if }j=i+1;\\
f_{j}^{\ast}(1+\frac{\epsilon}{\sum_{l=i+2}^{K}f_{l}^{\ast}}). & \text{if
}j\geq i+2;\\
f_{j}^{\ast}, & \text{otherwise;}%
\end{cases}
\]
where $\epsilon$ is small enough such that $f_{j}^{\ast}-\epsilon>0$ and
$\mathcal{C}(\tilde{f}_{i,2},\tilde{f}_{i+1,1})>0$. We prove this distribution
is feasible and generates higher entropy than $\{f^{\ast}\}.$

We first prove that $\{\tilde{f}\}$ is feasible. For all $j\not \in \{i,i+1\}$
we have that $C(\tilde{f}_{j,2},\tilde{f}_{j+1,1})$ is satisfied since the
constraint does not change relative to $\{f^{\ast}\}$. For $j=i$ we have that
$\mathcal{C}(\tilde{f}_{i,2},\tilde{f}_{i+1,1})>0$ because $\epsilon$ is small
enough. We thus check that $\mathcal{C}(\tilde{f}_{i+1,2},\tilde{f}%
_{i+2,1})\geq0$. If $\mathcal{C}(f_{i+1,2}^{\ast},f_{i+2,1}^{\ast})>0$, then
the constraint will be trivially satisfied when $\epsilon$ is close enough to
0. We now suppose that $\mathcal{C}(f_{i+1,2}^{\ast},f_{i+2,1}^{\ast})=0$, and
we prove that $\mathcal{C}(\tilde{f}_{i+1,2},\tilde{f}_{i+2,1})\geq0$. For
this, we note that:
\begin{align*}
-\frac{\frac{\partial C(f_{i+1,2}^{\ast},f_{i+2,1}^{\ast})}{\partial x}}%
{\frac{\partial C(f_{i+1,2}^{\ast},f_{i+2,1}^{\ast})}{\partial y}}= &
\frac{1}{2\sqrt{f_{i+1,2}^{\ast}}}-1;\\
\frac{\frac{\partial\tilde{f}_{i+2}}{\partial\epsilon}}{\frac{\partial
\tilde{f}_{i+1}}{\partial\epsilon}}= &  \frac{{\frac{f_{i+2}^{\ast}}%
{\sum_{j=i+2}^{K}f_{j}^{\ast}}}}{-1}=-f_{{i+2},2}=-\frac{\sqrt{f_{i+1,2}%
^{\ast}}-f_{i+1,2}^{\ast}}{1-f_{i+1,2}^{\ast}}.
\end{align*}
We thus have that:
\[
\frac{\frac{\partial\tilde{f}_{i+2}}{\partial\epsilon}}{\frac{\partial
\tilde{f}_{i+1}}{\partial\epsilon}}<\frac{\frac{\partial C(f_{i+1,2}^{\ast
},f_{i+2,1}^{\ast})}{\partial x}}{\frac{\partial C(f_{i+1,2}^{\ast}%
,f_{i+2,1}^{\ast})}{\partial y}}.
\]
We have that $\tilde{f}_{i+1,2}<f_{i+1,2}^{\ast}$ and $\tilde{f}%
_{i+2,1}>f_{i+2,1}^{\ast}$, so for a small enough $\epsilon$ the constraint is feasible.

We now prove that $\{\tilde{f}\}$ generates higher entropy. For this, we write
the derivative of the entropy:
\[
\frac{\partial}{\partial\epsilon}-\sum_{j=1}^{K}\tilde{f}_{j}\log_{2}%
(\tilde{f}_{j})=(\log_{2}(f_{i+1}^{\ast})+1)-\sum_{j=i+2}^{K}\frac{f_{j}%
^{\ast}}{\sum_{l=i+2}^{K}f_{l}^{\ast}}(\log_{2}(f_{j}^{\ast})+1).
\]
Recall that \eqref{fcvf} is satisfied, and hence, we have that:
\[
|\log_{2}(f_{i+1}^{\ast})|<|\log_{2}(f_{j}^{\ast})|,\text{ for all }%
j\in\{i+2,....\}.
\]
Hence, $\{\tilde{f}\}$ generates higher entropy.
\end{proof}

We now characterize the distribution that maximizes entropy for a fixed $N$.
For this, we describe the probabilities $\{f_{1,2}^{\ast},...,f_{\iota
+1,2}^{\ast}\}$. If we have these hazard rates we can recover the true
probability inductively using that:
\[
f_{i}^{\ast}=f_{i,2}^{\ast}(1-\sum_{j=1}^{i-1}f_{j}^{\ast}).
\]
Finally, $\{f_{\iota+2},...,f_{K}^{\ast}\}$ are found using that the total
probability must add up to 1 and these probabilities must all be the same.

We can find the different hazard rates $\{f_{1,2}^{\ast},...,f_{K-1,2}^{\ast
}\}$ inductively using that \eqref{eq:centrop} is satisfied with equality. For
this we define we first define:
\[
\mathcal{S}(x)\triangleq\frac{\sqrt{x}-x}{1-x}.
\]
We note that $\mathcal{C}(f_{i,2},f_{i+1,1})=0$ if and only if
\[
\mathcal{C}(f_{i,2},{f_{i+1,2}}{(1-f_{i,2})})=0.
\]
Solving for $f_{{i+1},2}$, we get that the constraint is satisfied if and only
if
\[
f_{{i+1},2}=\mathcal{S}(f_{i,2}).
\]
We denote by $\mathcal{S}^{i}$ the function composed with itself $i$ times and
$S^{0}$ is the identity function.

\begin{corollary}
[Characterization of Probabilities]\quad\newline For every $K$, $\{f_{1}%
^{\ast},...,f_{K-1}^{\ast},f_{K}^{\ast}\}$ maximizes \eqref{maximi} if and
only if, there exists $f_{0}\leq\frac{1}{2}\left(  3-\sqrt{5}\right)  $ and
$\iota\in\{K-5,...,K-2\}$ such that:
\begin{align*}
f_{i,2}^{\ast} &  =\mathcal{S}^{i-1}(f_{0}),\text{ for all $i\in
\{1,...,\iota+1\},$}\\
f_{\iota+2}^{\ast} &  =...=f_{K}^{\ast},\text{ for all $i\in\{\iota
+2,...,K\};$}%
\end{align*}
and $\{f_{\iota+2}^{\ast},...,f_{K}^{\ast}\}$ are determined by the constraint
that the probabilities must add up to one.
\end{corollary}

The Corollary shows how to find the probabilities that maximize entropy up to
two parameters $(f_{0},\iota)$. The Corollary described how to find the hazard
rates $\{f_{i,2}^{\ast}\}$, but we can recover the absolute probabilities as
follows:
\begin{equation}
f_{i}=(1-\sum_{j=1}^{i-1}f_{j})\mathcal{S}\left(  \frac{f_{i-1}}{(1-\sum
_{j=1}^{i-1}f_{j})}\right)  .\label{recurs}%
\end{equation}
For any $K$, most of the constraints will be binding (in fact, all of them
except at most 6). Motivated by this fact, we examine the entropy generated by
a distribution with infinite grid points in which every constraint binds:
\[
\widehat{E}(f_{1})=\sum_{n=1}^{\infty}f_{i}\log_{2}(f_{i}),\text{where
$\{f_{i}\}_{i>1}$ are found inductively as in \eqref{recurs}}.
\]
We now prove this is an appropriate bound for the total entropy $E^{\ast}%
$\ that can be attained by distributions that induce complete disclosure to be optimal.

\begin{lemma}
[Bound on Entropy with Infinite Support Distributions]\quad\newline As
$K\rightarrow\infty$,
\[
\lim_{K\rightarrow\infty}{E_{K}^{\ast}}=\sup_{f_{0}\in\lbrack0,1]}%
\widehat{E}(f_{0}),
\]
where possibly both sides of the equality are infinite. Furthermore the
supremum of $\widehat{E}(f)$ is attained at $f=0.$
\end{lemma}

\begin{proof}
We prove separately the cases in which $E_{K}^{\ast}\rightarrow\infty$ as
$K\rightarrow\infty$ and the case in which $E_{K}^{\ast}$ converges to a
finite number as $K\rightarrow\infty$.

\emph{(Case 1)} We first prove that if $E_{K}^{\ast}\rightarrow\infty$ as
$K\rightarrow\infty$, then $\widehat{E}$ also diverges. To prove this, we
first note that:
\[
E_{K}^{\ast}=-\sum_{i=1}^{K}f_{i}^{\ast}\log_{2}(f_{i}^{\ast})\leq-\sum
_{i=1}^{K-4}f_{i}^{\ast}\log_{2}(f_{i}^{\ast})+\log_{2}(5).
\]
Hence, if $K\rightarrow\infty$, we have that:
\[
\lim_{K\rightarrow\infty}\frac{-\sum_{i=1}^{K-4}f_{i}^{\ast}\log_{2}%
(f_{i}^{\ast})}{E_{K}^{\ast}}=1.
\]
We now note that:
\[
-\sum_{i=1}^{K-4}f_{i}^{\ast}\log_{2}(f_{i}^{\ast})\leq\widehat{E}(f_{1}%
^{\ast}).
\]
Hence, if $E_{K}^{\ast}\rightarrow\infty$ as $K\rightarrow\infty$, then we
must also have that:
\[
\sup_{f\in\lbrack0,1]}\widehat{E}(f)=\infty.
\]
Note that, if $\widehat{E}(f)$ diverges, then it must also diverge in the
limit $f\rightarrow0.$ To verify this, note that if $\widehat{E}(f)=\infty$
for some $f$, then $\widehat{E}(\mathcal{S}^{-1}(f))=\infty$. Iterating the
inverse of $\mathcal{S}$, we would get that $\widehat{E}(f)=\infty$ in the
limit $f\rightarrow0.$

\emph{(Case 2)} We now prove that, if
\begin{equation}
\lim_{k\rightarrow\infty}E_{K}^{\ast}<\infty,\label{dfcv}%
\end{equation}
then it converges to the supremum of $\widehat{E}(f)$ which is attained in the
limit $f\rightarrow0$. We first prove that $f_{K}^{\ast}\rightarrow0$ as
$K\rightarrow\infty$. To prove this, suppose the grid size increases by 1 grid
point so it is labelled as $\{1,...,K,K+1\}$. In this case, we can split
$f_{K}^{\ast}$, into two atoms, say $\widetilde{f}_{K}^{\ast}$ and
$\widetilde{f}_{K+1}^{\ast}$ with
\[
\mathcal{C}(\frac{f_{K-1}^{\ast}}{f_{K}^{\ast}+f_{K-1}^{\ast}},\frac
{\widetilde{f}_{K}^{\ast}}{f_{K}^{\ast}+f_{K-1}^{\ast}})=0\text{ and
}\widetilde{f}_{K}^{\ast}+\widetilde{f}_{K+1}^{\ast}=f_{K}^{\ast}.
\]
This would clearly be a feasible distribution. If $f_{K}^{\ast}$ is bounded
away from zero, the increase in entropy will be bounded away from zero.
However, if \eqref{dfcv} is satisfied we then must have that the entropy
increments as the grid size increases must converge to 0. Hence, we must have
that $f_{K}^{\ast}\rightarrow0$ as $K\rightarrow\infty$.

The fact that $f_{K}^{\ast}\rightarrow0$ as $K\rightarrow\infty$ implies
that:
\[
\lim_{K\rightarrow\infty}E_{K}^{\ast}=-\lim_{K\rightarrow\infty}\sum
_{i=1}^{\iota+1}f_{i}^{\ast}\log_{2}(f_{i}^{\ast}),
\]
where $\iota$ is the last binding constraint (as defined in \eqref{iota}).
That is, in the limit we can omit the contribution of the last atoms to the
entropy. However, we also have have that:
\[
\lim_{K\rightarrow\infty}\sum_{i=1}^{\iota}f_{i}^{\ast}\log_{2}(f_{i}^{\ast
})=\lim_{K\rightarrow\infty}\widehat{E}(f_{1}^{\ast}),
\]
where the limit $K\rightarrow\infty$ also includes how the probability
$f_{i}^{\ast}$ change with $K$. Thus, if \eqref{dfcv} is satisfied,
\[
\lim_{K\rightarrow\infty}E_{K}^{\ast}=\lim_{K\rightarrow\infty}\widehat{E}%
(f_{1}^{\ast}).
\]
We must also have that:
\[
\lim_{K\rightarrow\infty}\widehat{E}(f_{1}^{\ast})=\sup_{f\in\lbrack
0,1]}\widehat{E}(f),
\]
as the supremum of $\widehat{E}(f)$ can be attained by $E_{K}^{\ast}$ in the
limit $K\rightarrow\infty$ by the distribution $\{\widehat{f}_{k}(f)\}$.
Hence, the supremum of $\widehat{E}(f)$ is attained by the limit of
$f_{1}^{\ast}$, as $K\rightarrow\infty.$

Finally, we verify that $f_{1}^{\ast}\rightarrow0$ as $K\rightarrow\infty$. To
prove this, suppose the grid size increases by 1 grid point so it is labelled
as $\{0,1,...,K\}$. We can then construct the following distribution on the
new grid:
\[
\widetilde{f}_{i}^{\ast}=(1-\widetilde{f}_{0}^{\ast})f_{i}^{\ast},\text{ for
all $i\in\{1,...,K\}$}.
\]
Finally, we explain how to determine $\widetilde{f}_{0}^{\ast}$. We define
$x,y$ as follows:
\begin{align*}
x\triangleq &  \max\{w\in\lbrack0,1]:\mathcal{C}(z,f_{1}^{\ast})\geq0\text{
for all }z\leq w\};\\
y\triangleq &  \underset{z\in\lbrack0,1]}{\arg\max}\ \{z\log_{2}%
z+(1-z)\log_{2}(1-z)+(1-z)E_{K}^{\ast}\},
\end{align*}
and let
\[
\widetilde{f}_{0}^{\ast}\triangleq\min\left\{  x,y\right\}  .
\]
This generates a feasible distribution and if $f_{1}^{\ast}$ is bounded away
from zero, the entropy increment will be bounded away from zero. Hence, if
\eqref{dfcv} is satisfied we then have that $f_{1}^{\ast}\rightarrow0$ as
$K\rightarrow\infty$. So the supremum of $\widehat{E}(f)$ is attained at
$f=0.$ We thus obtain the result.
%To prove this
%For this, we note that:
%\begin{equation*}
%E^*_{K+1}>E^*_K+f_K^*\log_2(2).
%\end{equation*}
%We obtain this bound by considering the distribution on a grid of $K+1$
%points that consists in splitting the last atom $f_K^*$ into two atoms of
%equal probabilities (and all the other atoms remain the same). If $%
%f_{K-1}^*=f_{K}^*$, then this yields a feasible distribution and the entropy
%gains are exactly $f_K^*\log_2(2).$ We thus have that, if
%\begin{equation*}
%\lim_{K\to\infty}E^*_K<\infty,
%\end{equation*}
%then we must have that $f_K^*\to0$ as $K\to\infty$. Note that the same
%argument implies that $f_1^*\to0$ as $K\to\infty$, where we only need to
%``split'' the first atom instead of the last one. We thus have that:
%\begin{equation*}
%\lim_{K\to\infty}E^*_K=-\lim_{K\to\infty}\sum_{i=1}^{\iota+1}
%f_i^*\log_2(f_i^*),
%\end{equation*}
%where $\iota$ is the last binding constraint (as defined in \eqref{iota}).
%However, we also have have that:
%\begin{equation*}
%\lim_{K\to\infty}\sum_{i=1}^\iota f_i^*\log_2(f_i^*)=\lim_{K\to\infty}
%\widehat E(f_1^*),
%\end{equation*}
%where the limit as $K$ grows also includes the probability $f_i^*.$
%Furthermore, we have that $f_1^*\to0$ as $K\to\infty$, so the supremum of $%
%\widehat E(f)$ is attained at $f=0.$ We thus obtain the result.

\end{proof}

While are ultimately interested only in the maximum of $\widehat{E}(f)$, we
can find this maximum by characterizing the function $\widehat{E}\left(
\cdot\right)  \ $for all $f$. We can write the function $\widehat{E}\left(
\cdot\right)  $\ recursively as follows:
\begin{equation}
\widehat{E}(f)=-f\log_{2}(f)-(1-f)\log_{2}(1-f)+(1-f)\widehat{E}%
(\mathcal{S}(f)).\label{linf}%
\end{equation}
We can now prove that $\widehat{E}(0)$ is finite.

\begin{lemma}
[Properties of $\widehat{E}(f)$]\quad\newline The supremum of $\widehat{E}%
\left(  \cdot\right)  $ is finite.
\end{lemma}

\begin{proof}
We already proved the supremum of $\widehat{E}(f)$ is attained at $f=0.$ We
can write \eqref{recurs} alternatively as follows:
\[
\frac{\widehat{E}(f)-(1-f)\widehat{E}(\mathcal{S}(f))}{f-\mathcal{S}(f)}%
=\frac{-f\log_{2}(f)-(1-f)\log_{2}(1-f)}{{f-\mathcal{S}(f)}}.
\]
Taking limits $f\rightarrow0$, we obtain:
\begin{align*}
\lim_{f\rightarrow0}\frac{\widehat{E}(f)-(1-f)\widehat{E}(\mathcal{S}%
(f))}{f-\mathcal{S}(f)}= &  \widehat{E}^{\prime}(0);\\
\lim_{f\rightarrow0}\frac{-f\log_{2}(f)-(1-f)\log_{2}(1-f)}{{f-\mathcal{S}%
(f)}}=0. &
\end{align*}
We thus get that $\widehat{E}^{\prime}(0)=0$, which implies that
$\widehat{E}(0)<\infty$.
\end{proof}

We can now conclude the proof of Theorem \ref{thentrop} by computing
$\widehat{E}(0).$ We observe that \eqref{linf} is a linear functional and it
is thus easy to verify numerically that $\widehat{E}(0)\approx3.04$ while
$\log_{2}\ 9\approx3.17$ and $\log_{2}\ 8=3.$ We thus obtain the result of
Theorem \ref{thentrop}.
\end{proof}

\begin{proof}
[Proof of Theorem \ref{th:22}]We establish Theorem \ref{th:22} through a
sequence of optimal menus for successively richer environments. We begin with
the optimal single-item menu.

Suppose that the seller were constrained to only offer a single item, which
item would he then offer and at what price? The optimal single-item menu is
found by solving the following problem:
\begin{equation}
(q^{\ast},v^{\ast})\in\underset{q,v\in\mathbb{R}}{\arg\max}\ \mathbb{P}%
[v^{\prime}\geq v](\mathbb{E}[v^{\prime}\mid v^{\prime}\geq
v]q-c(q)).\label{opt}%
\end{equation}
We denote by $\mu^{\ast}$ the expectation of $v$ conditional on $v$\ exceeding
$v^{\ast}$:
\[
\mu^{\ast}\triangleq\mathbb{E}[v^{\prime}\mid v^{\prime}\geq v^{\ast}].
\]
The single-item mechanism consists of selling quality $q^{\ast}$ at a price
$p^{\ast}=\mu^{\ast}q^{\ast}$, which is sold to all values higher than
$v^{\ast}$. The buyer is only informed whether he should buy the good. Note
that the buyer is left with no surplus.

\begin{lemma}
[Single Item Menu]\label{single} \quad\newline The optimal single item menu
satisfies the following first-order conditions:
\begin{equation}
\mu^{\ast}=c^{\prime}(q^{\ast})\quad\text{ and }\quad{c(q^{\ast})}=v^{\ast
}q^{\ast},\label{foc}%
\end{equation}

\end{lemma}

\begin{proof}
The first-order condition is obtained by taking the derivative of \eqref{opt}
with respect to $v$ and $q$ and equating these to zero.
\end{proof}

The first condition states that the quality is efficiently supplied given that
the (expected) value of the buyer who buys the good is $\mu^{\ast}$. The
second condition states that the threshold $v^{\ast}$ is also efficiently
chosen: given that $q^{\ast}$ quality is going to be supplied, it is efficient
to sell to a buyer with value $v$ if and only if the utility he obtains from
this quality is larger than the cost of producing it. We note that the second
equality might eventually be satisfied by some $v^{\ast}<v_{l}$, which means
there is no exclusion. There are no distortions in the quality supplied (given
the threshold $v^{\ast}$) and there is no distortion in the threshold $v$
(given the quality $q^{\ast}$) because in a single-item mechanism there is
zero buyer surplus. So these quantities are not distorted to reduce consumer
surplus. In general, when the optimal mechanism is a multi-item mechanism,
both the thresholds and the qualities provided are distorted to reduce the
consumer surplus.

The rest of the proof proceeds as follows. We first show that when the cost is
quadratic and the density is linearly decreasing, the optimal mechanism is a
single-item mechanism. We then show that the optimal mechanism is a
single-item mechanism when the marginal cost is convex and the density is
linearly decreasing. Finally, we show that the distributions \eqref{eq:qc} are
mean-preserving contractions of an (appropriately constructed)
linearly-decreasing density, which we use to prove that the optimal mechanism
is a single-item mechanism.

Before we begin, we introduce some notation. An optimal mechanism is described
by cutoffs $\{w_{1},...,w_{K-1}\}$, with $w_{K}\triangleq v_{h}$, such that:
\[
\bar{f}_{k}=\int_{w_{k-1}}^{w_{k}}dF(v);\quad\bar{v}_{k}=\int_{w_{k-1}}%
^{w_{k}}vdF(v).
\]
The qualities offered are denoted by $\{q_{1},...,q_{K}\}$.
%\[q_k={c^{\prime}}^{-1}\left(\overline v_k-\frac{(\overline v_k-\overline v_{k-1})\sum_{j=k+1}^K\overline f_j}{\overline f_k}\right).\]
To make the notation more compact, we define: $\Delta q_{k}=q_{k}-q_{k-1}$.
Recall that in a finite-item menu, the revenue can be written as in
\eqref{cdca}. We consider the optimality conditions of the highest two
intervals of an optimal mechanism. For this, we define the profit from the
highest two items:
\begin{align}
\Pi_{K-1,K}(w_{K-1},\Delta q_{K-1},\Delta q_{K})  & \triangleq\big(({\overline
{f}}_{K-1}+{\overline{f}}_{K})\Delta q_{K-1}\overline{v}_{K-1}-{\overline{f}%
}_{K-1}c(q_{K-2}+\Delta q_{K-1})\label{eq:lt}\\
& +{\overline{f}}_{K}\left(  \Delta q_{K}\overline{v}_{K}-c(q_{K-2}+\Delta
q_{K-1}+\Delta q_{K})\right)  \big),\nonumber
\end{align}
which are the last two terms of the summations in \eqref{cdca} and subtracting
the respective cost. If the optimal mechanism is a multi-item mechanism, the
solution to the following problem:
\[
\Pi_{K-1,K}^{\ast}=\max_{{\footnotesize
\begin{matrix}
w_{K-1}\in\lbrack w_{K-2},v_{h}],\\
\Delta q_{K-1},\Delta q_{K}\geq0
\end{matrix}
}}\ \Pi_{K-1,K}(w_{K-1},\Delta q_{K-1},\Delta q_{K})
\]
must satisfy $\Delta q_{K-1},\ \Delta q_{K}>0$ and $w_{K-2}<w_{K-1}<v_{h}$,
where $q_{K-2}$ and $w_{K-2}$ are parameters that are kept fixed in the
optimization problem. We show the optimal mechanism is a single-item mechanism
in a specific parametrization of the model.

\begin{proposition}
[Linear Density and Quadratic Cost Environment]\label{qq}\quad\newline The
optimal menu is always a single-item menu when the density is
linearly-decreasing and the cost is linear-quadratic.
\end{proposition}
\end{proof}

\begin{proof}
We analyze the optimal mechanism when the distribution of values is given by:
\[
{L}(v;v_{l},v_{h})\triangleq\frac{(v-v_{l})(2v_{h}-v_{l}-v)}{(v_{h}-v_{l}%
)^{2}}.
\]
The density of this distribution, which we denote by $l(v;v_{l},v_{h})$, is
linearly-decreasing with zero density at the top of the support:
\[
l(v_{h};v_{l},v_{h})=0.
\]
We begin by proving that a single-item mechanism is optimal when the cost is
linear-quadratic:
\begin{equation}
{c}(q)\triangleq\alpha q+\frac{\beta q^{2}}{2}+\gamma.\label{eq:c}%
\end{equation}
The fixed cost $\gamma$ plays no role in the analysis and is added to the cost
function only to simplify the exposition of some arguments.
%so it corresponds to the maximization of $w_{K-1},\Delta q_{K-1},\Delta q_{K}$ (while omitting the terms that do not depend on these variables).

Given the quadratic cost function, the optimality conditions for $q_{K-1}$ and
$q_{K}$ are:
\begin{equation}
\Delta q_{K-1}=\max\left\{  \frac{\overline{v}_{K-1}-\alpha-\beta q_{K-2}%
}{\beta}-\frac{(\overline{v}_{K}-\overline{v}_{K-1}){\overline{f}}_{K}}%
{\beta{\overline{f}}_{K-1}},0\right\}  \text{ and }\Delta q_{K}=\frac
{\overline{v}_{K}-\alpha-\beta q_{K-1}}{\beta}.\label{eq:dif}%
\end{equation}
Hence, $\Delta q_{K-1}>0$ only if
\begin{equation}
\frac{\overline{v}_{K}-\alpha-\beta q_{K-2}}{\overline{v}_{K-1}-\alpha-\beta
q_{K-2}}<\frac{{\overline{f}}_{K-1}+{\overline{f}}_{K}}{{\overline{f}}_{K}%
}.\label{eq:const}%
\end{equation}
%
%Throughout the rest of the proof, we assume that $\alpha=q_{K-2}=0$ and $\beta=1$. This does not change the maximization problem we solve, as the general case can be analyzed the same way but translating and rescaling values so that $\tilde v=(v-\alpha-\beta q_{K-2})/\beta$.
To write expressions that are compact, we define:
\[
z\triangleq\frac{v_{h}-w_{K-1}}{v_{h}-w_{K-2}};\quad\kappa\triangleq
3\frac{{w_{K-2}-\alpha-\beta q_{K-2}}}{v_{h}-w_{K-2}}.
\]
%
%Using this change of variables, we have that:
%\begin{equation*}
%w_{K}=\frac{(\bar{v}-w_{K-2})(3-2z+\kappa )}{3};\quad w_{K-1}=\frac{(\bar{v}%
%-w_{K-2})(1-\frac{2z^{2}}{1+z}+\kappa )}{3};\quad \frac{({\overline f}_{K-1}+{\overline f}_{K})}{%
%{\overline f}_{K}}=\frac{1}{z^{2}}.
%\end{equation*}%
Note that in an optimal mechanism $\kappa\geq0$, as otherwise the mechanism
would be offering a quality $q_{K-2}$ whose marginal cost is higher than the
value for all values $v\in\lbrack w_{K-3},w_{K-2}]$, which is clearly
suboptimal. Using these definitions, we have that \eqref{eq:const} is
satisfied if and only if:
\begin{equation}
\frac{3-2z+\kappa}{1-\frac{2z^{2}}{1+z}+\kappa}<\frac{1}{z^{2}}%
.\label{eq:const2}%
\end{equation}
And, for every $z$ satisfying \eqref{eq:const2}, \eqref{eq:lt} can be written
as follows:
%{\footnotesize
%\begin{equation*}
%\Pi _{K,K-1}=\max_{v\in[w_{K-2},\bar v]}\frac{ {(2 v+\bar v) (v-\bar v)^2
%\left(2 v-\bar v-\frac{4 (v-\bar v)^2}{v-2 \bar v+w_{K-2}}-4 w_{K-2}\right)}%
%+(\bar v-w_{K-2})^2\left(\frac{2 (v-\bar v)^2}{v-2 \bar v+w_{K-2}}+\bar v+2
%w_{K-2}\right)^2}{18 \left((\bar v-w_{K-2})^2- (v-\bar v)^2\right)}.
%\end{equation*}}
%\begin{align*}
%\Pi _{K,K-1}=\max_{v\in[w_{K-2},\bar v]}&\frac{1}{\beta}\frac{ {(2 v+\bar v-3\alpha) (v-\bar v)^2
%\left(2 v-\bar v-\frac{4 (v-\bar v)^2}{v-2 \bar v+w_{K-2}}-4 w_{K-2}+3\alpha\right)}}{18 \left((\bar v-w_{K-2})^2- (v-\bar v)^2\right)}
%\\&+\frac{1}{\beta}\frac{ (\bar v-w_{K-2})^2\left(\frac{2 (v-\bar v)^2}{v-2 \bar v+w_{K-2}}+\bar v+2
%w_{K-2}-3\alpha\right)^2}{18 \left((\bar v-w_{K-2})^2- (v-\bar v)^2\right)}.
%\end{align*}%
\begin{equation}
\Pi_{z}\triangleq\frac{({\overline{f}}_{K-1}+{\overline{f}}_{K})(v_{h}%
-w_{K-2})^{2}}{18\beta\left(  1-z^{2}\right)  }\left(  {{(3-2z+\kappa
)z^{2}\left(  1-2z+\frac{4z^{2}}{1+z}-\kappa\right)  }}+{\left(
1+\kappa-\frac{2z^{2}}{1+z}\right)  ^{2}}\right)  .\label{eq:vz}%
\end{equation}
%
%\begin{align*}
%\Pi _{K,K-1}=\max_{z\in[0,1]}& \ \ \frac{(\bar v-w_{K-2})^2}{\beta18 \left(1- z^2\right)}\left( { {(3-2z+\kappa)z^2
%\left(1-2z+\frac{4 z^2}{1+z}-\kappa\right)}}{}+ { \left(-\frac{2 z^2}{1+z}+1+\kappa\right)^2}{ }\right);
%\\&\text{subject to: \ }\frac{ (3-2z+\kappa)}{ (1-\frac{2z^2}{1+z}+\kappa)}\leq\frac{1}{z^2}.
%\end{align*}%
%Here the constraint guarantees that:
%\[\Delta q_{K-1}=\frac{w_{K-1}-\alpha}{\beta}-\frac{(w_K-w_{K-1}){\overline f}_K}{\beta {\overline f}_{K-1}},\]
%i.e., instead of being 0 in \eqref{eq:dif}.
More precisely, we have that
\[
\Pi_{z}=\Pi_{K-1,K}+({\overline{f}}_{K-1}+{\overline{f}}_{K})c(q_{K-2}),
\]
when \eqref{eq:const} is satisfied. The last term on the right-hand-side is
constant from the optimization perspective, so we can just focus on optimizing
$\Pi_{z}$ over $z.$ If the optimal mechanism is a multi-item mechanism, there
must exist $z\in(0,1)$ satisfying \eqref{eq:const2} that maximizes \eqref{eq:vz}.

If $z^{\ast}$ maximizes \eqref{eq:vz} and satisfies \eqref{eq:const2} with
strict inequality, then $z^{\ast}$ must satisfy the first- and second-order
conditions. However, there is no $z^{\ast}\in(0,1)$ that satisfies the first-
and second-order conditions:
%\footnote{The objective function is a rational expression and we can verify that thereis no $v$ that satisfies the first- and second-order conditions.}
\[
\frac{\partial\Pi_{z}}{\partial z}\bigg|_{z=z^{\ast}}=0\quad\text{and}%
\quad\frac{\partial^{2}\Pi_{z}}{\partial z^{2}}\bigg|_{z=z^{\ast}}\leq0.
\]
Hence, there is no interior solution. This is a contradiction, so in the
optimal mechanism $\Delta q_{K-1}=0$.
\end{proof}

We now deploy the argument for the optimality of a single-item menu beyond the
quadratic model. Towards this end, we define the solution to a restricted
optimization problem for a linear-quadratic cost function with parameters
$\alpha$ and $\beta$:
\begin{equation}
(v^{\ast}(\alpha,\beta),\Delta q_{K}^{\ast}(\alpha,\beta))\triangleq
\underset{0\leq\Delta q,w_{K-2}\leq w\leq v_{h}}{\arg\max}\Pi_{K-1,K}%
(w,0,\Delta q),\label{eq:opt1}%
\end{equation}
where we define the optimal quality for the last interval:
\[
q_{K}=q^{\ast}(\alpha,\beta)\triangleq q_{K-2}+\Delta q_{K}^{\ast}%
(\alpha,\beta).
\]
Thus, we consider a restricted optimization problem where the seller takes as
given the first $K-2$ intervals and allocations. The restricted problem
(\ref{eq:opt1}) is then to find an interval $(w_{K-1},v_{h}]=(v^{\ast}%
(\alpha,\beta),v_{h}]$ and an allocation $q_{K}^{\ast}(\alpha,\beta)$ so as to
maximize the profit from all values in the given interval $(w_{K-2},v_{h}]$.
This restricted maximization problem allows the interval $(w_{K-1},v_{h}]$ to
be a strict inclusion of $(w_{K-2},v_{h}]$: that is, $(w_{K-1},v_{h}%
]\subsetneq(w_{K-2},v_{h}]$. In this case, all the values in the interval
$(w_{K-2},w_{K-1})$ will receive the allocation $q_{K-2}$. Now, from
Proposition \ref{qq}, we know that when the cost is linear-quadratic:
\[
\Pi_{K-1,K}(w_{K-1},\Delta q_{K-1},\Delta q_{K})<\Pi_{K-1,K}(v^{\ast}%
(\alpha,\beta),0,\Delta q_{K}^{\ast}(\alpha,\beta)),
\]
%
%\[\begin{matrix}({\overline f}_{K-1}+{\overline f}_{K})\Delta q_{K-1}w_{K-1}-{\overline f}_{K-1}  c(q_{K-2}+\Delta q_{K-1}) \\ +{\overline f}_{K}\left( \Delta q_{K}w_{K}-  c(q_{K-2}+\Delta q_{K-1}+\Delta q_{K})\right)\end{matrix}<\begin{matrix}g^*_{K}\left( \Delta q^*_{K}w^*_{K}-c(q_{K-2}+ \Delta q^*_{K})\right)\\-g^*_{K-1}c(q_{K-2}+\Delta q_{K-1})\end{matrix}\]
for every $\Delta q_{K-1},\ \Delta q_{K}>0$. We add $(\alpha,\beta)$ as an
argument because we will eventually vary these parameters; we don't add
$\gamma$ because the solution $\left(  v^{\ast}(\alpha,\beta),\Delta
q_{K}^{\ast}(\alpha,\beta)\right)  $\ evidently does not depend on the
constant $\gamma$.

We now analyze the entire class of convex cost functions with $c^{\prime
\prime\prime}\left(  q\right)  \geq0$. We assume that the optimal mechanism
consists of multiple items and reach a contradiction. We denote by
$\widehat{c}(q)$ a linear-quadratic cost function (as in \eqref{eq:c}). We
note that $c(q)$ and $\widehat{c}(q)$ intersect at most three times.
Furthermore, if $c\left(  q\right)  $ and $\widehat{c}(q)$\ are equal at
qualities $q_{1},\ q_{2},\ q_{3},$ then the difference $\widehat{c}%
(q)-c(q)$\ satisfies:
\[
\widehat{c}(q)-c(q)\geq0\iff q\in(-\infty,q_{1}]\cup\lbrack q_{2},q_{3}].
\]
We use this for the following result.

\begin{lemma}
[Convex Marginal Cost Functions]\label{lem:costs}\quad\newline For every
convex cost function with $c^{\prime\prime\prime}\left(  q\right)  \geq0$ and
for every $\left(  q_{K-2},q_{K-1},q_{K}\right)  $ with $q_{K-2}\leq
q_{K-1}\leq q_{K}$, there exists $(\alpha,\beta,\gamma)$ satisfying
$c(q_{K-2})=\widehat{c}(q_{K-2})$ and one of the following three conditions:

\begin{enumerate}
\item $c(q_{K})=\widehat{c}(q_{K});\ c(q_{K-1})=\widehat{c}(q_{K-1}%
);\ c(q^{\ast}(\alpha,\beta))<\widehat{c}(q^{\ast}(\alpha,\beta))$;

\item $c(q_{K})>\widehat{c}(q_{K});\ c(q_{K-1})=\widehat{c}(q_{K-1}%
);\ c(q^{\ast}(\alpha,\beta))=\widehat{c}(q^{\ast}(\alpha,\beta))$;

\item $c(q_{K})=\widehat{c}(q_{K});\ c(q_{K-1})>\widehat{c}(q_{K-1}%
);\ c(q^{\ast}(\alpha,\beta))=\widehat{c}(q^{\ast}(\alpha,\beta))$.
\end{enumerate}
\end{lemma}

\begin{proof}
We begin by considering $\alpha,\beta,\gamma$ chosen such that:
\begin{equation}
\widehat{c}(q_{K-2})=c(q_{K-2})\text{; }\widehat{c}(q_{K-1})=c(q_{K-1}%
);\ \widehat{c}(q_{K})=c(q_{K}).\label{eq:condit}%
\end{equation}
For this, we need to set the parameters $\alpha,\beta,\gamma$ as follows:
%\begin{align}
%\beta =& c^{\prime \prime }(q_{K-1});  \label{eq:abg} \\
%\alpha =& \frac{c(q_{K})-c(q_{K-1})}{q_{K}-q_{K-1}}-\frac{\beta }{2}%
%(q_{K-1}+q_{K});  \notag \\
%\gamma =& \frac{\beta q_{K}q_{K-1}}{2}+\frac{q_{K}c(q_{K-1})-c(q_{K})q_{K-1}%
%}{(q_{K}-q_{K-1})}.  \notag
%\end{align}%
\begin{align*}
\alpha= &  \frac{c(q_{K})\left(  q_{K-2}^{2}-q_{K-1}^{2}\right)
+c(q_{K-1})(q_{K}^{2}-q_{K-2}^{2})+c(q_{K-2})\left(  q_{K-1}^{2}-q_{K}%
^{2}\right)  }{(q_{K}-q_{K-1})(q_{K}-q_{K-2})(q_{K-1}-q_{K-2})},\\
\beta= &  \frac{2(c(q_{K})(q_{K-1}-q_{K-2})+c(q_{K-1})(q_{K-2}-q_{K}%
)+c(q_{K-2})(q_{K}-q_{K-1}))}{(q_{K}-q_{K-1})(q_{K}-q_{K-2})(q_{K-1}-q_{K-2}%
)},\\
\gamma= &  \frac{c(q_{K})q_{K-1}q_{K-2}(q_{K-1}-q_{K-2})+c(q_{K-1}%
)q_{K}q_{K-2}(q_{K-2}-q_{K})+c(q_{K-2})q_{K}q_{K-1}(q_{K}-q_{K-1})}%
{(q_{K}-q_{K-1})(q_{K}-q_{K-2})(q_{K-1}-q_{K-2})}.
\end{align*}
These are the coefficients one obtains from the interpolation of a
second-degree polynomial. Since $\widehat{c}$ is a linear-quadratic cost
function and since $c^{\prime\prime\prime}\geq0$, we have that for all $q\geq
q_{K-2}$:
\[
c(q)\leq\widehat{c}(q)\iff q\in\lbrack q_{K-1},q_{K}].
\]
In other words, $\widehat{c}$ is equal to $c$ at the qualities implemented by
the mechanism and exhibits higher costs at qualities that are in between these
two qualities and lower cost outside this interval. If
\[
q^{\ast}(\alpha,\beta)\in\lbrack q_{K-1},q_{K}],
\]
then we are in Case 1 of Lemma \ref{lem:costs}. We now show that, if $q^{\ast
}(\alpha,\beta)\not \in \lbrack q_{K-1},q_{K}]$, then we can find different
$\alpha,\beta,\gamma$ such that we are in Case 2 or 3 of Lemma \ref{lem:costs}.

Suppose that:
\begin{equation}
q^{\ast}(\alpha,\beta)<q_{K-1},\label{eq:case1}%
\end{equation}
where $(\alpha,\beta)$ satisfy \eqref{eq:condit}. We then need to find
different parameters $\alpha,\beta$. We consider parameters $\alpha,\beta$ as
a function of $q$ implicitly defined as follows:
\[
\widehat{c}(q_{K-2})=c(q_{K-2})\text{; }\widehat{c}(q_{K-1})=c(q_{K-1}%
);\ \widehat{c}(q)=c(q).
\]
We can write $\alpha,\beta,\gamma$ explicitly as before but replacing
$c(q_{K})$ with $c(q)$ and $q_{K}$ with $q$. Since $\alpha,\ \beta,\ \gamma$
are functions of $q$, we write $\alpha(q),\ \beta(q),\gamma(q)$ and observe
that they are continuous functions of $q$ (while some of the denominators
converge to 0 as $q\rightarrow q_{K-1},$ the limits exist). We also have
that:
\[
q^{\ast}(\alpha(q_{K}),\beta(q_{K}))-q_{K}<0\text{ and }q^{\ast}%
(\alpha(q_{K-2}),\beta(q_{K-2}))-q_{K-2}\geq0,
\]
where the first inequality follows from \eqref{eq:case1} and the second
inequality follows from the fact that $q^{\ast}$ by definition is larger than
$q_{K-2}$ (see \eqref{eq:opt1}) . Thus, following the intermediate value
theorem, there exists a $\widehat{q}\in\lbrack q_{K-2},q_{K}]$ such that:
$q^{\ast}(\alpha(\widehat{q}),\beta(\widehat{q}))=\widehat{q}.$ Furthermore,
note that $q_{K}>\max\{\widehat{q},q_{K-1}\}$, so we have that $\widehat{c}%
(q_{K})<c(q_{K})$. Thus, we are in Case~2 of Lemma \ref{lem:costs}.

Finally, if $q^{\ast}(\alpha,\beta)>q_{K},$\ we can find $\alpha,\beta,\gamma$
such that Case 3 is satisfied in an analogous way to the case when
\eqref{eq:case1} was satisfied. In particular, we consider parameters
$\alpha,\beta$ as functions of $q$ implicitly defined as follows:
\[
\widehat{c}(q_{K-2})=c(q_{K-2})\text{; }\widehat{c}(q)=c(q);\ \widehat{c}%
(q_{K})=c(q_{K}).
\]
And we can show there exists $\widehat{q}$ such that $q^{\ast}(\alpha
(\widehat{q}),\beta(\widehat{q}))=\widehat{q}$ and:
\[
c(q_{K})=\widehat{c}(q_{K});\ c(q_{K-1})>\widehat{c}(q_{K-1});\ c(q^{\ast
}(\alpha,\beta))=\widehat{c}(q^{\ast}(\alpha,\beta)).
\]
This concludes the proof.
\end{proof}

With Lemma \ \ref{lem:costs} we can extend the optimality result to convex
cost functions.

\begin{proposition}
[Optimality of Single-item Menu with Linear Density and Convex Cost]%
\label{lc}The optimal menu with linear decreasing density and $c^{\prime
\prime\prime}\left(  q\right)  \geq0$ is always a single-item menu.
\end{proposition}

\begin{proof}
We now suppose that the optimal mechanism satisfies $\Delta q_{K-1},\Delta
q_{K}>0$ and reach a contradiction. In the same manner as \eqref{eq:lt}, we
define:
\begin{align*}
& \widehat{\Pi}_{K-1,K}\triangleq\\
& ({\overline{f}}_{K-1}+{\overline{f}}_{K})\Delta q_{K-1}\overline{v}%
_{K-1}-{\overline{f}}_{K-1}\widehat{c}(q_{K-2}+\Delta q_{K-1})+{\overline{f}%
}_{K}\left(  \Delta q_{K}\overline{v}_{K}-\widehat{c}(q_{K-2}+\Delta
q_{K-1}+\Delta q_{K})\right)  .
\end{align*}
Now $c\left(  \cdot\right)  $ is the true cost function, which satisfied
$c^{\prime\prime\prime}\left(  \cdot\right)  \geq0$, and $\widehat{c}\left(
\cdot\right)  $ is a linear-quadratic cost. So $\widehat{\Pi}_{K-1,K} $ is
computed as ${\Pi}_{K-1,K}$ but using the linear-quadratic cost instead of the
true cost. With a linear-quadratic cost the optimal mechanism is a single-item
menu and thus:
%\begin{equation*}
%\widehat{\Pi}_{K,K-1}<({\overline f}_{K-1}+{\overline f}_{K})(\frac{({\overline f}_{K-1}w_{K-1}+{\overline f}_{K}w_{K})}{%
%({\overline f}_{K-1}+{\overline f}_{K})}-\widehat{c}(q^{\ast }(\alpha ,\beta )))
%\end{equation*}%
\[
\widehat{\Pi}_{K-1,K}(w_{K-1},\Delta q_{K-1},\Delta q_{K})<\widehat{\Pi
}_{K-1,K}(w_{K-1}^{\ast}(\alpha,\beta),0,\Delta q_{K}^{\ast}(\alpha,\beta)).
\]
We now consider the three cases in Lemma \ref{lem:costs}.

If we take $(\alpha,\beta)$ so that the first case in Lemma \ref{lem:costs}
holds, then we have:
\begin{align}
\Pi_{K-1,K}(w_{K-1},\Delta q_{K-1},\Delta q_{K})= &  \widehat{\Pi}%
_{K-1,K}(w_{K-1},\Delta q_{K-1},\Delta q_{K});\label{cost1}\\
{\Pi}_{K-1,K}(w_{K-1}^{\ast}(\alpha,\beta),0,\Delta q_{K}^{\ast}(\alpha
,\beta))> &  \widehat{\Pi}_{K-1,K}(w_{K-1}^{\ast}(\alpha,\beta),0,\Delta
q_{K}^{\ast}(\alpha,\beta)).\label{cost2}%
\end{align}
We thus have that:
\[
\Pi_{K-1,K}(w_{K-1},\Delta q_{K-1},\Delta q_{K})<{\Pi}_{K-1,K}(w_{K-1}^{\ast
}(\alpha,\beta),0,\Delta q_{K}^{\ast}(\alpha,\beta)),
\]
which contradicts the assumption that the multi-item mechanism is optimal.

If we consider $(\alpha,\beta)$ that satisfy the cases 2 or 3 of Lemma
\ref{lem:costs}, then the argument is analogous but \eqref{cost2} will hold
with equality and \eqref{cost1} will hold with strict inequality.
%\end{proof}
%\end{proof}

\end{proof}

We now analyze distributions with modest tails. We begin with an important
property of the optimal single-item mechanism when the distribution has a
linearly-decreasing density. For these distributions, the first-order
conditions \eqref{foc} are necessary and sufficient conditions for optimality
when $c^{\prime\prime\prime}\left(  \cdot\right)  \geq0.$

\begin{proposition}
[Sufficient Conditions for Optimality]\label{prop:suff}\quad\newline If
$c^{\prime\prime\prime}\left(  q\right)  \geq0$, the distribution is
$L(v;v_{l},v_{h})$, and $(\widehat{q},\widehat{v})$ satisfy the first-order
condition \eqref{foc}, then $(\widehat{q},\widehat{v})$ solves \eqref{opt},
i.e. $(\widehat{q},\widehat{v})=(q^{\ast},v^{\ast})$.
\end{proposition}

\begin{proof}
When the distribution is linearly decreasing, we have that:
\[
\mathbb{E}[v\mid v\geq\widehat{v}]=\frac{2\widehat{v}+v_{h}}{3}.
\]
Hence, if $(\widehat{q},\widehat{v})$ satisfy the first-order condition
\eqref{foc} we have that:
\[
\frac{2\widehat{v}+v_{h}}{3}=c^{\prime}(\widehat{q})\quad\text{ and }%
\quad\widehat{v}=\frac{c(\widehat{q})}{\widehat{q}}.
\]
We have that:
\[
v_{h}=3c^{\prime}(\widehat{q})-2\frac{c(\widehat{q})}{\widehat{q}}.
\]
We now note that:
\[
\frac{d}{dq}\left(  3c^{\prime}(q)-2\frac{c(q)}{q}\right)  =\frac{2}{q}\left(
c^{\prime\prime}(q)q-c^{\prime}(q)+\frac{c(q)}{q}\right)  +c^{\prime\prime
}(q).
\]
If $c^{\prime\prime\prime}(q)\geq0$ we have that $c^{\prime\prime}(q)q\geq
c^{\prime}(q)$ and hence:
\[
\frac{d}{dq}\left(  3c^{\prime}(q)-2\frac{c(q)}{q}\right)  >0.
\]
Thus, there is a unique pair $(\widehat{v},\widehat{q})$ such that the
first-order condition is satisfied.

We verify that the first-order condition is sufficient for optimality. For
this, we check that the solution is always interior, and since there is only
one point that satisfies the first-order condition, this must be the optimum.
We first note that $\widehat{q}\in\{0,\infty\}$ is clearly never optimal. It
is also easy to see that $v=v_{h}$ cannot be an optimum as then the objective
function of \eqref{opt} is 0. We finally note that $v^{\ast}=0 $ is never
optimal, which can be checked by noting that the first-order condition with
respect to the cutoff gives $c(q^{\ast})\leq v^{\ast}q^{\ast}$. Hence, the
solution is always interior and it must be the only point that satisfies the
first-order conditions.
\end{proof}
%We now generalize the result in the previous section to all distributions that have a modest right tail. For this we denote by $v^*$ the cutoff of the optimal single-item mechanism. The optimal single-item menu solves:
%\begin{equation*}
%(q^{\ast },v^{\ast })\in \underset{v,q}{\arg \max }\mathbb{P}[v^{\prime
%}\geq v](\mathbb{E}[v^{\prime }\mid v^{\prime }\geq v]q-\frac{1}{2}q^{2}).
%\end{equation*}%
%Due to the quadratic cost, the first-order condition with respect to $q$
%implies that the optimal quality is given by: $q^{\ast }=\mathbb{E}[v\mid
%v\geq v^{\ast }].$ Hence, we get that $v^{\ast }$ solves:
%\begin{equation*}
%v^{\ast }\in \underset{v,q}{\arg \max }\ \mathbb{P}[v^{\prime }\geq v]\frac{%
%\mathbb{E}[v^{\prime }\mid v^{\prime }\geq v]}{2}.
%\end{equation*}%
%The first-order condition with respect to $v^{\ast }$ is given by:
%\begin{equation*}
%(\mathbb{E}[v\mid v\geq v^{\ast }]-2v^{\ast })\leq 0,
%\end{equation*}%
%with equality if $v^{\ast }>\underline{v}$. We summarize the results in the
%following proposition.
%\begin{proposition}[Optimal Single Item Mechanism]
%\label{lem:single}\quad \newline
%If the cost is $c(q)=q^{2}/2$, the optimal single-item menu is:
%\begin{equation*}
%(q^{\ast },p^{\ast })=(\mathbb{E}[v\mid v\geq v^{\ast }],\mathbb{E}[v\mid
%v\geq v^{\ast }]^{2}).
%\end{equation*}%
%Whenever $v^{\ast }>\underline{v}$, the optimal cutoff satisfies $v^{\ast }$
%satisfies:
%\begin{equation}
%2v^{\ast }=\mathbb{E}[v\mid v\geq {v}^{\ast }]  \label{eq:foc}
%\end{equation}
%\end{proposition}
%We now consider the following distribution with linearly decreasing density:
For a given distribution $F$, we now introduce two related distributions, one
generated by a linear decreasing density, and the other by a truncated version
of the former. These latter two distributions are constructed in such a way as
to allow us to compare the profit from the optimal mechanism under $F$ (which
we do not know) to the optimal mechanism under these two related
distributions. Jointly with a cost-dominating argument, we can then establish
the optimality of a single-item menu in a large class of environments.

Towards this end, we consider a distribution ${L}(v;\underline{z},\overline
{z})$ with a linearly-decreasing density where the lower and upper bounds of
the distribution $L$, namely $\underline{z},\overline{z}$, are chosen to
satisfy the following properties relative to the distribution $F$ and the
optimal single-item threshold $v^{\ast}$\ under $F$ given by (\ref{opt}):
\begin{equation}
{L}(v^{\ast};\underline{z},\overline{z})=F(v^{\ast})\text{ and }%
\mathbb{E}_{{L}}[v\mid v\geq v^{\ast}]=\mathbb{E}_{F}[v\mid v\geq v^{\ast
}],\label{cond}%
\end{equation}
where the subscripts in the expectation indicate the distribution used to
compute the expectation. Namely, at the threshold $v^{\ast}$, $L$ and $F$
obtain the same quantile, and the conditional expectation above the threshold
$v^{\ast}$ are identical. To satisfy these conditions, it is necessary to
set:
%\[\bar v'=2 \mathbb{E}_{F}[v\mid v\geq v^*]-v^*\text{ and }\underline v'=\frac{v^*+F(v^*)(\mathbb{E}_{F}[v\mid v\geq v^*]-2v^*)}{1-F(v^*)}.\]%
\begin{align*}
\underline{z}= &  3\mathbb{E}_{F}[v\mid v\geq v^{\ast}]-2v^{\ast}%
-\frac{3(\mathbb{E}_{F}[v\mid v\geq v^{\ast}]-v^{\ast})}{\sqrt{1-F(v^{\ast})}%
};\\
\overline{z}= &  3\mathbb{E}_{F}[v\mid v\geq v^{\ast}]-2v^{\ast}\text{.}%
\end{align*}
We also consider the following distribution $\widehat{F}\left(  v\right)  $:
\begin{equation}
\widehat{F}\left(  v\right)  =%
\begin{cases}
L(\widehat{v};\underline{z},\overline{z}), & \text{if }v\in\lbrack
0,\widehat{v}];\\
L(v;\underline{z},\overline{z}), & \text{if }v\in\lbrack\widehat{v}%
,\overline{z}];
\end{cases}
\label{eq:hf}%
\end{equation}
where $\widehat{v}$ is chosen such that:
\[
\int_{0}^{\infty}vdF(v)=\int_{0}^{\infty}vd\widehat{F}(v).
\]
In the proof of Lemma \ref{lem:cop} we will show that indeed such a
$\widehat{v}$ exists. Thus, $\widehat{F}\left(  v\right)  $ is constructed by
taking the mass of the lower tail of $L(v;\underline{z},\overline{z})$ and
moving it to $0$. In other words, $\widehat{F}$ is equal to $L(v;v_{l},v_{h})
$ for $v\geq\widehat{v}$, and $\widehat{F}$ has an atom of size $L(\widehat{v}%
;\underline{z},\overline{z})$ at $v=0$.

We can now relate these three distributions in terms of stochastic orders.

\begin{lemma}
[Distribution Comparison]\label{lem:cop}\quad\newline Distribution
$\widehat{F}$ is a mean-preserving spread of $F$ and $\widehat{F}$ is
first-order stochastically dominated by $L(v;\underline{z},\overline{z})$.
\end{lemma}

\begin{proof}
We first compare $L(v,\bar{z},\underline{z})$ with $F.$ Since $f$ satisfies
\eqref{eq:qc} and $L^{\prime}(v,\bar{z},\underline{z})$ is linearly
decreasing, we must have that $f$ and $L^{\prime}(v,\bar{z},\underline{z})$
intersect at most twice. However, by construction $L$ is constructed to
satisfy \eqref{cond}, so they intersect exactly twice at two values
$v_{1},v_{2}\geq v^{\ast}$.
%Note that:
%\begin{align*}
%\frac{\int_{q}^{1}L^{-1}(v;\overline{z},\underline{v})dv}{1-q}=& \mu ^{\ast
%}+2\left( 1-\sqrt{\frac{1-q}{1-F(v^{\ast })}}\right) (\mu ^{\ast }-v^{\ast
%}); \\
%\frac{\int_{q}^{1}F^{-1}(v)dv}{1-q}=& \mathbb{E}[v\mid F(v)\geq q].
%\end{align*}%
Hence, \eqref{eq:qc} implies that for all $v^{\prime}\in\lbrack0,\infty)$:
\begin{equation}
\int_{F(v^{\prime})}^{1}F^{-1}(v)dv\leq\int_{F(v^{\prime})}^{1}L^{-1}%
(v;\overline{z},v_{l})dv.\label{eq:sd}%
\end{equation}
If the inequality is satisfied with equality for $v^{\prime}=0$, we have that
$\widehat{v}=\underline{z}$ and, otherwise, $\widehat{v}>\underline{z}$ (where
$\widehat{v}$ is used to construct $\widehat{F}$ in \eqref{eq:hf}). Since
$\widehat{F}$ is constructed by taking the mass of the lower tail of
$L(v;\underline{z},\overline{z})$ and moving it to 0, it is transparent that
$\widehat{F}$ is first-order stochastically dominated by $L(v;\underline{z}%
,\overline{z})$. We have that \eqref{eq:sd} implies that for all $v^{\prime
}\geq\widehat{v}$:
\[
\int_{F(v^{\prime})}^{1}F^{-1}(v)dv\leq\int_{F(v^{\prime})}^{1}\widehat{F}%
^{-1}(v)dv.
\]
We also have that by construction $\widehat{F}$ has the same mean as $F$. It
then follows that for all $v^{\prime}$
\[
\int_{F(v^{\prime})}^{1}F^{-1}(v)dv\leq\int_{F(v^{\prime})}^{1}\widehat{F}%
^{-1}(v)dv,
\]
with equality for $v^{\prime}=0.$ Hence, $\widehat{F}$ is a mean-preserving
spread of $F$ (see Theorem 3.A.5 in \cite{shsh07}).
%Since $\ell(v;\bar v^{\prime },\underline{v}^{\prime })$ is linear and $f$
%is quasi-concave and concave on the decreasing part, we must have that $%
%\ell(v;\bar v^{\prime },\underline{v}^{\prime })$ and $f$ intersect at most
%twice. However, \eqref{cond} implies that $\ell(v;\bar v^{\prime },%
%\underline{v}^{\prime })$ and $f$ intersect exactly twice. More precisely,
%there exists $v^*<v_1<v_2$ such that:
%\begin{equation}  \label{eq:sign}
%sign(\ell(v;\bar v^{\prime },\underline{v}^{\prime })-f(v))=%
%\begin{cases}
%- & \text{if }v\in[v_1,v_2]; \\
%+ & \text{if }v\not\in[v_1,v_2],%
%\end{cases}%
%\end{equation}
%where $sign()$ is the sign of the difference. This implies that $\widehat
%f(v)\leq \ell(v;\bar v^{\prime },\underline{v}^{\prime })$ for all $v$, so $%
%L(v;\bar v;\underline v^{\prime })$ first-order stochastically dominates $F$
%(see Theorem 1.A.12 in \cite{shsh07}). And, this implies that $\widehat F$ has
%the same mean as $F$ and the difference between their densities change sign
%twice (as in \eqref{eq:sign}) so $\widehat F$ is a mean-preserving spread of $F$
%(see Theorem 3.A.44 in \cite{shsh07}).

\end{proof}

\medskip

We can now conclude the proof by establishing Theorem \ref{th:22}.

\medskip

\begin{proof}
[Final Step of the Proof of Theorem \ref{th:22}]We first verify that the
optimal single-item mechanism when the distribution is $L(v;\underline{z}%
,\overline{z})$ is the same as when the distribution is $F$. By construction
of $L(v;\underline{z},\overline{z})$, the first-order condition that is
satisfied for $F$ is also satisfied for $L(v;\underline{z},\overline{z})$.
Following Proposition \ref{prop:suff}, for the linearly decreasing density the
first-order condition is sufficient for optimality, and thus $\left(  v^{\ast
},q^{\ast}\right)  $ given by (\ref{opt}) do in fact form the optimal
mechanism for $L$. We have that $L(v;\underline{z},\overline{z})$ generates at
least as much profit as $\widehat{F}$, and $\widehat{F}$ generates at least as
much profit as $F$. Since the optimal mechanism for distribution
$L(v;\underline{z},\overline{z})$ is a single-item mechanism, and this
mechanisms generates the same profit (by construction) under distribution $F$,
this must also be the optimal mechanism under distribution $F$.
\end{proof}

\bigskip

\newpage

\bibliographystyle{econometrica}
\bibliography{general}

\end{document}